\documentclass[traditabstract]{aa} 
\usepackage[pdftex]{graphicx}
\usepackage{breqn}
\usepackage[flushleft]{threeparttable}

\def\la{\lower.5ex\hbox{$\; \buildrel < \over \sim \;$}}
\def\ga{\lower.5ex\hbox{$\; \buildrel > \over \sim \;$}}

\newcommand{\ratioo} {N({\rm H}_2) / I_{\rm CO}}
\begin{document}
   \title{Low star formation efficiency due to turbulent adiabatic compression in the Taffy bridge\thanks{Based on observations carried out with the IRAM Plateau de Bure Interferometer. IRAM is supported by INSU/CNRS (France), MPG (Germany) and IGN (Spain).}}

   \author{B.~Vollmer\inst{1}, J.~Braine\inst{2}, B.~Mazzilli-Ciraulo\inst{1,3}, B.~Schneider\inst{1,4}}

   \institute{Universit\'e de Strasbourg, CNRS, Observatoire astronomique de Strasbourg, UMR 7550, F-67000 Strasbourg, France \and 
          Laboratoire d'astrophysique de Bordeaux, Univ. Bordeaux, CNRS, B18N, all\'e Geoffroy Saint-Hilaire, 33615 Pessac, France \and
          LERMA, Observatoire de Paris, PSL Research University, CNRS, Universit\'e de Sorbonne, UPMC, Paris, 75014, France \and
   AIM, CEA, CNRS, Universit\'e Paris-Saclay, Universit\'e Paris Diderot, Sorbonne Paris Cit\'e, F-91191, Gif-sur-Yvette, France}

   \date{Received ; accepted }


  \abstract
{The Taffy system (UGC~12914/15) consists of two massive spiral galaxies which had a head-on collision about $20$~Myr ago.
It represents an ideal laboratory to study the reaction of the interstellar medium to a high-speed ($\sim 1000$~km\,s$^{-1}$) gas-gas collision.
New sensitive, high-resolution ($2.7''$ or $\sim 800$~pc) CO(1-0) observations of the Taffy system with the IRAM Plateau de Bure Interferometer are presented. 
The total CO luminosity of the Taffy system detected with the PdBI is $L_{\rm CO,tot}=4.8 \times 10^9$~K\,km\,s$^{-1}$pc$^2$, 
60\,\% of the CO luminosity found with the IRAM 30m telescope.
About $25$\,\% of the total interferometric CO luminosity stems from the bridge region.
Assuming a Galactic $\ratioo$ conversion factor for the galactic disks and a third of this value for
the bridge gas, about $10$\,\% of the molecular gas mass is located in the bridge region.
The giant H{\sc ii} region close to UGC~12915 is located at the northern edge of the high-surface brightness giant molecular
cloud association (GMA), which has the highest velocity dispersion among the bridge GMAs. 
The bridge GMAs are clearly not virialized because of their high velocity dispersion.
Three dynamical models are presented and while no single model reproduces all of the observed features, they are
all present in at least one of the models. Most of the bridge gas detected in CO does not form stars.
We suggest that turbulent adiabatic compression is responsible for the exceptionally high velocity dispersion of the molecular ISM 
and the suppression of star formation in the Taffy bridge. In this scenario the turbulent velocity dispersion of the largest eddies and
turbulent substructures/clouds increase such that giant molecular clouds are no longer in global virial equilibrium.
The increase of the virial parameter leads to a decrease of the star formation efficiency.
The suppression of star formation caused by turbulent adiabatic compression was implemented in the dynamical simulations
and decreased the star formation rate in the bridge region by $\sim 90$\,\%. 
Most of the low-surface density, CO-emitting gas
will disperse without forming stars but some of the high-density gas will probably collapse and form dense
star clusters, such as the luminous H{\sc ii} region close to UGC~12915.
We suggest that globular clusters and super star clusters formed and still form through the gravitational
collapse of gas previously compressed by turbulent adiabatic compression during galaxy interactions.
}


   \authorrunning{Vollmer et al.}

   \maketitle
%

\section{Introduction\label{sec:introduction}}

\begin{figure}[!ht]
  \centering
  \hspace*{1.5cm}
  \resizebox{7cm}{!}{\includegraphics{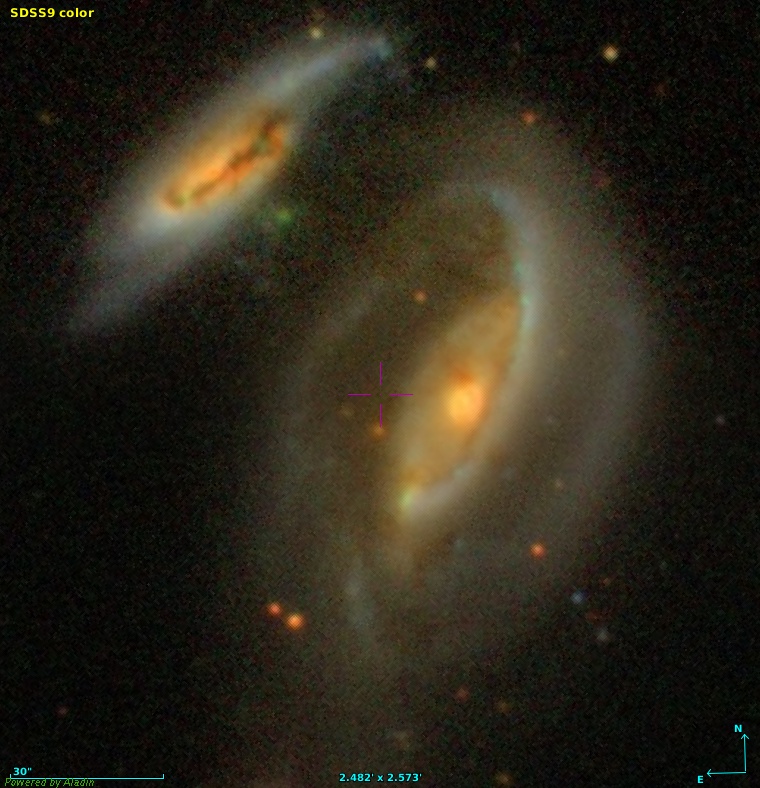}}
  \resizebox{\hsize}{!}{\includegraphics{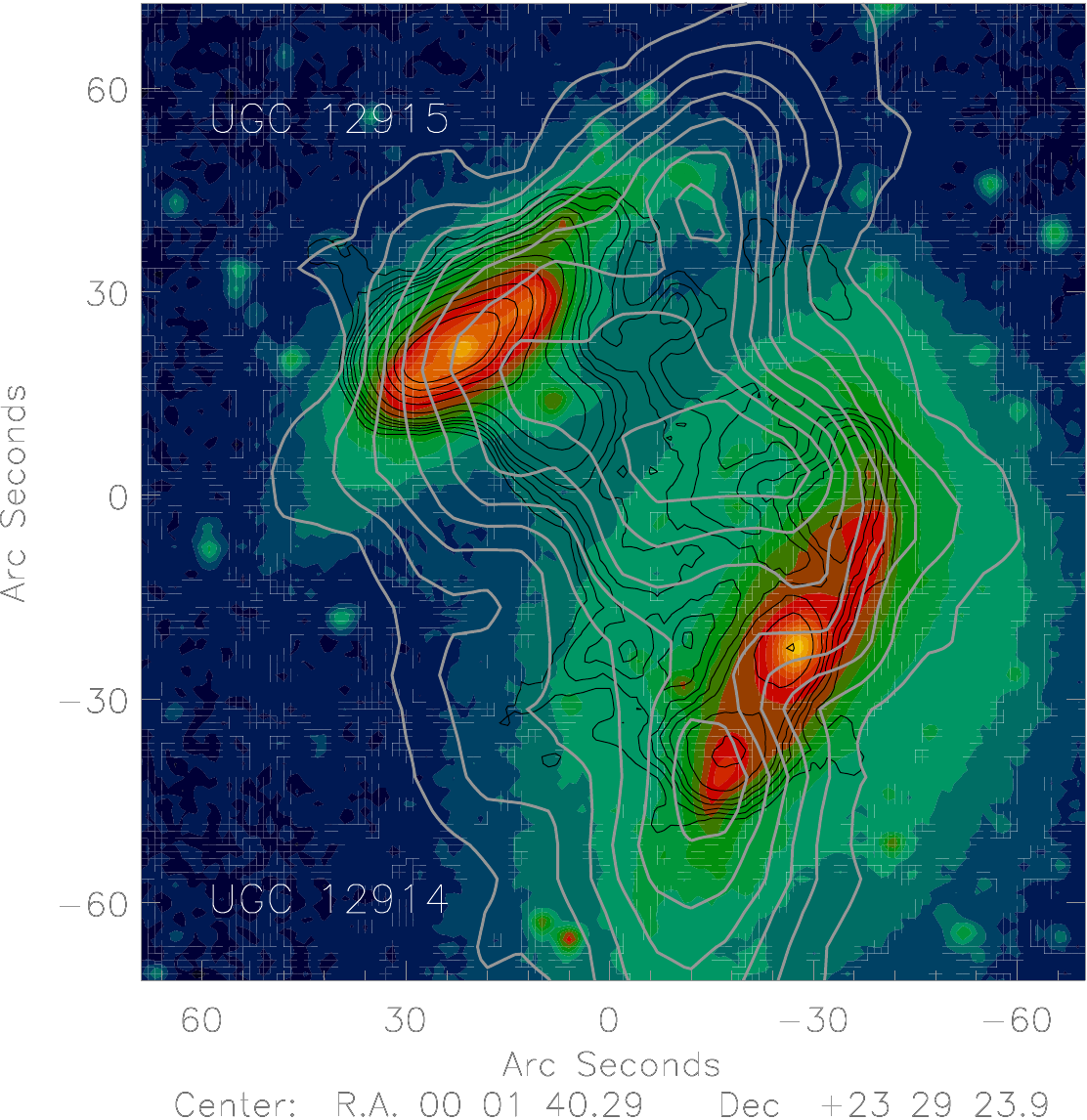}}
  \caption{The Taffy system UGC~12914/15. Upper panel: SDSS color image. Lower panel: color: stellar content (Spitzer 3.6~$\mu$m emission), 
    light grey contours: H{\sc i} emission (Condon et al. 1993), black contours: CO emission (Gao et al. 2003).
  \label{fig:cohistars}}
\end{figure}

Head-on collisions between spiral galaxies represent an ideal laboratory to study the behavior of the interstellar medium (ISM)
under extreme conditions. During the collision the interstellar media of both galactic disks collide, heat up, and exchange
momentum. In merging galaxy pairs, an ISM-ISM collision occurs towards the end of the interaction process (see, e.g., Renaud et al. 2015
or di Matteo et al. 2008). The Taffy system (UGC~12914/15; Fig.~\ref{fig:cohistars}) is a special case because both spiral galaxies are 
particularly massive, 
were gas-rich before the collision, and collided at high speed ($\sim 1000$~km\,s$^{-1}$; Condon et al. 1993, Vollmer et al. 2012).
We observe the galaxy pair about $20$~Myr after the impact that occurred in the plane of the sky.
The transverse velocity difference at the present time is $650$~km\,s$^{-1}$.

The Taffy system attracted attention through its strong radio synchrotron bridge (Condon et al. 1993), a very 
unusual feature.  The bridge is H{\sc i}-rich and was subsequently found to be rich in molecular 
gas as well through CO observations (Gao et al. 2003, Braine et al. 2003).  Dust appears to be underabundant
with respect to gas in the bridge (Zink et al. 2000, Zhu et al. 2007), presumably due to grain ablation
during the collision. The system contains about $1.5 \times 10^{10}$~M$_{\odot}$ of H{\sc i} and a similar quantity of 
molecular gas, dependent on the $\ratioo$ conversion factor from CO emission to H$_2$ column
density.  Some 10 -- 20\,\% of the gas is in the bridge, making it at least as rich in molecular gas as the 
entire Milky Way.
The ionized gas is highly disturbed kinematically, with gas spread in two main filaments between the two galaxies. 
Hot, X-ray emitting gas that has presumably been shock heated during the collision, is also present in the bridge region (Appleton et al. 2015).
This hot and tenuous gas is spatially more correlated with the low density atomic gas and seems to avoid the high density 
molecular gas.

The head-on collision of the Taffy system was simulated by Vollmer et al. (2012) with a model which includes a collisionless 
(halo and stellar particles) and a collisional (gas) component. 
A wealth of observational characteristics are available for the comparison with simulations:
a distorted stellar distribution, a prominent H{\sc i} and CO gas bridge with large linewidths and H{\sc i} double-line profiles,
and a large-scale magnetic field with projected field vectors parallel to the bridge.
Since these authors could not find a single simulation which reproduces all observed characteristics, they presented two ``best-fit'' simulations.
The first simulation better reproduced the H{\sc i} and CO line profiles of the bridge region (Braine et al. 2003), whereas the second simulation
better reproduced the stellar distribution of UGC~12915, the symmetric gas velocity fields of the galactic disks,
the projected magnetic field vectors in the bridge region, and the distribution of the 6~cm polarized radio continuum emission (Condon et al. 1993).
The stellar distribution of the model secondary galaxy is more
distorted than that of UGC~12915. These models were successful in producing (1) the prominent H{\sc i} and CO gas bridge,
(2) the offset of the CO emission to the south with respect to the H{\sc i} emission in the bridge region,
(3) the gas symmetric velocity fields in the galactic disks, (4) the isovelocity contours of the CO velocity field which are parallel to the bridge,
(5) the H{\sc i} double-line profiles in the disk region, (6) the large gas linewidths ($100$-$200$~km\,s${-1}$) in the bridge region, 
(7) the velocity separation between the double lines ($\sim 330$~km\,s$^{-1}$), (8) the high field strength of the regular magnetic field in the 
bridge region, (9) the projected magnetic field vectors, which are parallel to the bridge, (10) the offset of the maximum of the 6~cm polarized radio 
continuum emission to the south of the bridge, (11) and the strong total power emission from the disk.
The structure of the model gas bridge was found to be bimodal: a dense ($\sim 0.01$~M$_{\odot}$pc$^{-3}$) component
with a high velocity dispersion $> 100$~km\,s$^{-1}$ and a less dense ($\sim 10^{-3}$~M$_{\odot}$pc$^{-3}$)
component with a smaller, but still high velocity dispersion $\sim 50$~km\,s$^{-1}$. The synchrotron lifetime of
relativistic electrons is only long enough to be consistent with the existence of the radio continuum bridge (Condon et al. 1993) 
for the less dense component. On the other hand, only the high-density gas undergoes a high enough mechanical energy input to produce
the observed strong emission of warm H$_{2}$ (Peterson et al. 2012).

The star formation efficiency of the molecular gas in the bridge region is at least two to three times smaller than
that of the molecular gas located within the galactic disks (Vollmer et al. 2012). There is one exception: a compact region of high star formation
is located about $15''$ or $4.4$~kpc\footnote{We use a distance of 60 Mpc for the Taffy galaxy system.} southwest of the center of UGC~12915. 
Despite low star formation rates in the bridge, the [C II] emission appears to be enhanced (Peterson et al. 2018)
consistent with shock and turbulent gas heating (Joshi et al. 2019).

In this article we present new high-resolution CO(1--0) observations of the Taffy system to better understand the distribution
and kinematics of the dense molecular gas. In addition, we investigate why the star formation efficiency with respect to the
molecular gas $(SFR/M_{\rm H_2}$) is so low in the gas bridge. To do so, the dynamical model of Vollmer et al. (2012) was modified to
include the effects of turbulent adiabatic compression and expansion. Both effects are able to temporarily suppress star formation in the
dense gas.

\section{Observations\label{sec:observations}}

Observations of the $^{12}$CO(1–-0) emission were carried out with the IRAM Plateau de Bure Interferometer (PdBI) in summer 2014
using all six antennas in C and D configuration. The system was covered by a mosaic of $11$ PdBI primary beams.
Each position was observed during $55$~min.
The bandpasses calibration was on 3C454.3 on May 30th and Nov 21st and on 1749+096 on May 29th.
Phase and amplitude calibrations were performed on 2319+272 (every day), 0007+171 (21 nov), and 0006+243 (May 29th and 30th).
The absolute flux scale was checked on MWC 349 every day. A total bandwidth of  $640$~MHz  with  a  spectral resolution of $2.5$~MHz was used. 
We reach an rms of $\sim 5$~mJy in $6.5$~km\,s$^{-1}$ wide velocity channels. Applying robust weighting in the mapping process, 
a beam size of $2.7''$ ($\sim 800$~pc) was derived. 

\section{Results\label{sec:results}}

\begin{figure}[!ht]
  \centering
  \resizebox{\hsize}{!}{\includegraphics{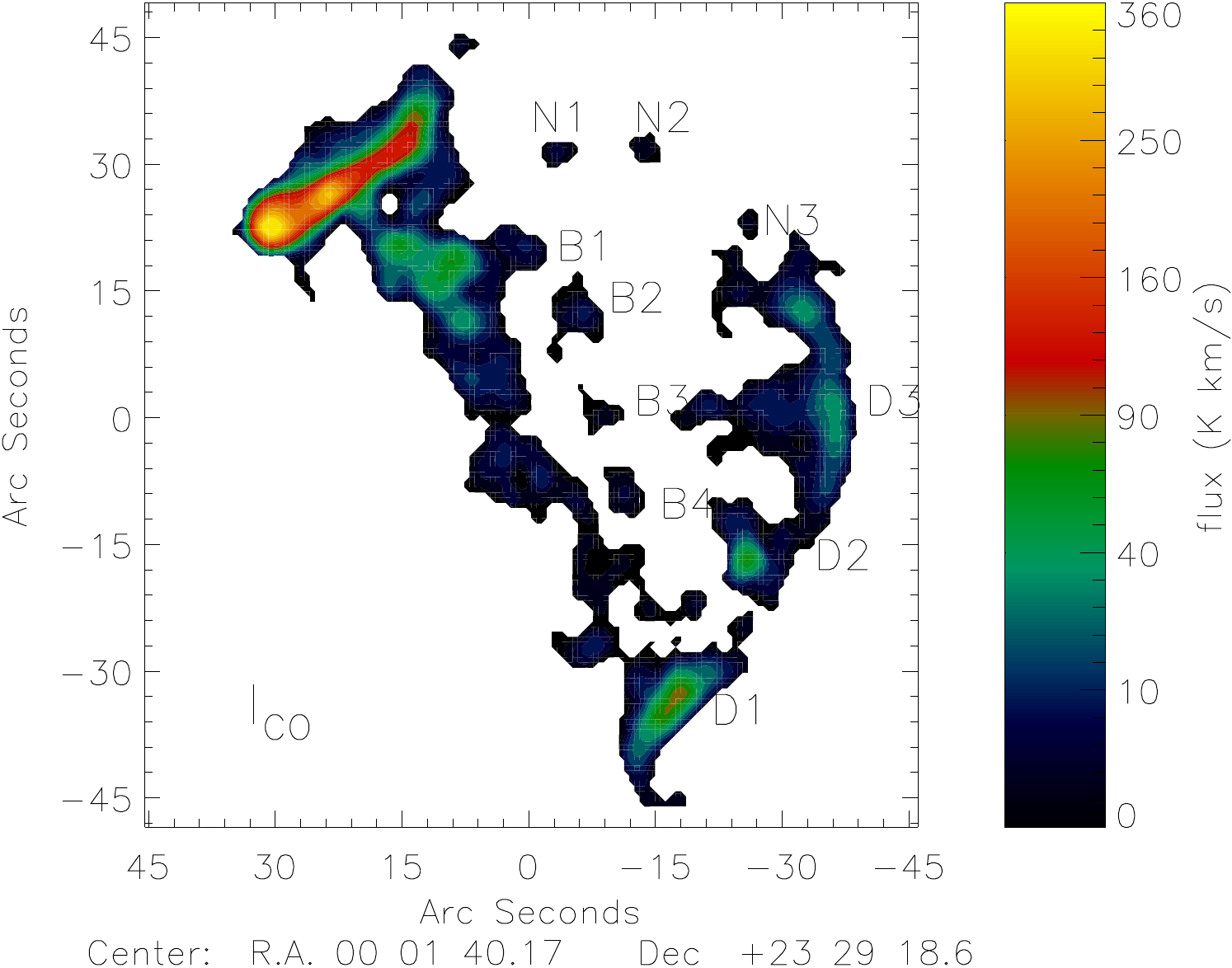}}
  \resizebox{\hsize}{!}{\includegraphics{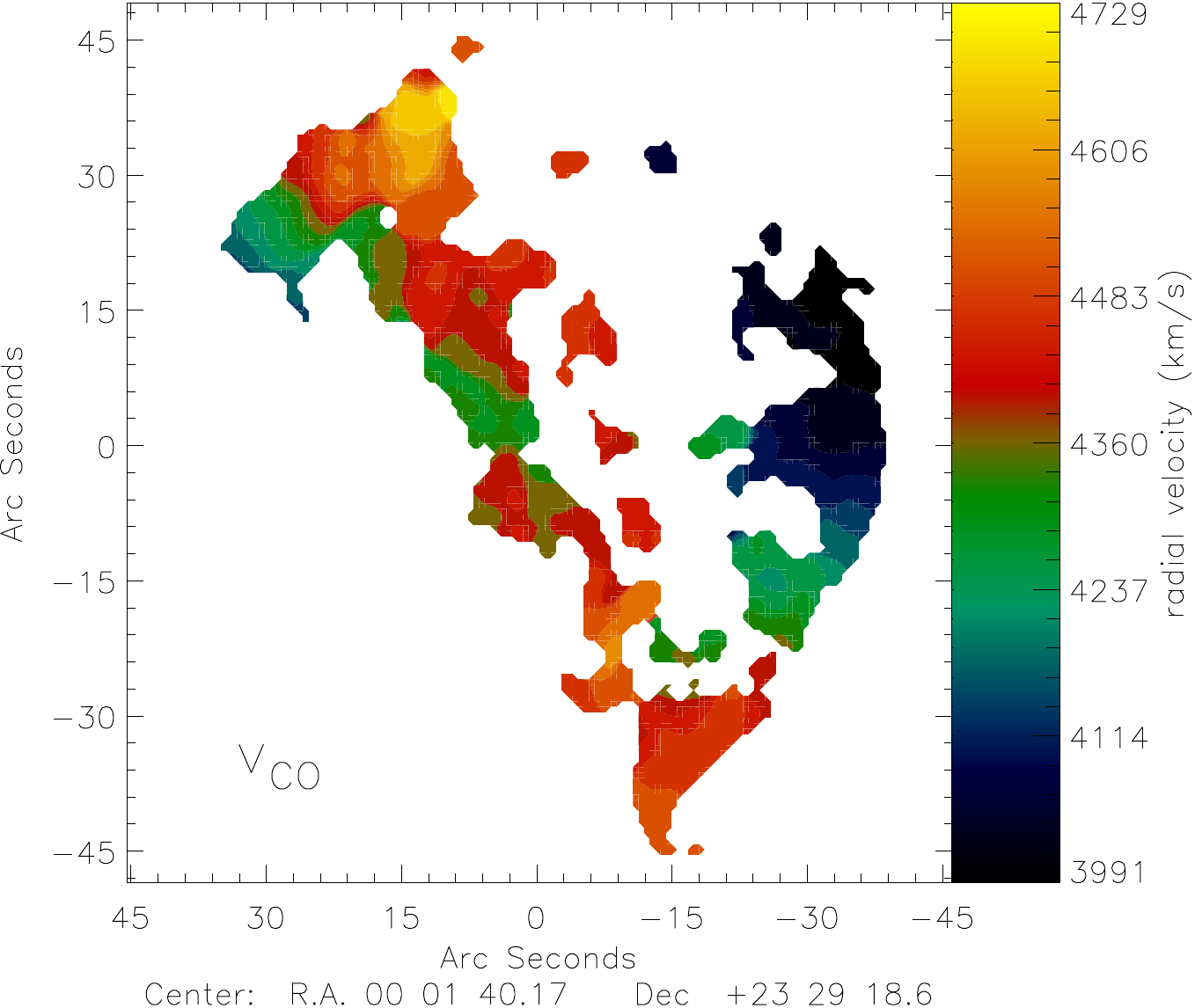}}
  \resizebox{\hsize}{!}{\includegraphics{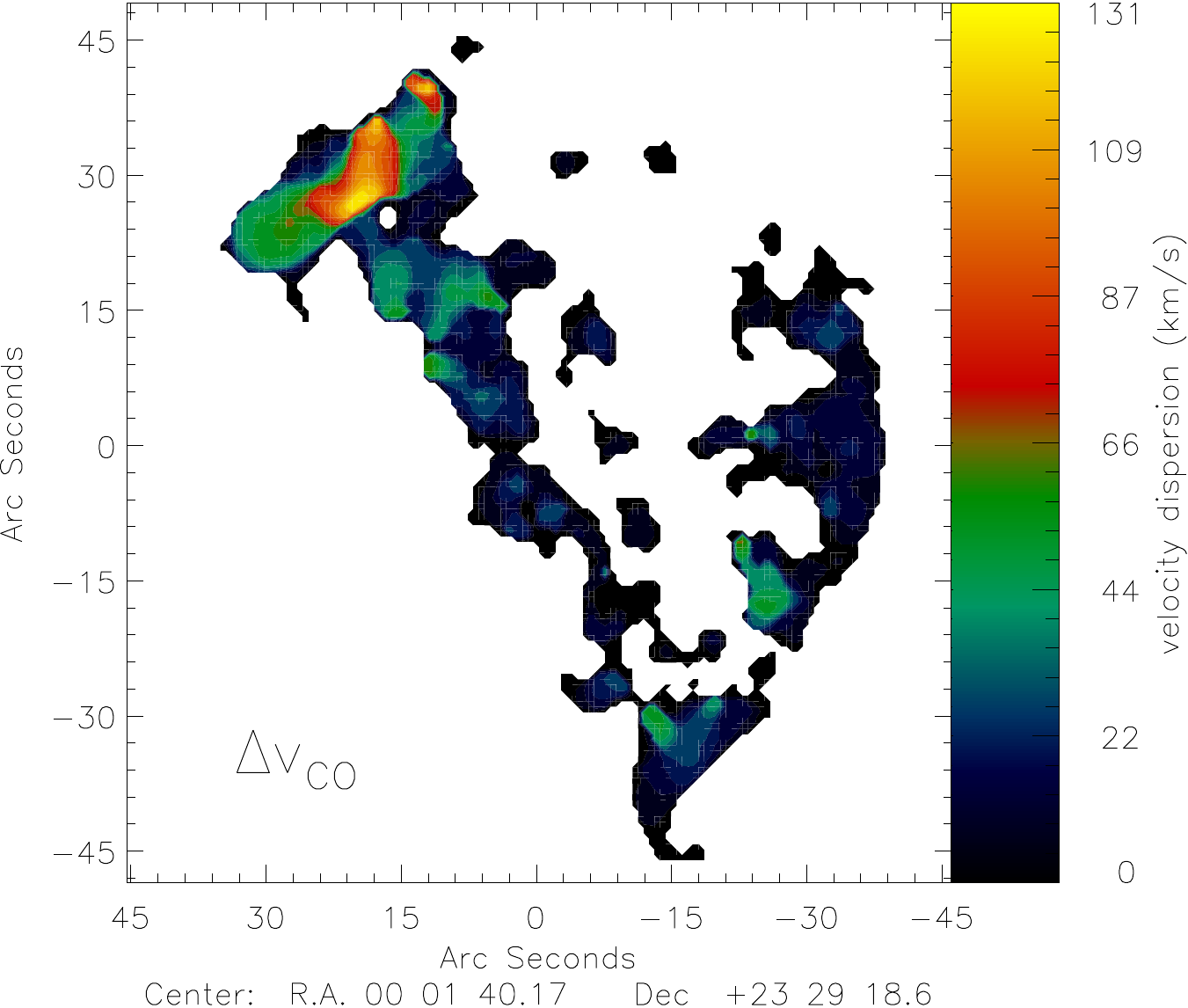}}
  \caption{CO(1-0) moment maps based on detections identified by CPROPS. Disk, bridge, and northern emission regions are labelled.
  \label{fig:taffy_obs_mom_final_mom0}}
\end{figure}

The CPROPS (CloudPROPertieS) software (Rosolowsky \& Leroy 2006) was used to identify and measure the properties 
(size, flux, velocity dispersion) of molecular
cloud associations (GMAs) in the $2.7''$ datacube. The CPROPS program first assigns contiguous regions of the datacube to individual clouds
and then computes the cloud properties (flux, radius, and velocity width) from the identified emission.
The algorithm ignores clouds smaller than a resolution element and does not decompose clouds smaller than two resolution elements.  
We used the modified CLUMPFIND\footnote{For the CLUMPFIND algorithm see Williams et al. (1994).} algorithm (ECLUMP) and required a peak of at 
least $1.5\,\sigma$ in every distinct cloud and at least two channels.
The CPROPS decomposition was used to produce the moment maps which are presented in Fig.~\ref{fig:taffy_obs_mom_final_mom0}.
As a consistency check, we cleaned the datacube with a velocity channel width of $6.5$~km\,s$^{-1}$ by iteratively (i)
boxcar averaging of each spectrum (width=$4$ channels), (ii) fitting Gaussians to the boxcar-averaged spectrum ($v_0$ is the central velocity),
(iii) all corresponding voxels in a 3D mask that are located between $v_0-$FWHM and $v_0+$FWHM are set to one, (iv) the Gaussian is subtracted 
from the boxcar-averaged spectrum, (v) the next Gaussian is fitted to the spectrum until its amplitude is smaller than 
$5\,\sigma$ of the boxcar-averaged spectrum, (vi) the 3D mask is applied to the initial datacube.
Moment maps were produced without clipping the datacube (Figs.~\ref{fig:taffy_obs_mom_final_mom0c}).
The moment maps based on CPROPS and the ``cleaned'' moment maps are consistent, the former being deeper as expected from the lower
CLUMPFIND limit of $1.5\,\sigma$. The cloud or GMA properties derived by CPROPS are shown in Table~\ref{tab:molent}.

\begin{table*}[!ht]
      \caption{Molecular entities from CPROPS.}
         \label{tab:molent}
      \[
         \begin{array}{lccccccc}
           \hline
               {\rm number} & {\rm total\ number} & {\rm RA} & {\rm DEC}  & {\rm radius} & {\rm velocity\ dispersion}  & {\rm flux} & {\rm region} \\
            & {\rm of\ pixels} &  ({\rm degrees)} & {\rm (degrees)} &  {\rm (pc)} & {\rm (km\,s}^{-1})  & {\rm (K\,km\,s}^{-1}{\rm pc}^2) & \\
           \hline
            1 &  545 &  0.42114 & 23.4988 &  511.47 & 14.18 & 4.29E+07 & {\rm UGC\ 12915} \\
            2 & 1011 &  0.42180 & 23.4981 &  568.03 & 16.88 & 1.02E+08 & {\rm UGC\ 12915} \\
            3 & 1425 &  0.42218 & 23.4973 & 1011.25 & 15.93 & 1.62E+08 & {\rm UGC\ 12915} \\
            4 & 2297 &  0.42189 & 23.4975 & 1023.43 & 31.56 & 1.89E+08 & {\rm UGC\ 12915} \\
            5 & 2558 &  0.42401 & 23.4964 &  772.34 & 44.94 & 2.43E+08 & {\rm UGC\ 12915} \\
            6 & 1271 &  0.42102 & 23.4959 &  928.04 & 21.25 & 8.21E+07 &     {\rm bridge} \\
            7 &  238 &  0.41837 & 23.4943 &    <300 & 12.99 & 1.28E+07 &     {\rm bridge} \\
            8 & 1117 &  0.42283 & 23.4968 &  647.71 & 29.10 & 8.99E+07 & {\rm UGC\ 12915} \\
            9 & 2228 &  0.42013 & 23.4936 &  764.19 & 50.21 & 1.66E+08 &     {\rm bridge} \\
            10 & 1361 &  0.42100 & 23.4929 &  634.85 & 30.11 & 1.06E+08 &     {\rm bridge} \\
            11 &  465 &  0.41774 & 23.4939 &  728.04 & 14.64 & 2.55E+07 &     {\rm bridge} \\
            12 &  108 &  0.41811 & 23.4906 &    <300 & 18.04 & 6.71E+06 &     {\rm bridge} \\
            13 & 1994 &  0.42213 & 23.4939 &  952.07 & 30.94 & 1.51E+08 &     {\rm bridge} \\
            14 &  251 &  0.41844 & 23.4903 &  415.32 & 15.90 & 1.39E+07 &     {\rm bridge} \\
            15 & 1789 &  0.41984 & 23.4917 &  970.35 & 23.51 & 1.29E+08 &     {\rm bridge} \\
            16 & 4934 &  0.42521 & 23.4956 & 1117.62 & 65.40 & 6.44E+08 & {\rm UGC\ 12915} \\
            17 &  266 &  0.41870 & 23.4900 &  664.70 & 11.04 & 1.50E+07 &     {\rm bridge} \\
            18 & 1710 &  0.41928 & 23.4896 & 1126.03 & 38.68 & 8.90E+07 &     {\rm bridge} \\
            19 &  241 &  0.41960 & 23.4931 &  540.82 & 19.70 & 1.46E+07 &     {\rm bridge} \\
            20 &  703 &  0.42252 & 23.4937 &  803.97 & 19.81 & 3.74E+07 &     {\rm bridge} \\
            21 & 2203 &  0.42356 & 23.4960 & 1167.63 & 44.71 & 1.77E+08 & {\rm UGC\ 12915} \\
            22 &   91 &  0.42224 & 23.4925 &    <300 & 24.26 & 6.46E+06 &     {\rm bridge} \\
            23 & 4616 &  0.42654 & 23.4948 &  826.16 & 50.50 & 7.31E+08 & {\rm UGC\ 12915} \\
            24 &  216 &  0.41495 & 23.4819 &    <300 & 10.53 & 1.02E+07 &     {\rm bridge} \\
            25 &  173 &  0.41409 & 23.4835 &  547.85 &  7.18 & 1.01E+07 &     {\rm bridge} \\
            26 &   61 &  0.41532 & 23.4828 &   81.03 &  7.20 & 2.78E+06 &     {\rm bridge} \\
            27 &  100 &  0.41535 & 23.4831 &  747.03 &  7.08 & 7.17E+06 &     {\rm bridge} \\
            28 &  290 &  0.41527 & 23.4808 &    <300 & 26.86 & 1.79E+07 &     {\rm bridge} \\
            29 & 2111 &  0.41181 & 23.4795 &  954.00 & 32.61 & 2.17E+08 & {\rm UGC\ 12914} \\
            30 &  266 &  0.41543 & 23.4809 &  649.93 & 18.41 & 1.76E+07 &     {\rm bridge} \\
            31 &  187 &  0.41557 & 23.4838 &    <300 & 16.24 & 1.08E+07 &     {\rm bridge} \\
            32 & 1744 &  0.41296 & 23.4783 &  996.61 & 33.99 & 2.03E+08 & {\rm UGC\ 12914} \\
            33 &  329 &  0.41342 & 23.4797 &  593.83 & 16.89 & 2.41E+07 & {\rm UGC\ 12914} \\
            34 &   80 &  0.40977 & 23.4816 &    <300 &  9.99 & 5.00E+06 & {\rm UGC\ 12914} \\
            35 &   95 &  0.41991 & 23.5007 &    <300 & 11.19 & 6.79E+06 & {\rm UGC\ 12915} \\
            36 &  330 &  0.41602 & 23.4918 &  507.34 & 13.51 & 1.78E+07 &     {\rm bridge} \\
            37 &  196 &  0.41553 & 23.4918 &  506.89 & 10.65 & 1.14E+07 &     {\rm bridge} \\
            38 &  158 &  0.41528 & 23.4917 &    <300 & 10.93 & 8.12E+06 &     {\rm bridge} \\
            39 &  178 &  0.41645 & 23.4972 &    <300 & 16.71 & 1.12E+07 &     {\rm bridge} \\
            40 &  354 &  0.41408 & 23.4860 &  681.41 & 16.46 & 1.67E+07 &     {\rm bridge} \\
            41 &   90 &  0.41808 & 23.4858 &    <300 & 14.51 & 5.77E+06 &     {\rm bridge} \\
            42 &  134 &  0.41828 & 23.4870 &    <300 & 14.96 & 8.86E+06 &     {\rm bridge} \\
            43 &  115 &  0.41652 & 23.4860 &    <300 & 10.12 & 7.26E+06 &     {\rm bridge} \\
            44 &  387 &  0.41592 & 23.4857 &  751.19 & 13.80 & 2.04E+07 &     {\rm bridge} \\
            45 &  177 &  0.41833 & 23.4861 &  517.13 & 14.99 & 1.10E+07 &     {\rm bridge} \\
            46 &  522 &  0.41873 & 23.4870 &  933.92 & 23.21 & 3.01E+07 &     {\rm bridge} \\
            47 &  167 &  0.41854 & 23.4877 &  612.88 &  7.06 & 8.79E+06 &     {\rm bridge} \\
            48 &  558 &  0.41720 & 23.4867 &  659.95 & 17.79 & 3.10E+07 &     {\rm bridge} \\
            49 &  241 &  0.42128 & 23.4994 &    <300 & 14.52 & 2.13E+07 & {\rm UGC\ 12915} \\
            50 &  208 &  0.41488 & 23.4885 &  667.54 & 27.73 & 1.06E+07 &     {\rm bridge} \\
            51 &  276 &  0.40974 & 23.4832 &  639.13 & 17.27 & 2.40E+07 & {\rm UGC\ 12914} \\
            52 & 1519 &  0.40962 & 23.4839 & 1117.39 & 27.38 & 1.07E+08 & {\rm UGC\ 12914} \\
            53 &  159 &  0.40844 & 23.4840 &    <300 & 14.64 & 1.34E+07 & {\rm UGC\ 12914} \\
            54 &  621 &  0.40938 & 23.4841 & 1006.50 & 13.26 & 4.44E+07 & {\rm UGC\ 12914} \\
            55 &   46 &  0.40739 & 23.4846 &    <300 & 10.67 & 4.97E+06 & {\rm UGC\ 12914} \\
            56 &  270 &  0.40965 & 23.4839 &  296.58 & 10.49 & 1.73E+07 & {\rm UGC\ 12914} \\
            57 &   57 &  0.40968 & 23.4887 &    <300 & 17.80 & 4.59E+06 & {\rm UGC\ 12914} \\
            58 &  202 &  0.40989 & 23.4845 &    <300 & 14.72 & 1.33E+07 & {\rm UGC\ 12914} \\
            59 &  118 &  0.41046 & 23.4875 &  348.53 & 26.59 & 8.11E+06 & {\rm UGC\ 12914} \\
            60 &  144 &  0.40863 & 23.4875 &  300.83 & 12.13 & 9.22E+06 & {\rm UGC\ 12914} \\
        \noalign{\smallskip}
        \hline
        \noalign{\smallskip}
        \hline
        \end{array}
      \]
\begin{list}{}{}
\item[$^{\rm (a)}$ When CPROPS was unable to deconvolve the cloud size, we put $<300$~pc.]
\end{list}
\end{table*}

\setcounter{table}{1}
\begin{table*}[!ht]
      \caption{continued.}
         \label{tab:molent1}
      \[
         \begin{array}{lccccccc}
           \hline
           {\rm number} & {\rm total\ number} & {\rm RA} & {\rm DEC}  & {\rm radius} & {\rm velocity\ dispersion}  & {\rm flux} & {\rm region} \\
            & {\rm of\ pixels} &  ({\rm degrees)} & {\rm (degrees)} &  {\rm (pc)} & {\rm (km\,s}^{-1})  & {\rm (K\,km\,s}^{-1}{\rm pc}^2) & \\
           \hline
           61 &   90 &  0.40882 & 23.4890 &  160.47 & 11.36 & 6.39E+06 & {\rm UGC\ 12914} \\
           62 &  146 &  0.40985 & 23.4885 &    <300 & 18.75 & 7.45E+06 & {\rm UGC\ 12914} \\
           63 &   58 &  0.40761 & 23.4918 &    <300 &  6.85 & 3.86E+06 & {\rm UGC\ 12914} \\
           64 & 1585 &  0.40674 & 23.4871 & 1051.63 & 42.77 & 1.29E+08 & {\rm UGC\ 12914} \\
           65 &   72 &  0.40762 & 23.4868 &    <300 &  9.30 & 5.36E+06 & {\rm UGC\ 12914} \\
           66 &  506 &  0.40906 & 23.4888 &  771.13 & 19.15 & 2.91E+07 & {\rm UGC\ 12914} \\
           67 &  341 &  0.40788 & 23.4920 &  577.13 & 11.67 & 2.34E+07 & {\rm UGC\ 12914} \\
           68 &  550 &  0.40814 & 23.4888 &  765.39 & 13.50 & 3.29E+07 & {\rm UGC\ 12914} \\
           69 & 1118 &  0.40682 & 23.4891 &  873.53 & 16.26 & 1.05E+08 & {\rm UGC\ 12914} \\
           70 &  465 &  0.40992 & 23.4927 & 1181.87 & 19.69 & 2.81E+07 &     {\rm bridge} \\
           71 &  597 &  0.40682 & 23.4908 &  503.07 & 14.06 & 5.44E+07 & {\rm UGC\ 12914} \\
           72 & 1130 &  0.40775 & 23.4924 & 1028.35 & 18.64 & 9.86E+07 & {\rm UGC\ 12914} \\
           73 &   85 &  0.41281 & 23.4818 &  619.77 & 12.53 & 8.16E+06 & {\rm UGC\ 12914} \\
           74 &   72 &  0.41328 & 23.4824 &    <300 & 20.01 & 4.29E+06 & {\rm UGC\ 12914} \\
           75 &   82 &  0.42309 & 23.4982 &    <300 & 11.65 & 7.60E+06 & {\rm UGC\ 12915} \\
           76 &   79 &  0.41153 & 23.4822 &    <300 &  6.95 & 4.81E+06 & {\rm UGC\ 12914} \\
           77 &   82 &  0.41154 & 23.4824 &    <300 & 11.28 & 5.93E+06 & {\rm UGC\ 12914} \\
           78 &  477 &  0.41114 & 23.4887 &    <300 & 29.15 & 2.68E+07 &     {\rm bridge} \\
           79 &  120 &  0.41332 & 23.4973 &  224.35 & 11.56 & 9.96E+06 &     {\rm bridge} \\
           80 &   65 &  0.40959 & 23.4948 &    <300 &  7.42 & 4.48E+06 &     {\rm bridge} \\
       \noalign{\smallskip}
       \hline
       \noalign{\smallskip}
 \hline
        \end{array}
      \]
\end{table*}

\subsection{Moment maps \label{sec:maps}}

The optical image of UGC~12915 (upper panel of Fig.~\ref{fig:cohistars}) shows an asymmetric dust ridge or tilted ring 
visible in absorption and two symmetric stellar arms, the northern arm being brighter than the southern arm.
The moment~0 map (Fig.~\ref{fig:taffy_obs_mom_final_mom0}) shows a bright, asymmetric, and twisted thin molecular disk rather than a tilted ring in UGC~12915 
which corresponds to the asymmetric dust ridge.
The surface brightness distribution along the major axis is asymmetric. The second brightest maximum in this disk corresponds to the galaxy center.
The brightest maximum is located in the southeastern half of the disk. The northwestern half has a much lower
surface brightness and is approximately twice as extended as the southeastern half of the disk. Moreover, the most northwestern part of UGC~12915's 
molecular disk is bent to the north, away from UGC~12914 and the bridge region.

The optical image of UGC~12914 (upper panel of Fig.~\ref{fig:cohistars}) shows an inner lens structure with dust lanes
and a much fainter outer double-ring structure. In addition, a stellar arm starts from the northern tip of the stellar lens structure
joining the eastern faint outer stellar ring.
The CO emission distribution of UGC~12914 has three maxima along the
major axis: the galaxy center (D1) and the two elongated structures at a distance of $\sim 20''$ or $5.8$~kpc from the center (D1 and D3).
The latter structures correspond to the tips of the optical lens and are reminiscent of a limb-brightened molecular ring. 

\begin{figure}[!ht]
  \centering
  \resizebox{\hsize}{!}{\includegraphics{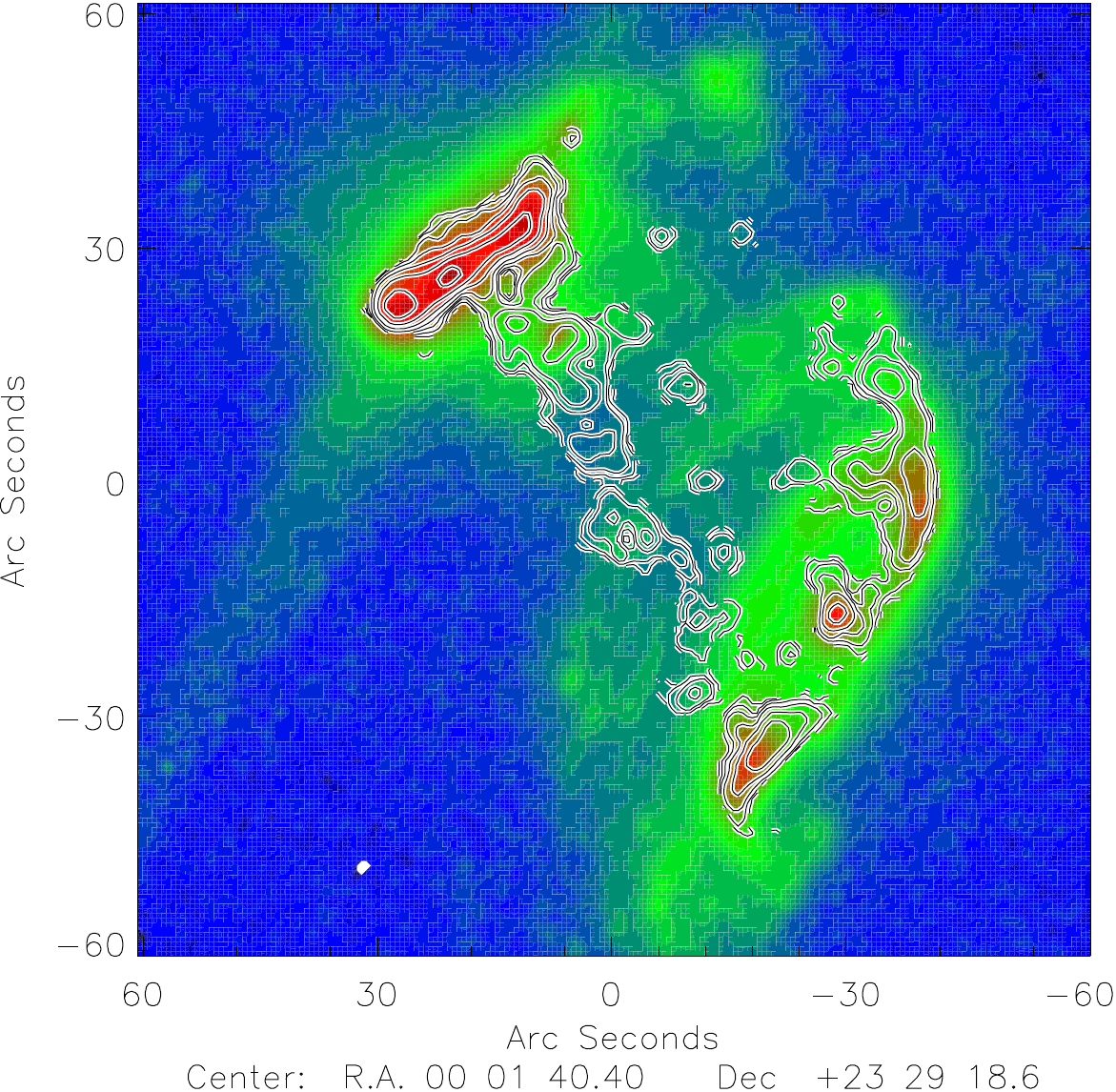}}
  \resizebox{\hsize}{!}{\includegraphics{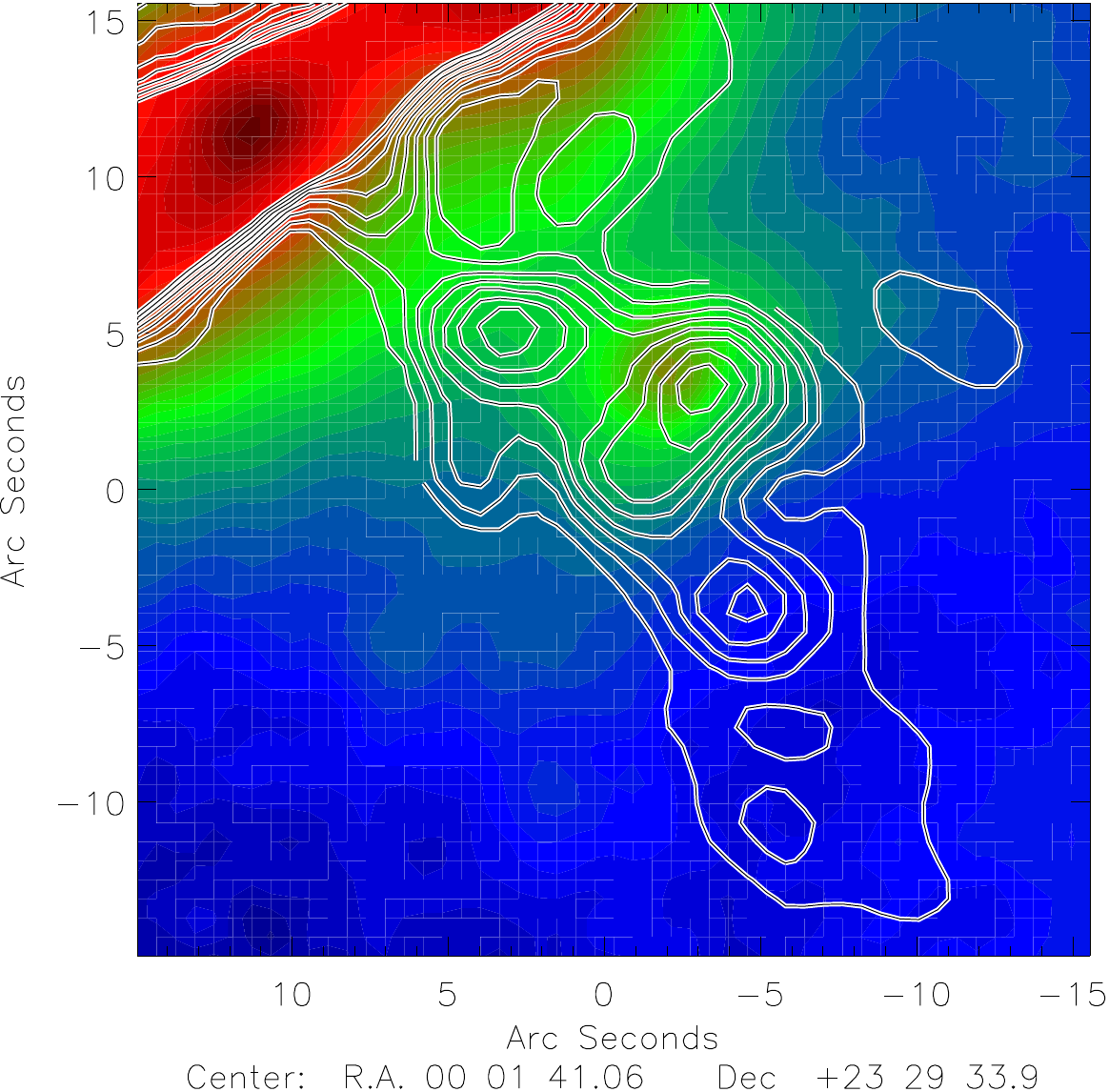}}
  \caption{CO(1-0) contours on the Spitzer $8$~$\mu$m PAH emission map. Upper panel: the whole Taffy system.
    The stripe starting from the southern end of the disk of UGC~12915 is an image artifact.
    Contour levels are $(2,4,8,16,32,64,128,256)$~K\,km\,s$^{-1}$.
    Lower panel: zoom on the compact extraplanar star forming region south of UGC~12915.
    Contour levels are $(10,20,30,40,50,60,70,80)$~K\,km\,s$^{-1}$.
  \label{fig:taffy_spitzer}}
\end{figure}

The northernmost part of UGC~12914's molecular disk is curved towards the bridge as is the stellar arm
(upper panel of Fig.~\ref{fig:cohistars} and upper left panel of Fig.~\ref{fig:taffy_sm2_mom_final_stars}), 
suggesting this is a tidal effect. Whereas the western border of the CO distribution, which corresponds to
the western dust lane within the optical lens structure
of UGC~12914, is sharp, the eastern border is disrupted showing east-west filaments elongated into the bridge direction.
These filaments are due to the ISM-ISM collision.
The molecular gas bridge connecting the two galaxies has a width of $\sim 10''$--$30''$ or $\sim 3$--$9$~kpc and shows a
maximum adjacent to the giant H{\sc ii} region close to UGC~12915 (upper panel of Fig.~\ref{fig:taffy_spitzer}).
It roughly connects the center of UGC~12915 and the southern CO maximum of UGC~12914. Four distinct
CO clouds (B1--B4) are located parallel to the bridge to the west (upper panel of Fig.~\ref{fig:taffy_obs_mom_final_mom0}). 
Finally, three CO clouds (N1--N3) seem to connect the northern part of UGC~12914 and the northern part of the disk of UGC~12915.

The total CO luminosity of the Taffy system identified by CPROPS is $L_{\rm CO,tot}=4.8 \times 10^9$~K\,km\,s$^{-1}$pc$^2$. 
This represents 60\,\% of the CO luminosity found by Braine et al. (2003) with the IRAM 30m telescope.
We divided the moment~0 into disk and bridge regions (Fig.~\ref{fig:bridge_separation}). The CO luminosity of
the bridge is $L_{\rm CO,bridge}=1.2 \times 10^9$~K\,km\,s$^{-1}$pc$^2$. Thus, $25$\,\% of the total CO luminosity stems from
the bridge region. Assuming a Galactic $\ratioo$ conversion factor for the galactic disks and a third of this value for
the bridge gas, we obtain the following H$_2$ masses: $M_{\rm H_2,tot}=1.7 \times 10^{10}$~M$_{\odot}$ and
$M_{\rm H_2,bridge}=1.7 \times 10^{9}$~M$_{\odot}$. Thus, about $10$\,\% of the molecular gas mass is located in the bridge region.

An overlay with the Spitzer $8$~$\mu$m PAH emission map is shown in Fig.~\ref{fig:taffy_spitzer}.
Within the galactic disks, the CO(1-0) emission closely follows the high surface brightness $8$~$\mu$m emission.
In the bridge region there is dense gas traced by CO emission which
is not forming stars, as shown by a lack of PAH emission, which is usually a tracer
of star formation. This implies that the bulk of the bridge
high-density gas does not form stars (see also Braine et al. 2003 and Gao et al. 2003). 
The luminous compact extraplanar H{\sc ii} region south of UGC~12915
represents the exception to that rule. A close-up of the region (lower panel of  Fig.~\ref{fig:taffy_spitzer})
shows that the H{\sc ii} region does not coincide with, but is located at the northern edge of a high-surface brightness GMA
(GMA~9 in Table~\ref{tab:molent}). This GMA has the highest velocity dispersion of the bridge GMAs. 

The velocity fields of UGC~12914 and UGC~12915 are dominated by rotation. The bridge shows a mixture of positive and
negative radial velocities with respect to the systemic velocities of the galaxies ($4350$~km\,s$^{-1}$).
The region of high surface brightness close to UGC~12915 has an overall positive velocity with respect to the systemic velocity.
The CO clouds aligned parallel to the bridge share this velocity range.
\begin{figure}[!ht]
  \centering
  \resizebox{\hsize}{!}{\includegraphics{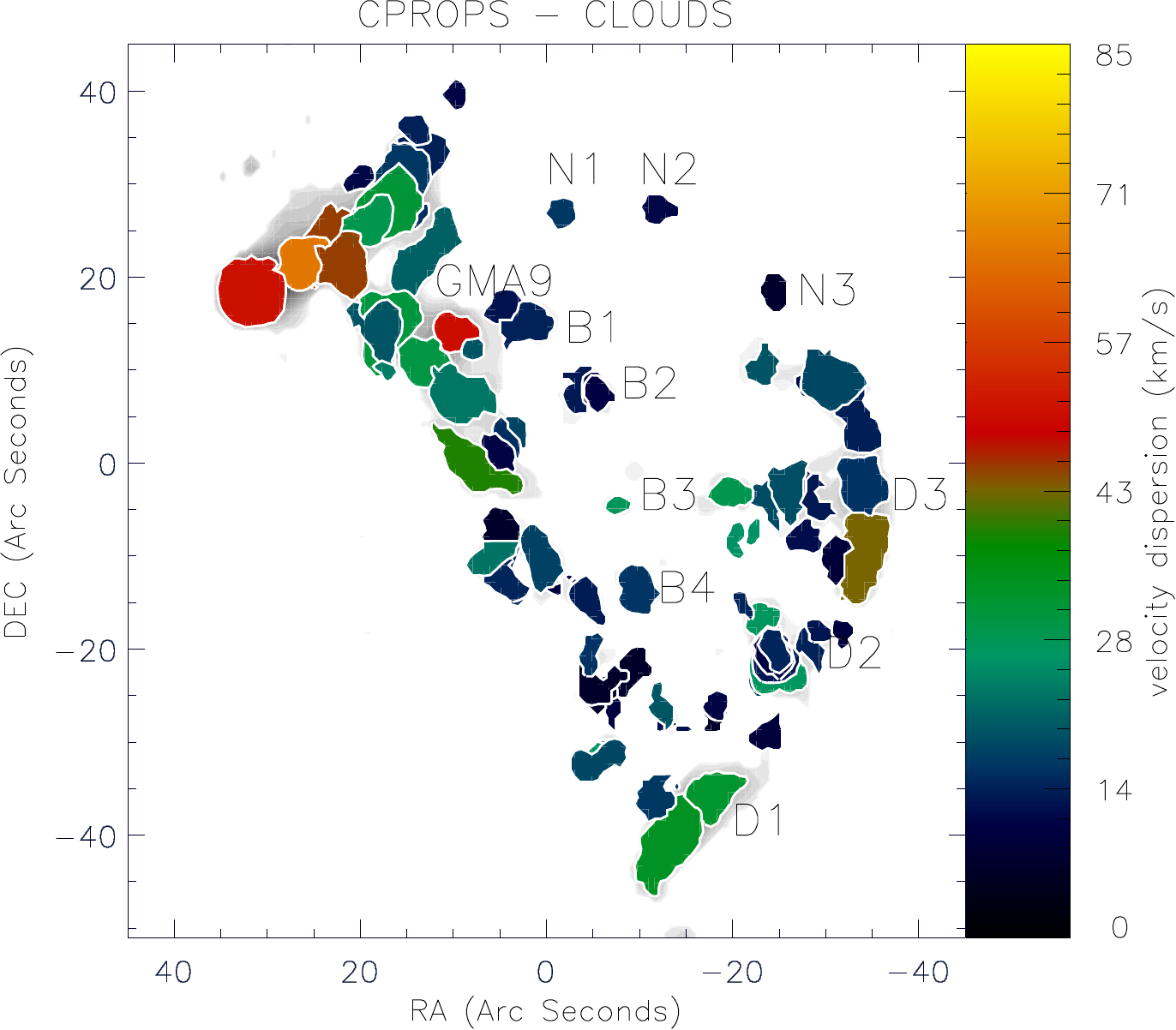}}
  \caption{Internal velocity dispersion of the CPROPS molecular clouds (color) on the moment~0 map (greyscale).
    Disk, bridge, and northern emission regions are labelled as in Fig.~\ref{fig:taffy_obs_mom_final_mom0}.
  \label{fig:mom0_taffysm2_cloud_eclump_bis}}
\end{figure}

The internal velocity dispersions of the CO clouds derived by CPROPS are shown in Fig.~\ref{fig:mom0_taffysm2_cloud_eclump_bis}.
The velocity dispersion of the inner parts of the molecular disk in UGC~12914 is about $30$~km\,s$^{-1}$, roughly normal for
an edge-on spiral galaxy at 800~pc resolution. 
The highest velocity dispersions are found in the southeastern disk of UGC~12915 and its center.
A cloud with a velocity dispersion of $\sim 50$~km\,s$^{-1}$ (GMA~9 in Table~\ref{tab:molent} and Fig.~\ref{fig:mom0_taffysm2_cloud_eclump_bis}) 
is found in the high surface brightness part of the bridge, close to the extraplanar H{\sc ii} region. 
Overall, the northern half of the gas bridge has significantly higher velocity dispersions than the southern half.

\subsection{Cloud properties}

\begin{figure}[!ht]
  \centering
  \resizebox{\hsize}{!}{\includegraphics{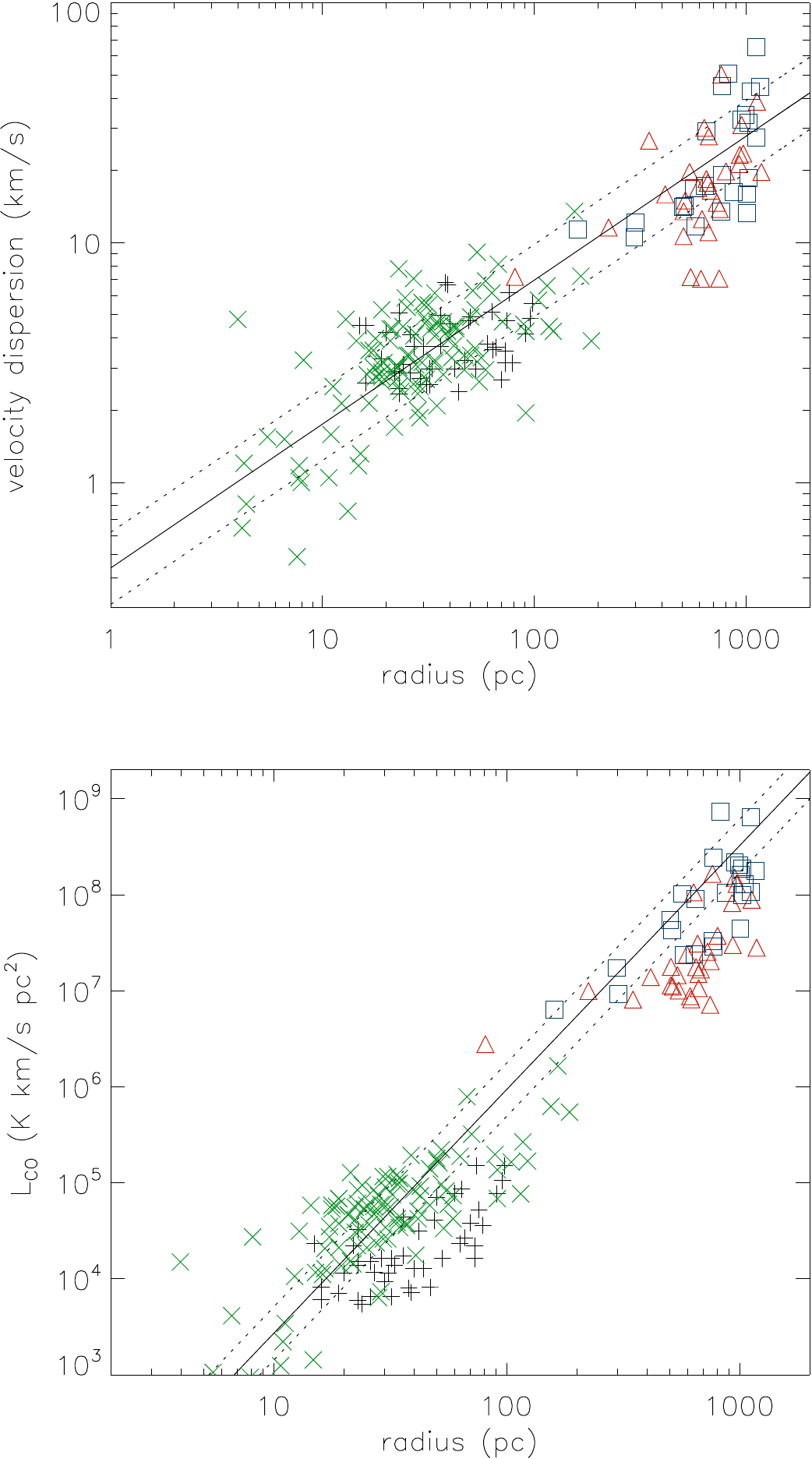}}
  \caption{Properties of CPROPS molecular entities (bridge: red triangles; galaxies: blue boxes) compared to data from 
    Bolatto et al. (2008; green crosses) and Gratier et al. (2012; black pluses).
    Upper panel: velocity dispersion as a function radius. Lower panel: CO line flux as a function of radius.
    The lines correspond to the relations determined by Bolatto et al. (2008).
  \label{fig:cloudproperties}}
\end{figure}

\begin{figure}[!ht]
  \centering
  \resizebox{\hsize}{!}{\includegraphics{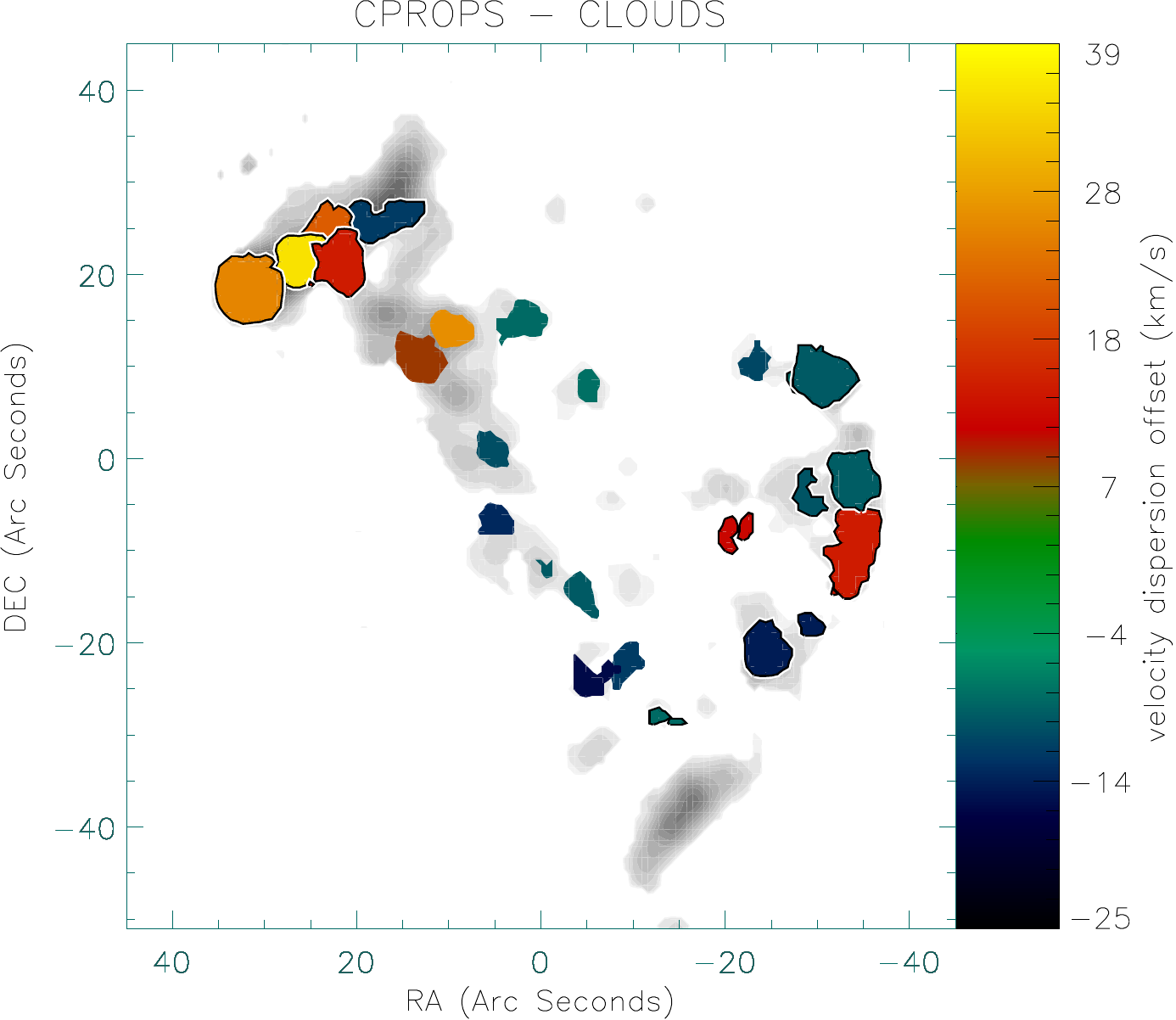}}
  \caption{Velocity dispersion offset of CPROPS molecular clouds with respect to the relation for extragalactic GMCs found by Bolatto et al. (2008).
    Only GMAs with linewidth outside $1\,\sigma$ of the size--linewidth relation shown in Fig.~\ref{fig:cloudproperties} are colored. 
    The disk GMAs are marked with a black contour.
  \label{fig:mom0_taffysm2_cloud_eclump_bisnew}}
\end{figure}

We separated the CO clouds identified by CPROPS into disk and bridge clouds according to Fig.~\ref{fig:bridge_separation}.
The resulting assignments are given in Table~\ref{tab:molent}. The cloud properties are compared to those
of extragalactic GMAs from Bolatto et al. (2008) and those of M~33 derived by Gratier et al. (2012) in Fig.~\ref{fig:cloudproperties}.
With a resolution of $2.7''$ or $800$~pc we can only detect associations of giant molecular clouds (GMAs).  
It is remarkable that the GMAs in the disk and bridge regions follow, as the molecular clouds in M~33,
the size--linewidth relation established by Bolatto et al. (2008)
which is valid for extragalactic and Galactic molecular clouds. The scatter around the relation is also comparable to that
of Bolatto et al. (2008) and Gratier et al. (2012). It is especially surprising that the disk GMAs follow the relation, because a significant
fraction of their linewidth is expected to be caused by large-scale motions, i.e. rotation and non-circular motions.
The offset between the velocity dispersion determined by CPROPS and that predicted by the size--linewidth relation is presented
in Fig.~\ref{fig:mom0_taffysm2_cloud_eclump_bisnew}. For clarity we only colored GMAs whose linewidths are outside $1\,\sigma$ of the size--linewidth relation.
Two regions with exceptionally high linewidths stand out from this figure: the southeastern half of UGC12915's disk and the region around
the extraplanar H{\sc ii} region close to UGC12915.

The size--luminosity relation of the GMAs in the bridge and disk regions is different. Whereas the clouds in the galactic disks follow the
relation established by  Bolatto et al. (2008), the majority of the bridge clouds show about three times lower CO luminosity than
expected from the relation. The molecular clouds in M~33 are also CO-underluminous by about a factor of two.
Gratier et al. (2012) argued that this is due to a two times higher $\ratioo$ conversion factor.
For the Taffy bridge region Braine et al. (2003) excluded a higher $\ratioo$ conversion factor based on their $^{13}$CO measurements.
On the contrary, Braine et al. (2003) and Zhu et al. (2007) argued that
the $\ratioo$ conversion factor is several times lower in the bridge than in the galactic disk. 

\section{Comparison to dynamical models \label{sec:compmodel}}

Vollmer et al. (2012) calculated $17$ models of head-on collisions of two gas-rich spiral galaxies.
To the two ``best-fit'' models presented by these authors we added a third model with a higher
velocity between the two galaxies. The maximum impact velocity is $1200$~km\,s$^{-1}$ and the transverse velocity difference at the present time is 
$\sim 900$~km\,s$^{-1}$ versus $\sim 700$~km\,s$^{-1}$ for the previous simulations. We call this new simulation ``sim19fast''. 
Because of the high time resolution of our simulations
the cloud collisions are well resolved even for this enormous impact velocity.
The system is observed at the same lapse of time ($20$~Myr) after impact as the two simulations in Vollmer et al. (2012).

\subsection{Moment maps}

\begin{figure*}[!ht]
  \centering
  \resizebox{\hsize}{!}{\includegraphics{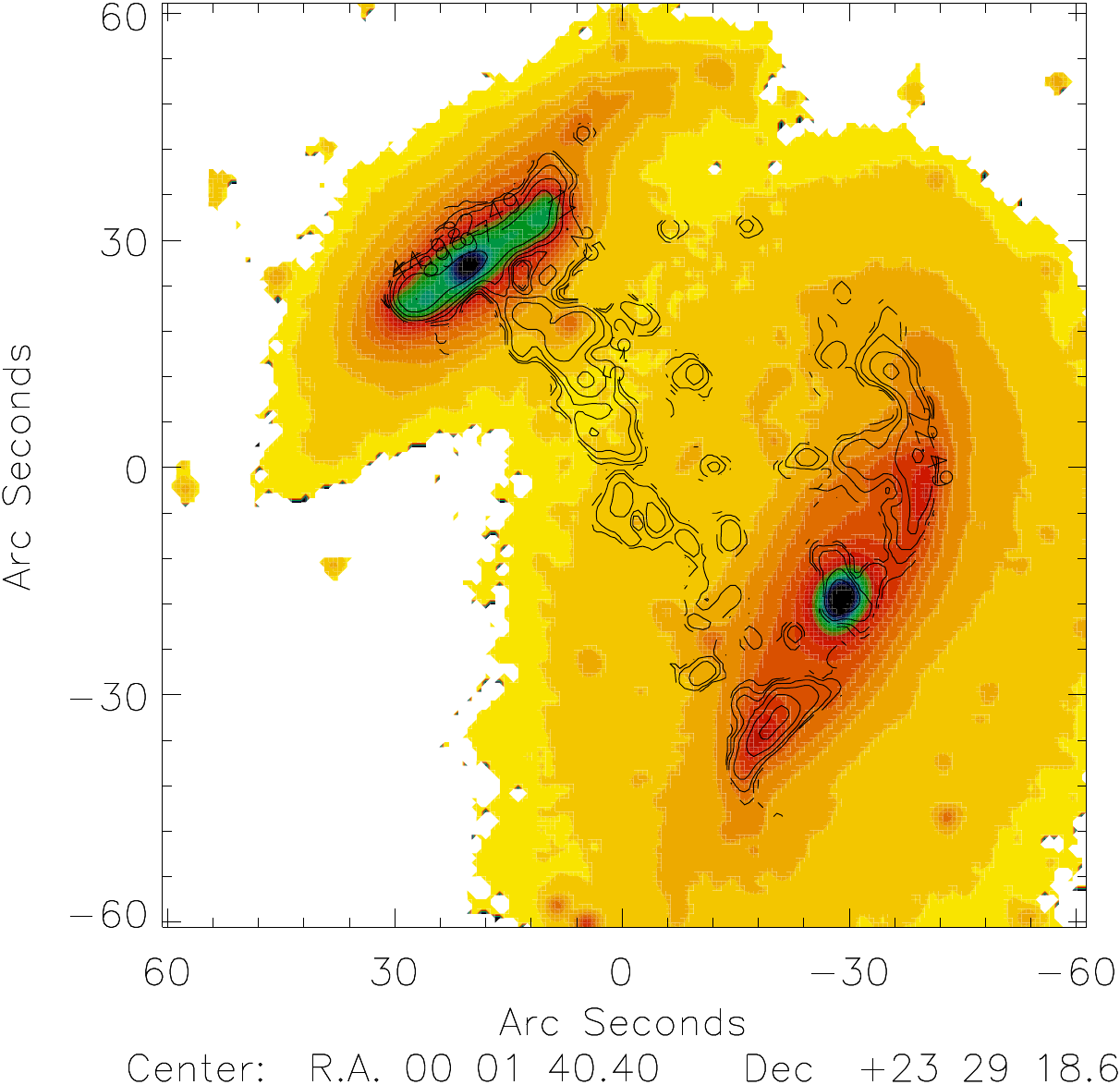}\includegraphics{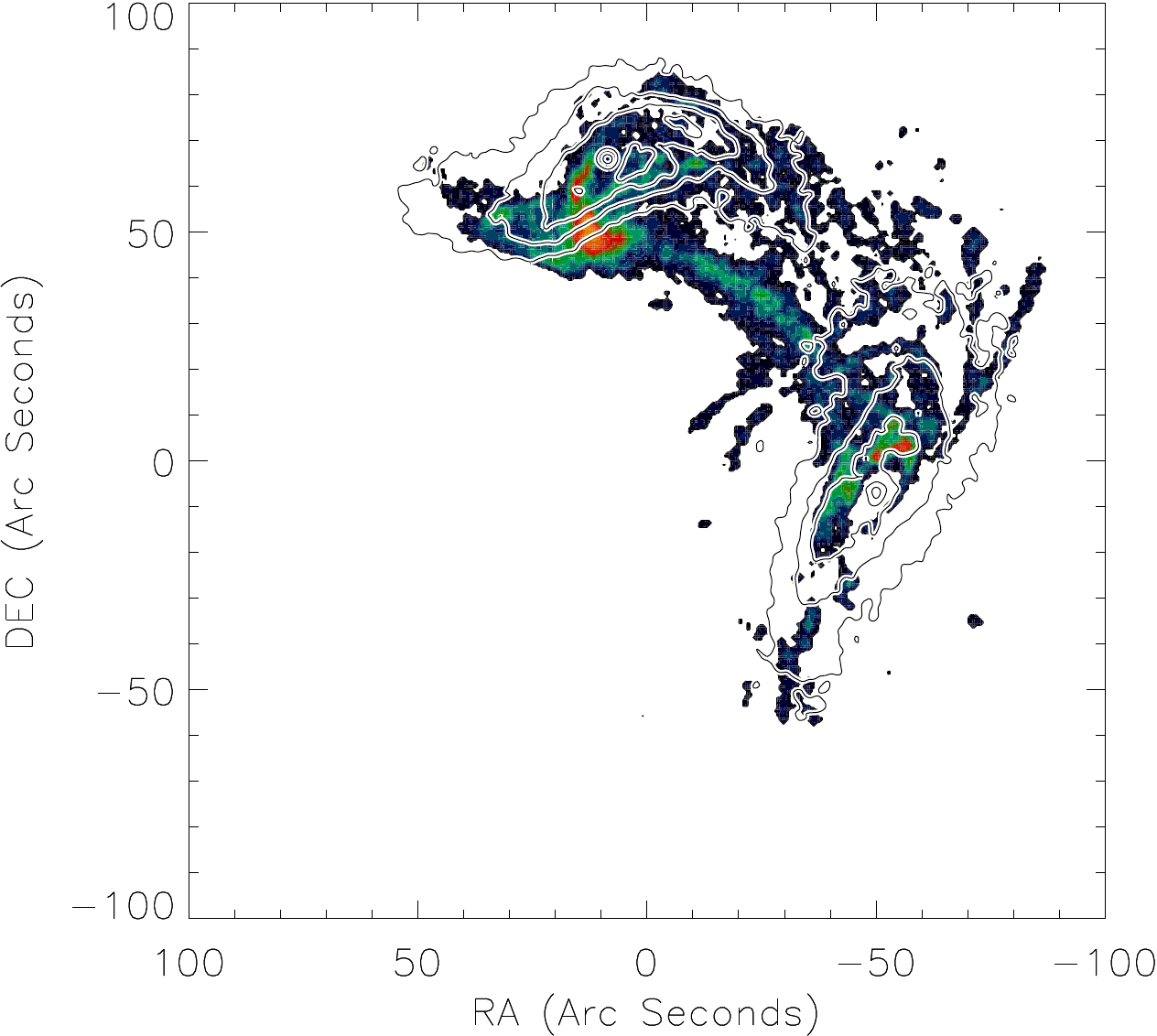}}
  \put(-470,45){\Large observations}
  \put(-210,40){\Large sim 19}
  \vspace{0cm}
  \resizebox{\hsize}{!}{\includegraphics{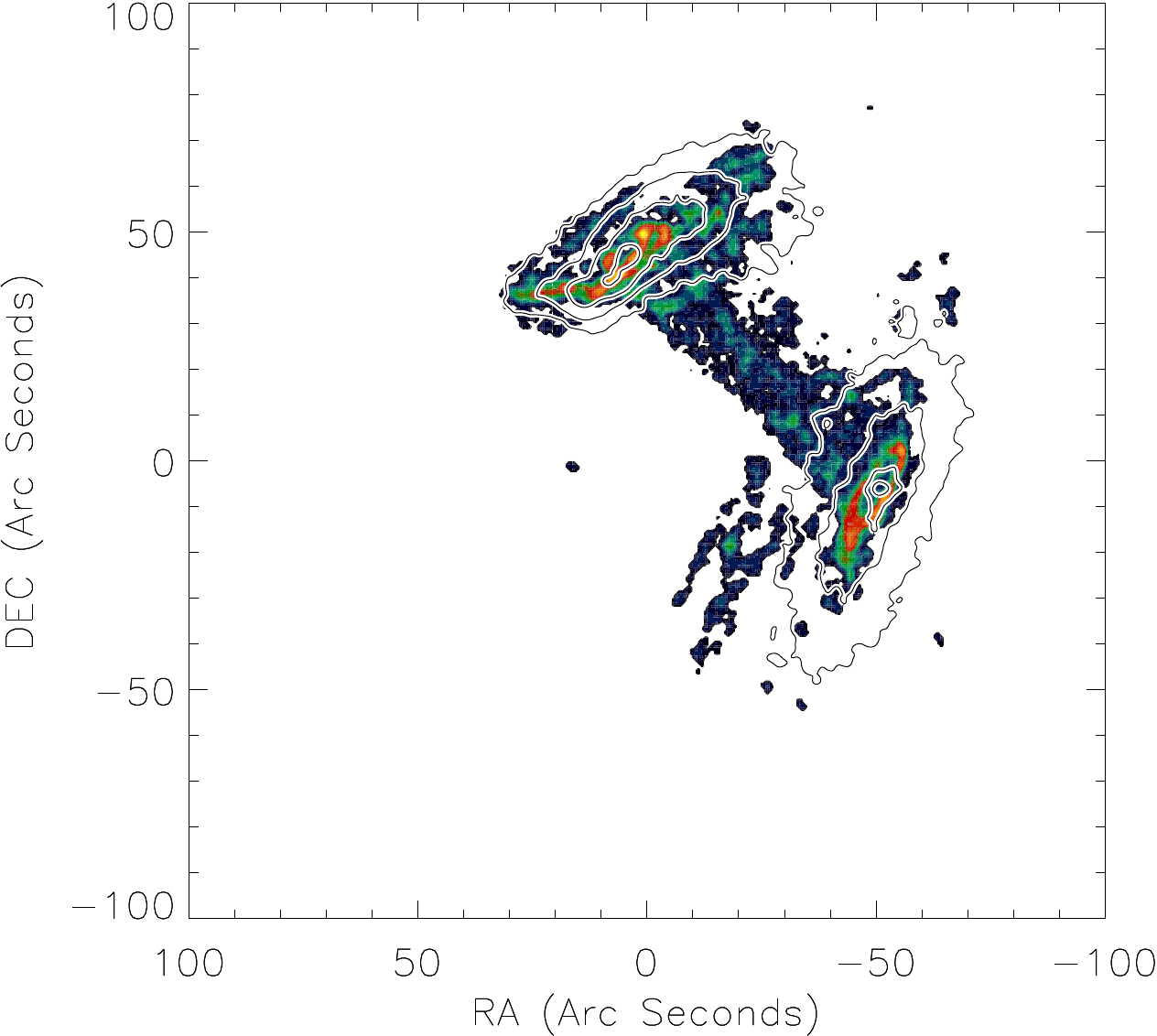}\includegraphics{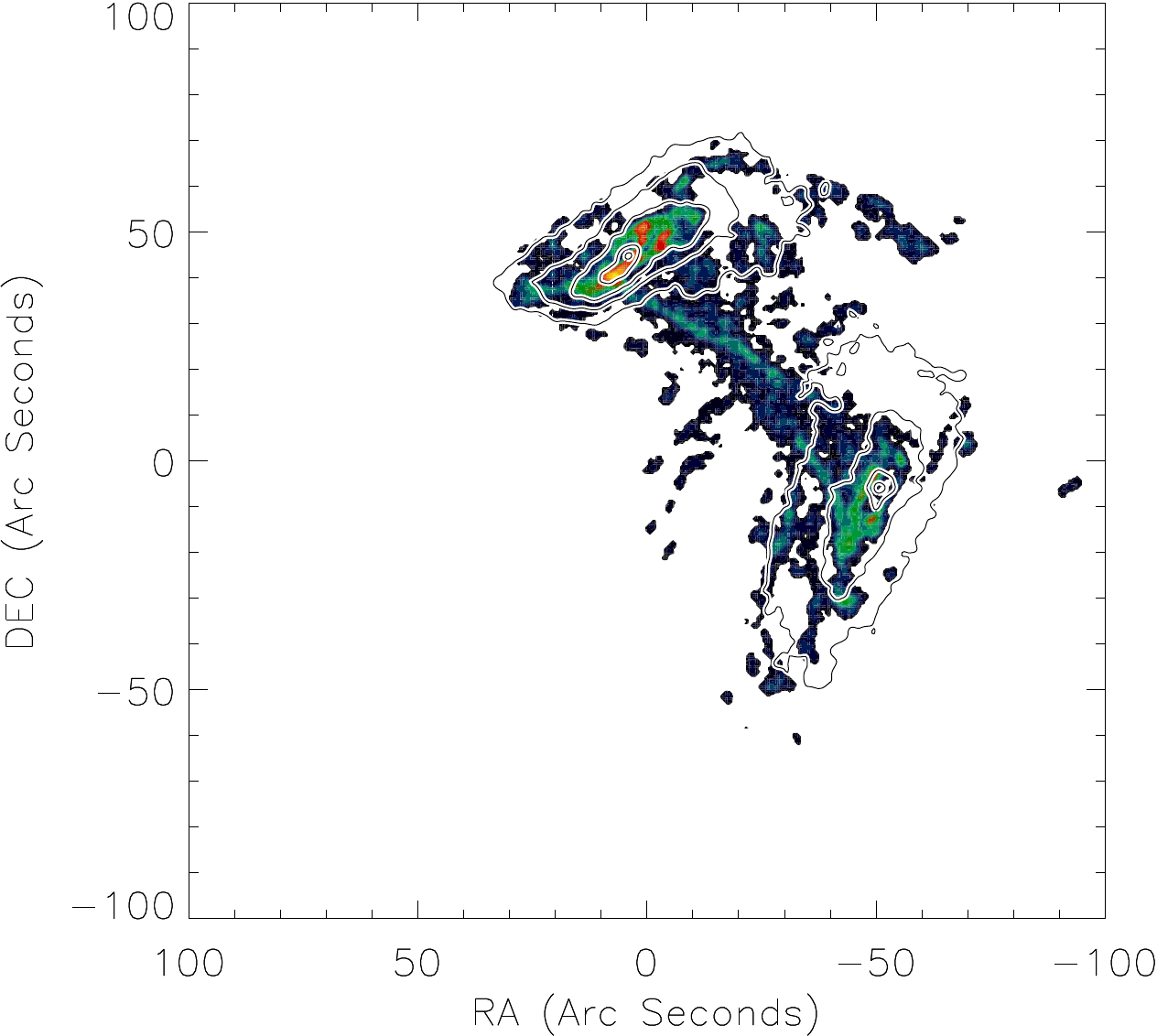}}
  \put(-470,40){\Large sim 19 fast}
  \put(-210,40){\Large sim 20}
  \caption{CO(1-0) moment~0 maps together with the model H$_2$ moment~0 maps. Upper left panel: PdBI observations (contour) together
    with the Spitzer $3.6$~$\mu$m map. The contour levels are $(0.35,0.7,1.4,2.8,5.6,11.2,22.5,45.0,90.0,180.0,360.0)$~K\,km\,s$^{-1}$.
    Other panels: contour: stellar distribution; color: molecular gas distribution. The color stretch is the same as the contours of
    the upper left panel.
  \label{fig:taffy_sm2_mom_final_stars}}
\end{figure*}

The comparison between the observed surface brightness distribution and the model moment~0 maps is shown in Fig.~\ref{fig:taffy_sm2_mom_final_stars}.
All three simulations develop a gas-rich bridge and show the observed sharp western border of gas distribution of UGC~12914
which is mainly a tidal feature.
As already stated in Vollmer et al. (2012), none of the models reproduce the detailed morphology of the system.
Whereas the model bridge starts close to the center of the northern galaxy, as is observed, it joins the
southern galaxy also close to its center. In the observations the bridge joins the disk of UGC~12914
further to the south. The edge-on projection of UGC~12915 is better reproduced by sim20.
On the other hand, the east--west asymmetry of its surface brightness is better reproduced by sim19 and
sim19fast. Contrary to observations, all models show a second bridge filament to the west of the main bridge.
This filament is brightest in sim19fast. The northern part of the disk of UGC~12914 with its filaments pointing
toward UGC~12915 is reproduced by sim19fast and to a much lesser degree by sim19. It is not reproduced by
sim20, because the northern galaxy passed through the southern galaxy at this location, removing all gas there.
Only in model sim19fast, the gas near the northern galaxy is much denser than that close to the southern galaxy,
as is observed.

The comparison between the observed velocity field and the model moment~1 maps is shown in Fig.~\ref{fig:taffy_sm2_mom_final_mom1}.
The velocity field of UGC~12914 is reasonably reproduced by sim19fast and to a lesser degree by sim19, 
whereas that of UGC~12915 is best reproduced by sim20.
The velocity field of the bridge with its positive and negative velocities with respect to the systemic velocity is best reproduced by
sim20 and to a much lesser degree by sim19fast. The model secondary bridge filaments to the north with their high velocities 
with respect to the systemic velocity are not observed.
We conclude that a single model among our limited set of simulations (see Vollmer et al. 2012) 
is not able to reproduce the observed characteristics of the Taffy system.
However, almost all characteristics can be found in one of the three models.
In many ways, this is to be expected as the initial gas distribution is not known.
The advantages and disadvantages of the models are summarized in Table~\ref{tab:comp}.

The comparison between the observed velocity dispersion and the model moment~2 maps is shown in Fig.~\ref{fig:taffy_sm2_mom_final_mom2}.
None of the models reproduce the extremely high velocity dispersion in the disk of UGC~12915. 
In sim19 and sim20 the regions of highest velocity dispersion are located close to the southern galaxy.
In sim20 another region of high velocity dispersion is located in the middle of the bridge where the two bridge
filaments cross. Only sim19fast shows a velocity dispersion in the bridge region close to the northern
galaxy which is comparable to the observed velocity dispersion.
We conclude that sim19fast is in rough agreement with the observed distribution of the velocity dispersion in the bridge.

\subsection{3D visualization of the datacubes}

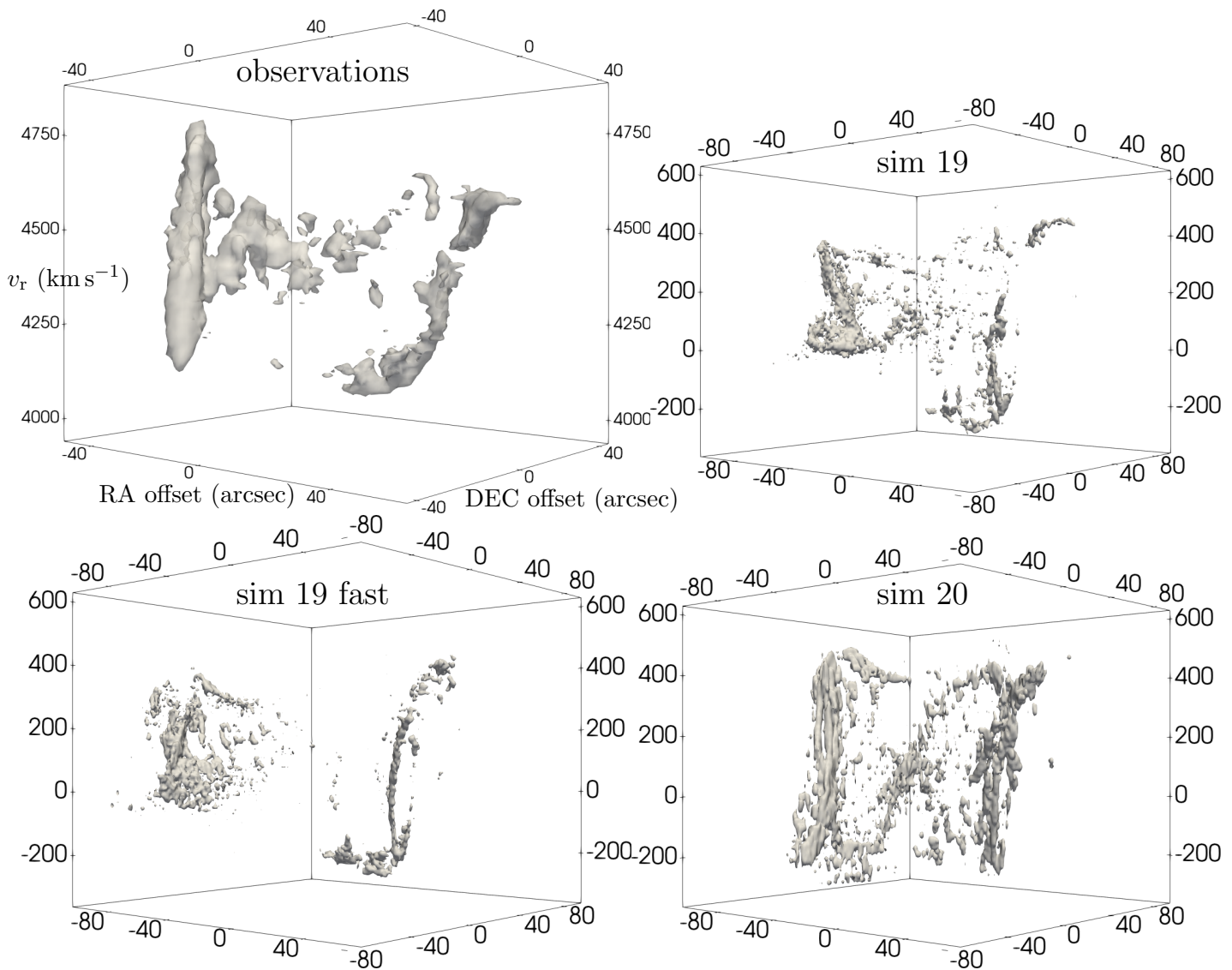
\begin{figure*}[!ht]
  \centering
  \resizebox{\hsize}{!}{\includegraphics{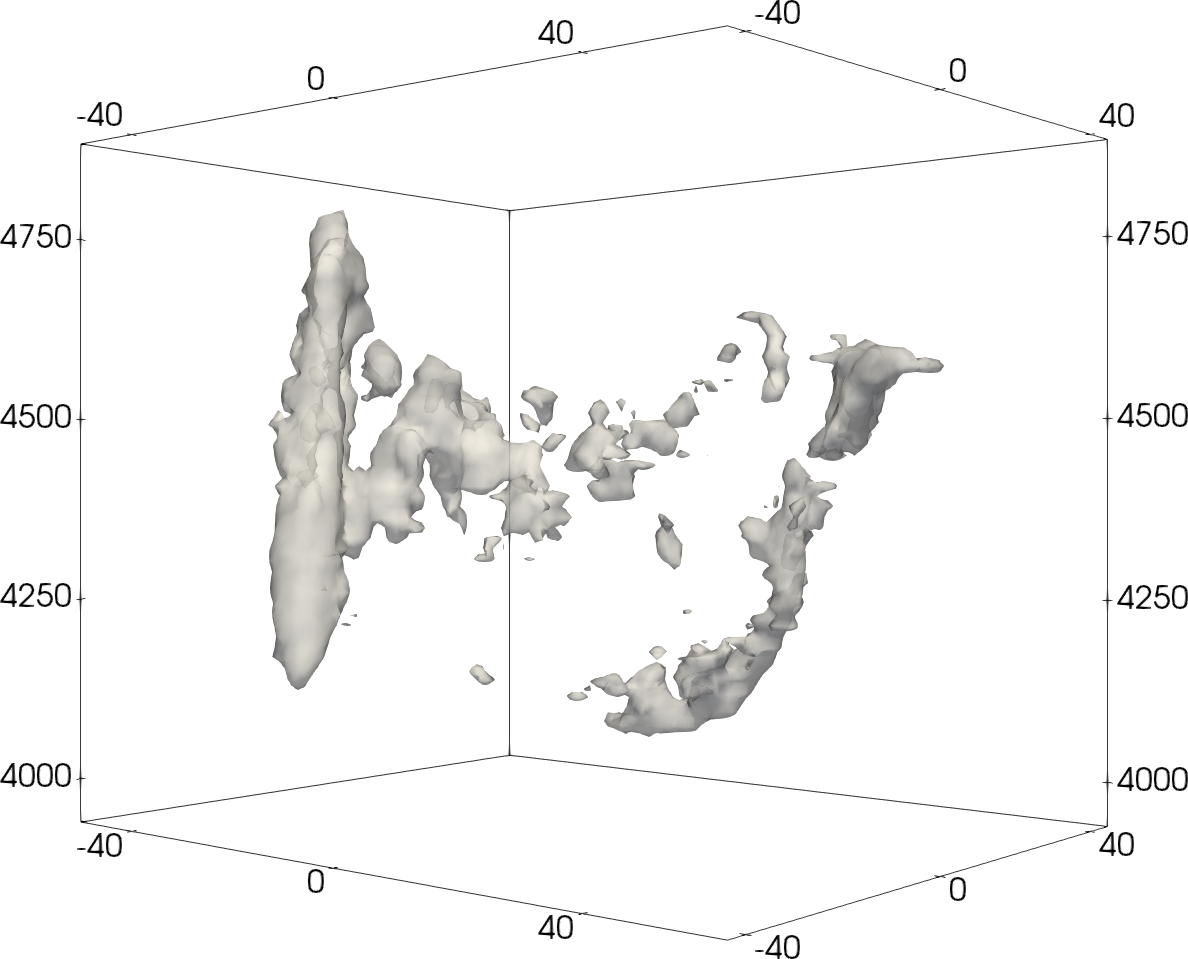}\includegraphics{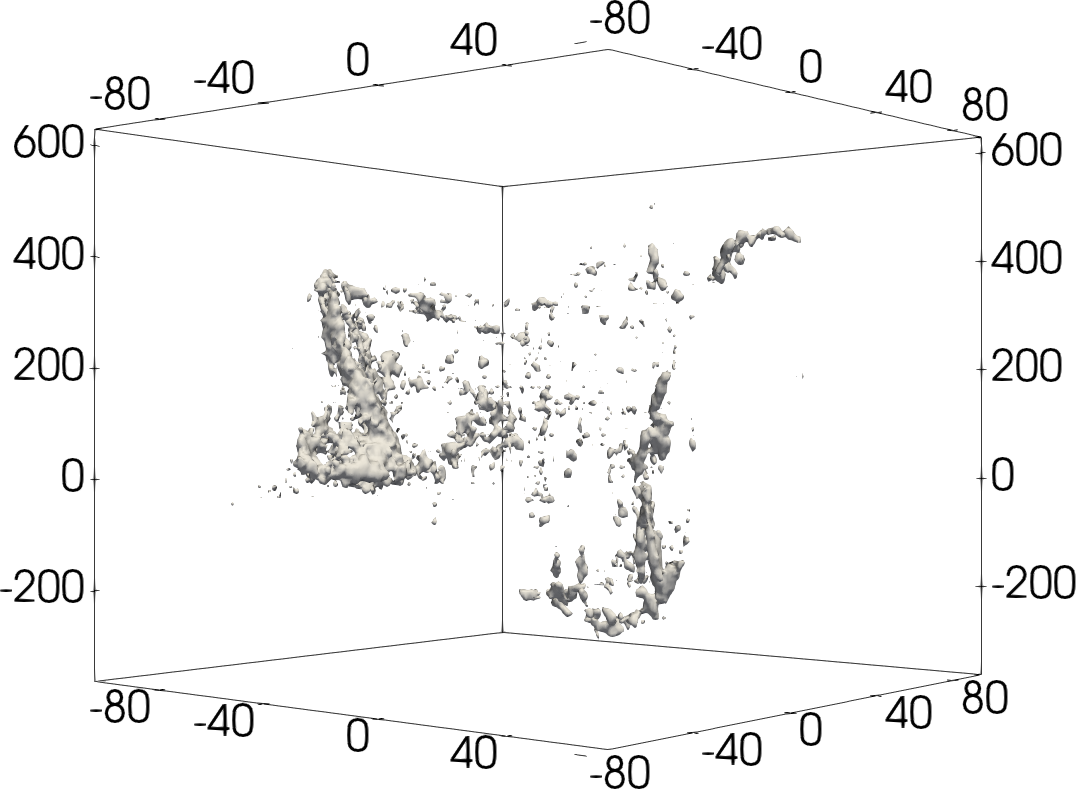}}
   \put(-430,190){\Large observations}
  \put(-150,150){\Large sim 19}
  \put(-530,100){\large $v_{\rm r}$ (km\,s$^{-1}$)}
  \put(-490,5){\large RA offset (arcsec)}
  \put(-330,3){\large DEC offset (arcsec)}
 \vspace{0cm}
  \resizebox{\hsize}{!}{\includegraphics{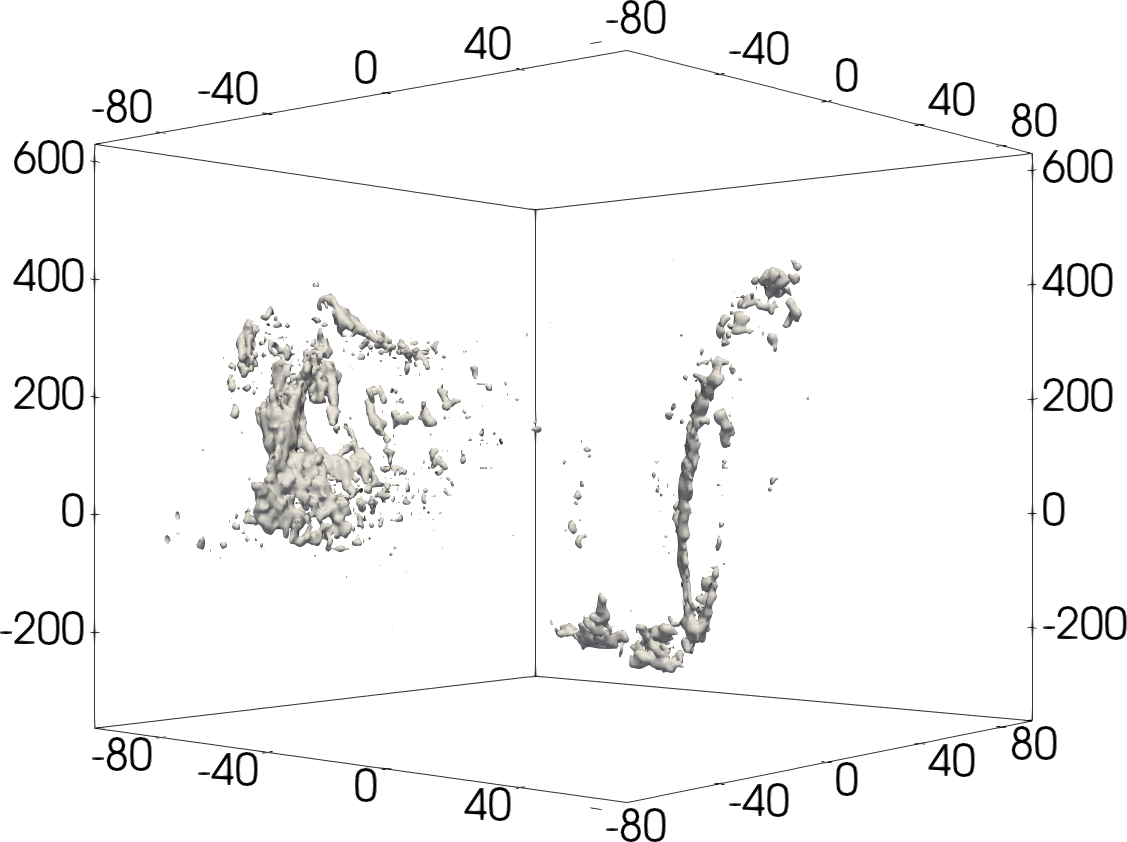}\includegraphics{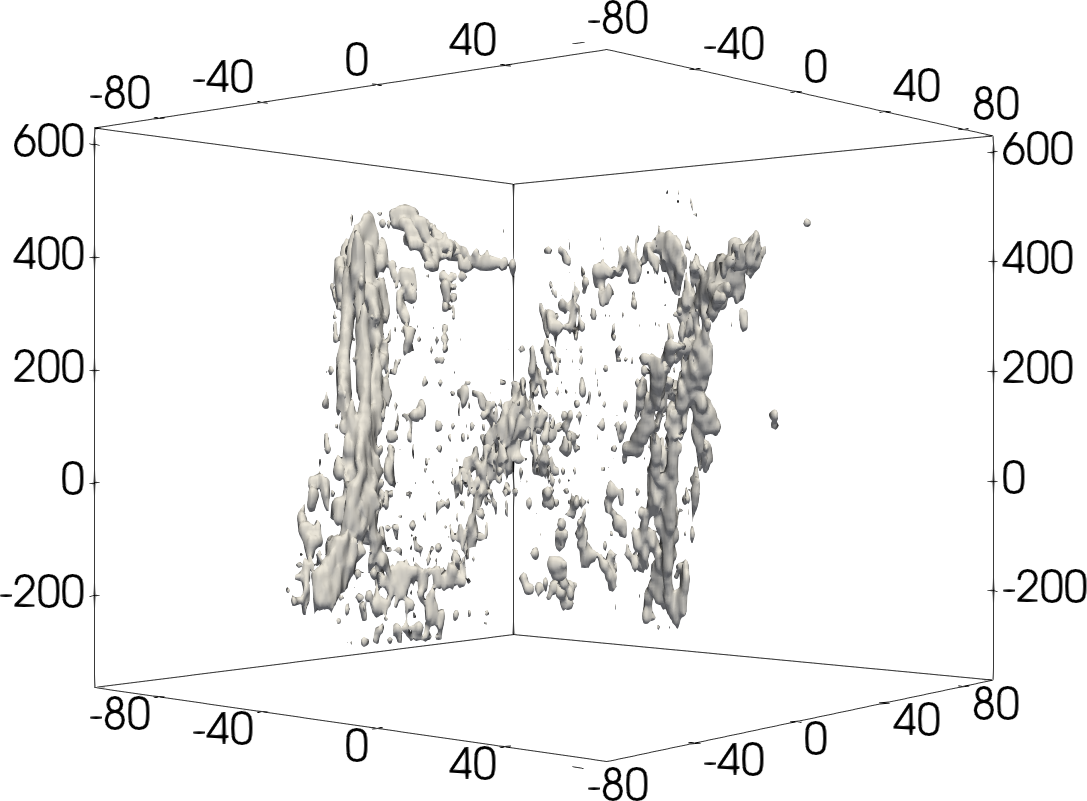}}
  \put(-430,160){\Large sim 19 fast}
  \put(-150,160){\Large sim 20}
  \caption{First 3D view of the observed CO(1-0) datacube and the model H$_2$ data cubes. These views correspond to a
    position-velocity diagram. The axis labels are only shown for the observations. For a better understanding of these views, 
    three 3D animations of the rotating datacube are
    attached to this figure (\texttt{taffy\_cube3D\_z.gif}, \texttt{taffy\_cube3D\_z1.gif}, and \texttt{taffy\_cube3D\_x.gif}).
  \label{fig:Taffy_sm2_3D_x_45d_eclump}}
\end{figure*}

To appreciate the full wealth of information provided by the datacubes, we decided to compare the observed and the
model datacubes by means of a 3D visualization. Here, we provide four different views of the datacubes rendered at
the same given intensity (Fig.~\ref{fig:Taffy_sm2_3D_x_45d_eclump} and Figs.~\ref{fig:Taffy_sm2_3D_x_85d_eclump}
to \ref{fig:Taffy_sm2_3D_z_180d_eclump}).
The total linewidth of UGC~12914 is significantly smaller than that of the southern model galaxy.
This can be due to an overestimated model inclination angle or an overestimated rotation velocity of the southern model galaxy.
The bridge region of high surface brightness and intensity near UGC~12915 has an inverted V-shape in the projection of
Fig.~\ref{fig:Taffy_sm2_3D_x_45d_eclump}. Moreover, it is confined to a relatively narrow velocity range around the
systemic radial velocity. A filament of low surface brightness and intensity emanating from this region  
smoothly joins the high radial velocity part of the disk of UGC~12914. The only model which reproduces these features is
sim20. However, the region of high intensities is further away from the northern galaxy than is observed for UGC~12915.
All models show emission emanating from the sides of highest and lowest radial velocities in the northern galaxy.
These features are not observed in UGC~12915. 
The velocity structure of the southern galaxy is rather well reproduced. As already mentioned, only the northern part of the
gas disk of the southern galaxy is missing in the model sim20, because the impact entirely removed the gas there.
Inspection of Fig.~\ref{fig:Taffy_sm2_3D_x_85d_eclump} to \ref{fig:Taffy_sm2_3D_z_180d_eclump} corroborates these conclusions.

\section{Star formation suppression caused by turbulent adiabatic compression \label{sec:sfrsuppress}}

What do the Circumnuclear Disk in the Galactic Center, a thick obscuring AGN torus, 
the ram-pressure stripped and tidally distorted
Virgo spiral galaxy NGC~4438, Stephan's Quintet, and the Taffy galaxies all have in common? At the first glance, all these systems
are very different. First of all, the spatial scales and timescales differ enormously. The CND and AGN tori have spatial extents
of about $10$~pc and rotation timescales of $10^{4}$~yr, whereas the relevant scales and timescales in NGC~4438,  Stephan's Quintet,
and the Taffy galaxies are of the order of tens of kpc and $100$~Myr. 
The common property of all these systems is that they are undergoing gas-gas collisions with
high energy injection rates. In these collisions, one gaseous body is the turbulent clumpy multi-phase ISM, 
while the other can be of different mean density and temperature (e.g. ISM, intragroup or intracluster gas):
NGC~4438 is affected by ongoing ram pressure caused by its rapid motion through the Virgo intracluster medium 
(Vollmer et al. 2005, 2009), and the intragroup gas of the Stephan's Quintet is compressed by a high-velocity intruder galaxy
(Appleton et al. 2017).
We suggest that the common theme of all these gas-gas interactions is adiabatic large-scale compression of the ISM leading to an increase of 
the turbulent velocity dispersion of the gas (Robertson \& Goldreich 2012; Mandal et al. 2020). 

It is generally assumed that within the disks of isolated galaxies turbulence is driven
by energy injection through stellar feedback (SN explosions). In an equilibrium state a balance 
between turbulent pressure and gravity is reached leading to a global virial equilibrium state of the GMCs (Heyer et al. 2009). 
If the energy injection through large-scale gas compression exceeds that of stellar feedback deduced via the star formation rate,
the velocity dispersion of the largest eddies is expected to increase.
In this case, we presume that the velocity dispersion of the turbulent 
substructures/clouds also increases (Fig.~2 of Mandal et al. 2020). Such clouds were observed in the Galactic Center region by 
Oka et al. (1998, 2001). As a result, these GMCs will no longer be in global virial equilibrium.
Oka et al. (2001) argued that the high virial parameters ($\alpha_{\rm vir}=5\,\sigma_{\rm cl}^2 R_{\rm cl}/(G\,M_{\rm cl})$,
where $\sigma_{\rm cl}$, $R_{\rm cl}$, and $M_{\rm cl}$ are the cloud 1D velocity dispersion, radius, and mass) 
of the Galactic Center GMCs may explain the paucity of star formation activity in this region.
Indeed, analytical and numerical models of turbulent star-forming gas clouds predict a decreasing star 
formation efficiency per free fall timescale with the virial parameter of a GMC (Fedderath \& Klessen 2012;
Padoan et al. 2012, 2017).

Following Robertson \& Goldreich (2012) and Mandal et al. (2020), we expect turbulent adiabatic heating,
i.e., an increase of the turbulent velocity dispersion due to the p\,dV work,  to occur if the timescale 
of large-scale gas compression 
\begin{equation}
\label{eq:tcomp1}
t_{\rm comp}=\rho/(d\rho/dt) 
\end{equation}
is smaller than the dissipation timescale of turbulence $t_{\rm diss}$.
From the dynamical simulations of Vollmer et al. (2012) we derived a compression timescale within the bridge 
of $t_{\rm comp} \la 10$~Myr (Fig.~\ref{fig:tcomptdiss_TAFFY22new_10new1}).
The driving length in the bridge is somewhere between the average cloud size ($l_{\rm cl} \sim 1$~kpc from Table~\ref{tab:molent}) and the filament width
($\sim 3$~kpc), considerably longer than for GMCs in an unperturbed disk.  
The crossing time is then approximately $t_{\rm cross} \sim 2$~kpc\,$/ 50$~km\,s$^{-1} \sim 40$~Myr and this can be taken as $t_{\rm diss}$. 
A detailed comparison between the compression and dissipation timescales of the model is given in Appendix~\ref{sec:compmodel1}.
The $t_{\rm comp}$ is signicantly smaller than $t_{\rm diss}$ in the
bridge but not in the galaxies. Thus, we expect high virial parameters and weak star formation in the bridge gas.
Adiabatic compression and its effect on star formation are included in the dynamical model and the results are 
compared to observations in Appendix~\ref{sec:compmodel1}.

\section{Discussion}

The shorter the timescale, the more important the process is.
In this work, we compare the dissipation timescale, a few Myr as given in Eq.~\ref{eq:tdiss1} which assumes energy injection 
via star formation, to the compression timescale (Eq.~\ref{eq:tcomp1}).
In a disk environment, there is little compression, i.e. $d\rho/dt$ is small, and hence the compression time is long, 
such that dissipation is the dominant process (Fig.~\ref{fig:tcomptdissquiet1}).  
During the Taffy collision, and afterwards in the bridge region, extremely strong 
shocks are present (as witnessed by the H$_2$ emission observed by Peterson et al. 2018) and  $d\rho/dt$ becomes enormous, and 
thus $t_{\rm comp}$ short (Fig.~\ref{fig:tcomptdiss_TAFFY22new_10new1}). 
Furthermore, the dissipation timescale ($l_{\rm driv}/v_{\rm turb}$) in the bridge is higher due to the much 
longer driving scale (Sect.~\ref{sec:adcompmodel}).  These two factors result in $t_{\rm comp} < t_{\rm diss}$.  
The injected energy cannot be evacuated and this largely suppresses the star formation in the bridge.

A single model among our limited set of simulations cannot reproduce all observed characteristics of the Taffy system. However, all
characteristics are present in one of the models (Table~\ref{tab:comp}). The models
sim19 and sim19fast fail to reproduce the gas morphology of UGC~12915, because the model
inclination is significantly lower than the observed edge-on projection.
Since the parameter space for the head-on collision of both galaxies is vast, we did not try to
search for better initial conditions than those found in Vollmer et al. (2012) and thus a better reproduction of the Taffy system.
We could show that the observed detailed velocity structure of the gas bridge can be well reproduced by
one of our models (sim20; Fig.~\ref{fig:taffy_sm2_mom_final_mom1}).
The observed north--south surface brightness gradient of the gas bridge and the increased velocity dispersion of its
high surface brightness part can be reproduced by model sim19fast (Fig.~\ref{fig:taffy_sm2_mom_final_stars}). 
We are thus confident that such a model is in principle possible to account for all observed characteristics (Table~\ref{tab:comp}).
\begin{table*}[!ht]
      \caption{Comparison between our CO(1-0) and GALEX FUV observations and the models.}
         \label{tab:comp}
      \[
         \begin{tabular}{lccc}
           \hline
           feature & sim19 & sim19fast & sim20 \\
           \hline
           gas morphology of UGC~12915 & - & - & + \\
           gas morphology of UGC~12914 & + & + & $\sim$ \\
           morphology of the gas bridge & $\sim$ & $\sim$ & $\sim$ \\
           velocity field of UGC~12915 & $\sim$ & $\sim$ & + \\
           velocity field of UGC~12914 & + & + & + \\
           velocity field of the gas bridge & $\sim$ & $\sim$ & + \\
           velocity dispersion of the gas bridge & - & + & - \\
           global FUV morphology & $\sim$ & + & $\sim$ \\
           large-scale magnetic field$^{\rm (a)}$ & + &  & +  \\
  \noalign{\smallskip}
       \hline
       \noalign{\smallskip}
 \hline
        \end{tabular}
      \]
\begin{list}{}{}
\item[$^{\rm (a)}$ based on the results of Vollmer et al. (2012).]
\end{list}
\end{table*}

Based on our models, we could show that a high-velocity head-on encounter can lead to
a significant fraction of the bridge gas undergoing turbulent adiabatic compression $\sim 20$~Myr after impact.
We claim that the absence of star formation in bridge regions is due to 
turbulent adiabatic compression where the turbulent velocity dispersion of the largest eddies increases.
It is expected that the velocity dispersions of the turbulent substructures/clouds increase such that GMCs are no longer 
in global virial equilibrium. The increase of the virial parameter leads to a decrease of the star 
formation efficiency per free fall timescale in the turbulent ISM (Fedderath \& Klessen 2012; Padoan et al. 2012, 2017) and thus to 
the suppression of star formation.

\begin{figure}[!ht]
  \centering
  \resizebox{\hsize}{!}{\includegraphics{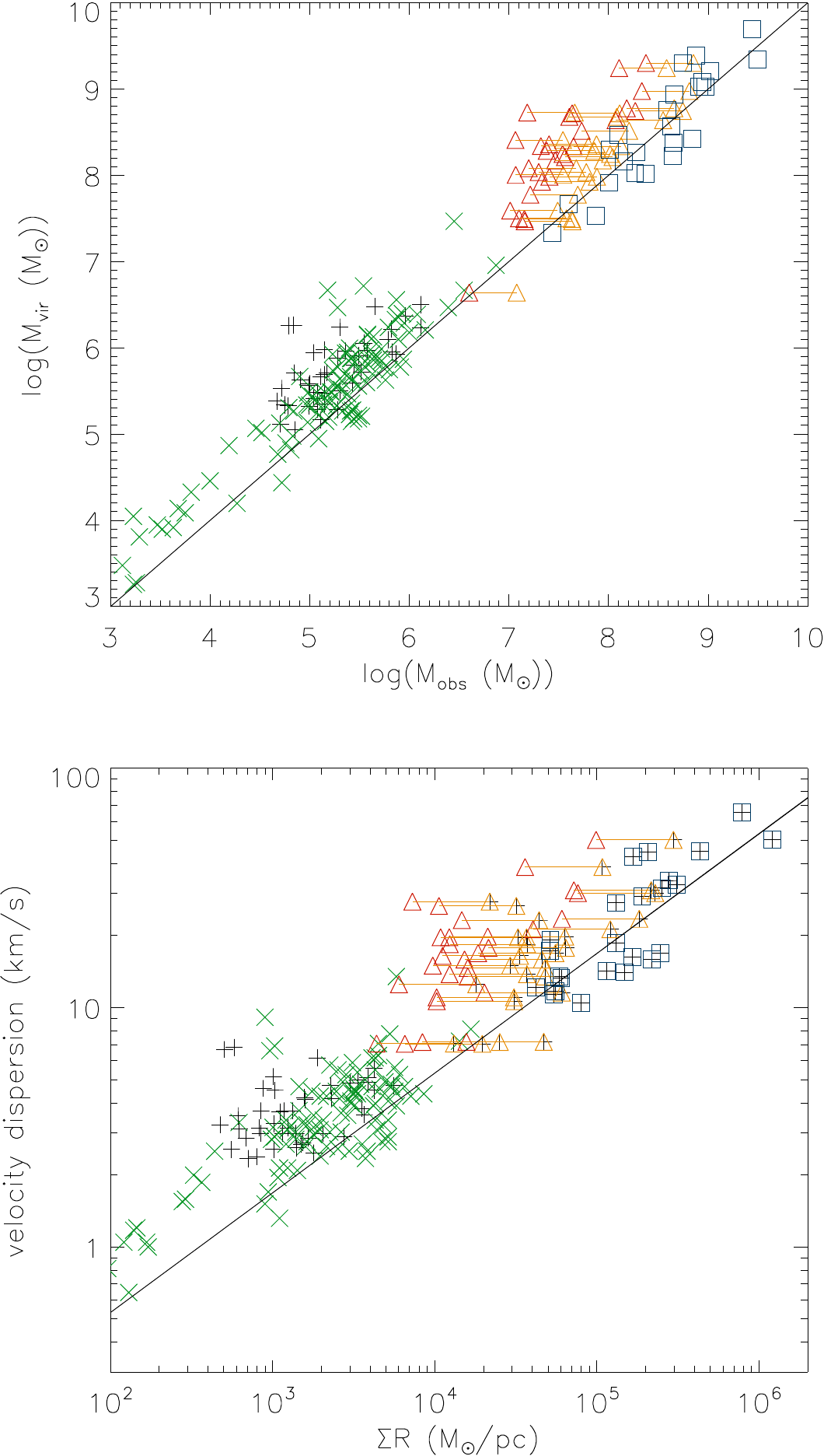}}
  \caption{CO cloud properties. Upper panel: cloud Virial mass as a function of the gas mass derived from 
    the CO luminosity. We applied a $\ratioo$ conversion factor which is one third of the Galactic value
    to the bridge clouds (red triangles). The orange triangles correspond to a Galactic  $\ratioo$ conversion factor.
    Lower panel: cloud velocity dispersion as a function of the product of size and mass surface density.
    CPROPS clouds (bridge: red triangles; galaxies: blue boxes) compared to data from 
    Bolatto et al. (2008; green crosses) and Gratier et al. (2012; black pluses).
    The orange triangles correspond to a Galactic  $\ratioo$ conversion factor.
    The solid line corresponds to Eq.~\ref{eq:virial}.
  \label{fig:virial}}
\end{figure}

Relating the Virial mass of a gas cloud to its CO-derived mass yields
\begin{equation}
\sigma=\sqrt{\pi/5\,G\,R\,\Sigma}\ ,
\label{eq:virial}
\end{equation}
where $R$ and $\Sigma$ are the radius and surface density of the cloud. 
For the disk clouds we applied the Galactic $\ratioo$ conversion factor, for the bridge clouds a three times lower
 $\ratioo$ conversion factor. The resulting relation is shown in Fig.~\ref{fig:virial}.
Whereas the Virial mass of the molecular clouds from Bolatto et al. (2008) and Gratier et al. (2012) are
higher than the gas masses derived from the CO luminosities, the Virial masses of the Taffy disk GMAs are consistent
with the gas masses derived from the CO luminosities. Again, this is surprising because a significant
fraction of their linewidth is expected to be caused by large-scale motions, i.e. rotation and non-circular motions.
Therefore, one should not expect a correlation and the one we found is most probably coincidental.

Based on the comparison between simulations and observations, Vollmer et al. (2012) concluded that the 
bridge extent along the line-of-sight is small compared to its extent in the plane of the sky and
the dominant component of the gas velocities follows the bridge geometry with small line-of-sight gradients.
Applying a Virial analysis and assuming a $\ratioo$ conversion factor
of a third of the Galactic Value, these GMAs have masses well below the Virial mass. They are thus far from being
self-gravitating. The same behavior is observed in the $\Sigma\,R$--$\sigma$
relation (Eq.~\ref{eq:virial}) and is expected in a scenario where the turbulent ISM is compressed adiabatically.

The gas in the bridge region has different phases: the molecular gas is mainly arranged in a filament
with a width of $\sim 3$~kpc, the maximum of the neutral hydrogen emission distribution is shifted to the northwest of
the CO filament (Condon et al. 1993), there are two distinct filaments of ionized gas (Fig.~6 of Joshi et al. 2019) 
and the X-ray emission (Appleton et al. 2015) both of which are also shifted to the northwest of the CO filament.
Thus, the dense molecular gas avoids the other gas phases, especially the diffuse warm and hot phases.
Could it be that the secondary gas tail which is present in all simulations (Fig.~\ref{fig:taffy_sm2_mom_final_stars}) 
is not molecular, but atomic and/or ionized? Based on the FUV image (Fig.~\ref{fig:taffy_sfr}), we argue that the H{\sc i} maximum
stems from gas which belongs to UGC~12914 than to the bridge (see sim19 in Fig.~\ref{fig:taffy_sm2_mom_final_stars}).
The ionized gas is prominent at negative velocities with respect to the systemic velocity. It thus belongs
kinematically more to UGC~12914. The morphology of the hot X-ray emitting gas is reminiscent of the gas distribution
of the northern bridge filament in sim19 and sim19fast. In our simulations this gas has mostly positive velocities with respect
to the systemic velocity. It is thus unlikely that the observed ionized gas coincides with the northern bridge filament.
The observed H{\sc i} in the bridge region (Condon et al. 1993) has a double line structure, as the ionized gas.
At low velocities ($4060$ to $4320$~km\,s$^{-2}$) the H{\sc i} channel maps show a northwest--southeast velocity gradient. At high velocities 
($4440$ to $4570$~km\,s$^{-1}$) there seems to be a southwest--northeast gradient present. The low-velocity part of the
H{\sc i} emission belongs to UGC~12914, whereas the high-velocity part belongs to UGC~12915. 

What is the fate of the bridge gas? Will the high surface density bridge region close to UGC~12915 collapse and
form stars or will it expand and disperse? We think that most of the low-surface density, CO-emitting gas
will disperse without forming stars. On the other hand, the high-density gas will probably have a different fate.
It is remarkable that the luminous extraplanar H{\sc ii} region 
close to UGC~12915 does not coincide with a bridge GMA (but there is a GMA close to it; lower panel of Fig.~\ref{fig:taffy_spitzer}). 
This implies that the gas cloud(s) from which the
H{\sc ii} region has formed has already been disrupted by stellar feedback (stellar wind and supernova explosions).
For this process we offer the following explanation: the compression timescale is proportional to the
gas density (Eq.~\ref{eq:tcomp}), where the dissipation timescale is proportional to the square root of the density
(Eq.~\ref{eq:tdiss1}). At the beginning of the phase of adiabatic compression the gas density is not too high
permitting $t_{\rm comp} < t_{\rm diss}$. During the phase of adiabatic compression the gas density increases until 
$t_{\rm comp} > t_{\rm diss}$ and the region collapses and forms stars. 
On the Spitzer $3.6$~$\mu$m $1.7''$-resolution image the H{\sc ii} region is round and has FWHM of $5''$ or
$\sim 1.5$~kpc. This is about the same size as GMA~9 (Table~\ref{tab:molent}). 
Compared to the extreme molecular cloud in the Antennae system (Johnson et al. 2015), GMA~9 has
an about ten times higher mass ($\sim 2 \times 10^8$~M$_{\odot}$ assuming a $\ratioo$ conversion factor which is one third
of the Galactic value), but a comparable velocity dispersion. The size of the Antennae cloud is only $24$~pc.
This implies that GMA~9 will certainly be resolved into several distinct clouds.
We can only speculate that single massive
high-velocity dispersion molecular clouds collapsed due to their high density and formed the H{\sc ii} region
composed of several dense star clusters.
High-resolution ALMA CO observations (Appleton et al., in prep.) will give further insight into
the formation scenario of this atypical H{\sc ii} region.

We suggest that star clusters with extreme stellar densities ($\ga 10^4$~stars\,pc$^{-3}$), such as
globular clusters and super star clusters (O'Connell et al. 1994), formed and still form through the gravitational
collapse of gas previously compressed by turbulent adiabatic compression during galaxy interactions.
During the compression phase the cloud accumulates mass and increases its velocity dispersion. The
high velocity dispersion prevents collapse  but once the
critical density reached the turbulent energy is dissipated rapidly and the cloud collapses and forms
an extremely dense and massive star cluster.

\begin{figure}[!ht]
  \centering
  \resizebox{\hsize}{!}{\includegraphics{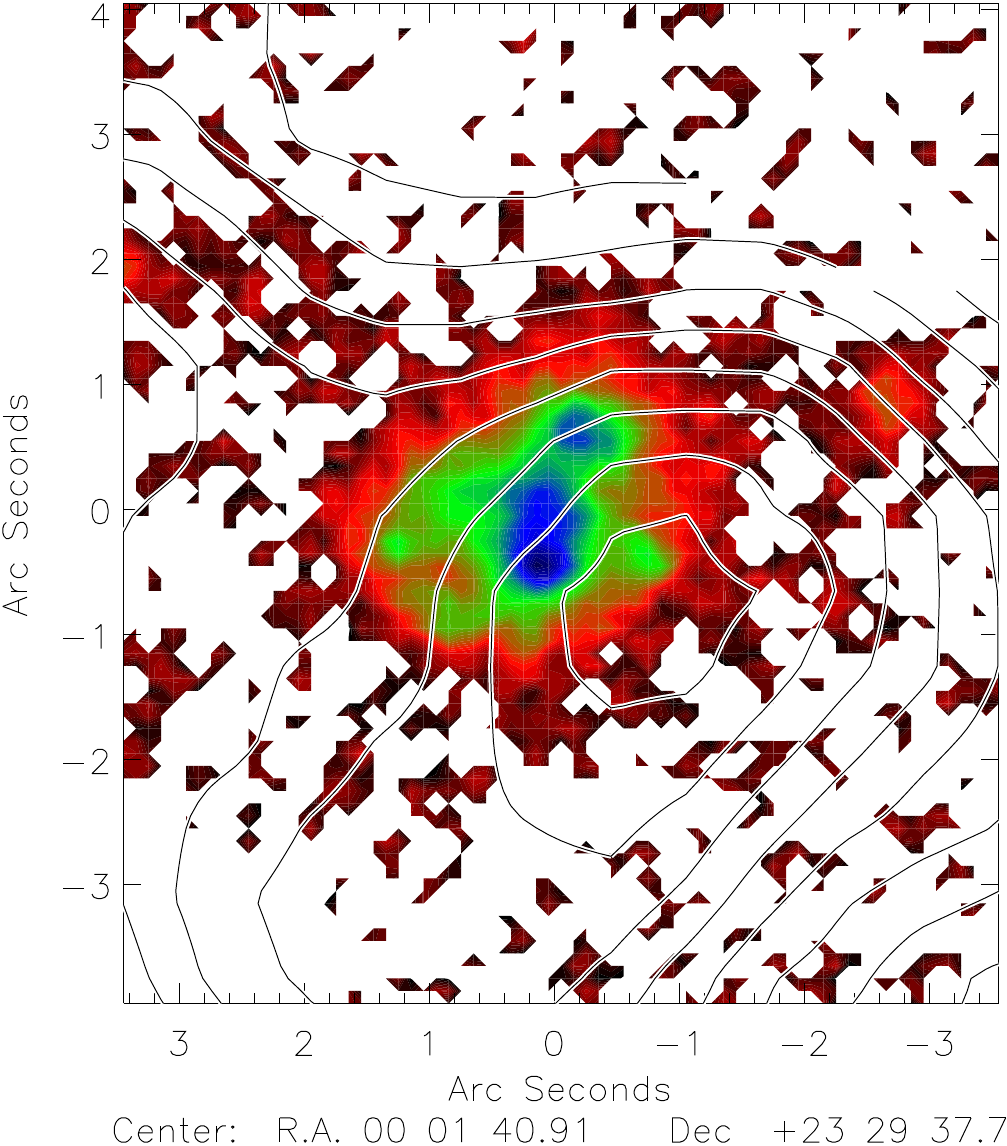}}
  \resizebox{\hsize}{!}{\includegraphics{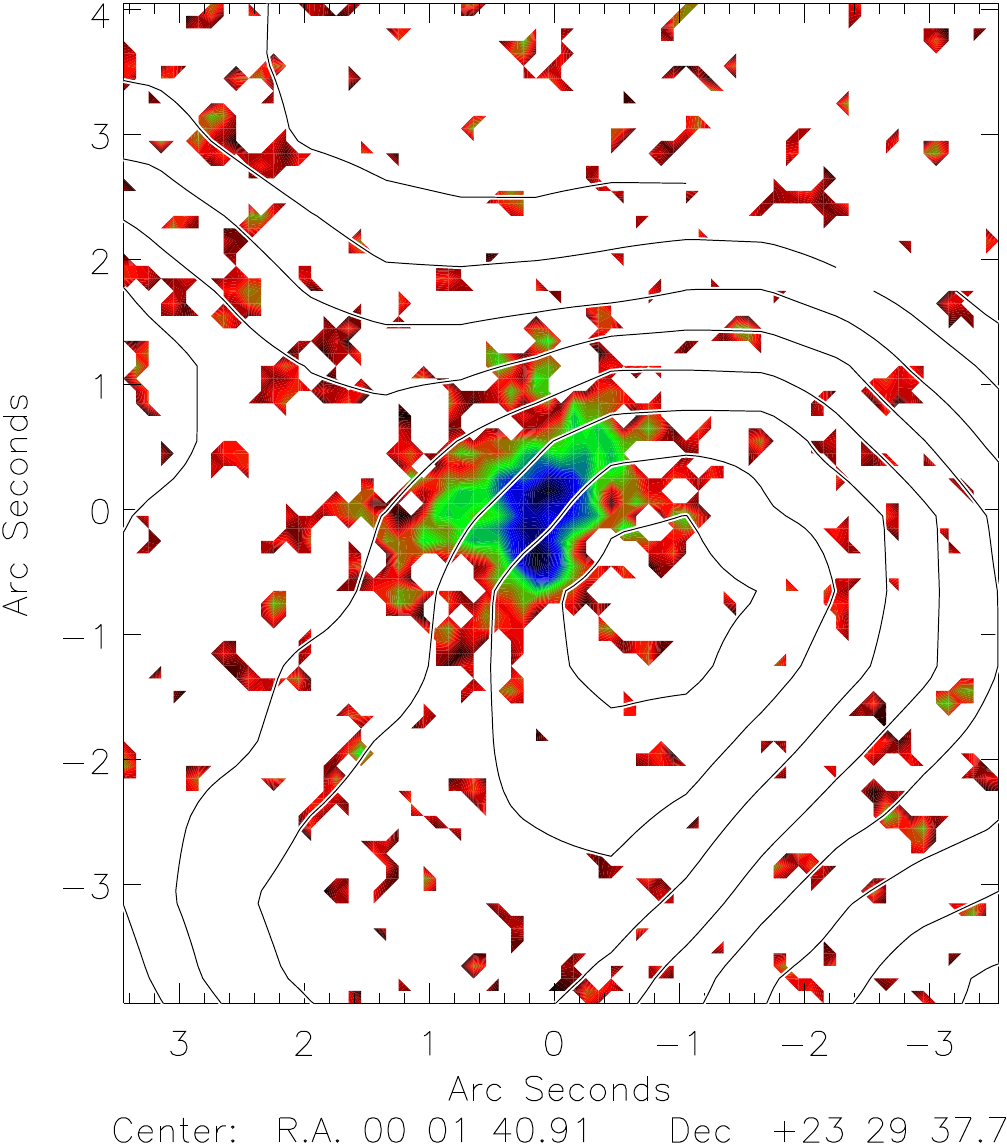}}
  \caption{Close-up of the giant bridge HII region, CO(1-0) contours on the HST NICMOS F190N (Pa-$\alpha$; upper panel) and F187N (off-band; lower panel) image
    (PropID~11080, PI: D. Calzetti). The HST astrometry was approximately aligned with the Spitzer astrometry. 
    Contour levels are $(10,20,30,40,50,60,70,80)$~K\,km\,s$^{-1}$. 
  \label{fig:taffy_hst}}
\end{figure}

This scenario probably applies to the  extragalactic H{\sc ii} region close to UGC~12915 (lower panel of Fig~\ref{fig:taffy_spitzer}):
the Pa-$\alpha$ emission of the H{\sc ii} region detected in the HST NICMOS\footnote{Retrieved from the MAST HLA database.}
 F190N filter (upper panel of Fig.~\ref{fig:taffy_hst})
has a complex structure within a circular region
of $\sim 600$~pc diameter: a central prominent compact source with a FWHM of $0.4''=120$~pc with a northern extension and three fainter
compact sources of about the same size. In the F187N off-band filter (lower  panel of Fig.~\ref{fig:taffy_hst}) 
only the prominent compact source and a second compact source
in the northern ionized extension are visible. The size of the compact source is in excess but comparable to the size of the
largest super star cluster in the Antennae galaxies  (SSC_B: FWHM of $1''=95$~pc and mass of $5 \times 10^6$~M$_{\odot}$; 
Gilbert \& Graham 2007). The F187N emission is either dominated by massive O stars if the super star clusters are younger than $\sim 8$~Myr or 
by red supergiants if they are older. The maximum age of the clusters is given by the time since the interaction $\sim 20$~Myr.
Based on these findings we suggest that super star clusters were and maybe still are formed within the bridge H{\sc ii} region 
close to UGC~12915. 

\section{Conclusions}

The Taffy system is composed of two massive spiral galaxies which had a head-on collision about $20$~Myr ago.
We present new high-resolution ($\sim 2.7''$) CO(1-0) observations with the Plateau de Bure Interferometer.
An rms of $\sim 5$~mJy in a $6.5$~km\,s$^{-1}$ channel was reached by our observations. The CPROPS software (Rosolowsky \& Leroy 2006)
was used to identify and measure the properties of giant molecular cloud associations (GMAs). 
The detected CO luminosity of the Taffy system is $L_{\rm CO,tot}=4.8 \times 10^9$~K\,km\,s$^{-1}$pc$^2$. 
We divided the CO intensity map into disk and bridge regions (Fig.~\ref{fig:bridge_separation}). The CO luminosity of
the bridge is $L_{\rm CO,bridge}=1.2 \times 10^9$~K\,km\,s$^{-1}$pc$^2$, $25$\,\% of the total CO luminosity. 
Assuming a Galactic $\ratioo$ conversion factor for the galactic disks and a third of this value for
the bridge gas, we obtain H$_2$ masses of $M_{\rm H_2,tot}=1.7 \times 10^{10}$~M$_{\odot}$ and
$M_{\rm H_2,tot}=1.7 \times 10^{9}$~M$_{\odot}$. Thus, about $10$\,\% of the molecular gas mass is located in the bridge region.

The bulk of the bridge high-density gas does not form stars (Braine et al. 2003, Gao et al. 2003). 
The luminous extraplanar H{\sc ii} region south of UGC~12915
represents the exception to that rule. A close-up of the region (lower panel of  Fig.~\ref{fig:taffy_spitzer})
shows that the H{\sc ii} region does not coincide with, but is located at the northern edge of a high-surface brightness GMA
(GMA~9 in Table~\ref{tab:molent}) with a flux of $1.7 \times 10^8$~K\,km\,s$^{-1}$pc$^2$ and a velocity dispersion of
$50$~km\,s$^{-1}$). This GMA has the highest velocity dispersion of the bridge GMAs.

We separated the CO clouds identified by CPROPS into disk and bridge clouds.
It is remarkable that the GMAs in the disk and bridge regions approximately follow
the size--linewidth relation established by Bolatto et al. (2008) for extragalactic and Galactic molecular clouds. 
The scatter around the relation is also comparable to that of Bolatto et al. (2008) and Gratier et al. (2012).
On the other hand, the size--luminosity relations of the GMAs in the bridge and disk regions are different:
the bridge GMAs have lower luminosities for their sizes than the disk GMAs and the bridge GMAs are clearly not virialized.

The CO(1--0) observations were compared to the dynamical models of Vollmer et al. (2012) together with a new simulation. 
None of the simulations reproduce all observed features of the Taffy system. However, all characteristics can be found in 
one of the three models. Table~\ref{tab:comp} lists the features reproduced (or not) by each of the models.
 
Rapid turbulent adiabatic compression induced by the $\sim 1000$~km\,s$^{-1}$ collision could explain the high velocity dispersions 
and the subsequent suppression of star formation (Fedderath \& Klessen 2012; Padoan et al. 2012, 2017) in the Taffy bridge.
In this scenario the turbulent velocity dispersions of the largest eddies and their substructures/clouds increase such that
GMCs are no longer in global virial equilibrium.

The suppression of star formation caused by turbulent adiabatic compression was implemented in the dynamical simulations:
once the gas compression timescale is shorter than the turbulent dissipation timescale, star formation is suppressed.
This mechanism decreased the model star formation in the bridge region by a factor of about three to five, consistent with observations.

The bulk of the bridge molecular gas is not gravitationally bound and will disperse.
The densest regions will probably become self-gravitating and form stars as in the giant bridge H{\sc ii} region. 
Because of their enhanced velocity dispersion these regions are much denser and more massive than common galactic GMCs.
This mechanism could explain the extreme stellar densities in globular clusters and super star clusters (O'Connell et al. 1994),
as observed in the Antennae.

\begin{acknowledgements}
We would like to thank the IRAM staff for the help observing the Taffy system with the PdBI.
Based on observations made with the NASA/ESA Hubble Space Telescope,  
and obtained from the Hubble Legacy Archive, which is a collaboration 
between the Space Telescope Science Institute (STScI/NASA), the Space 
Telescope European Coordinating Facility (ST-ECF/ESA) and the Canadian Astronomy Data Centre (CADC/NRC/CSA).
We thank Dominique Aubert for the creation of the 3D views.
\end{acknowledgements}

\appendix

\section{Moment maps and bridge separation \label{sec:momentmaps}}

\begin{figure}[!ht]
  \centering
  \resizebox{\hsize}{!}{\includegraphics{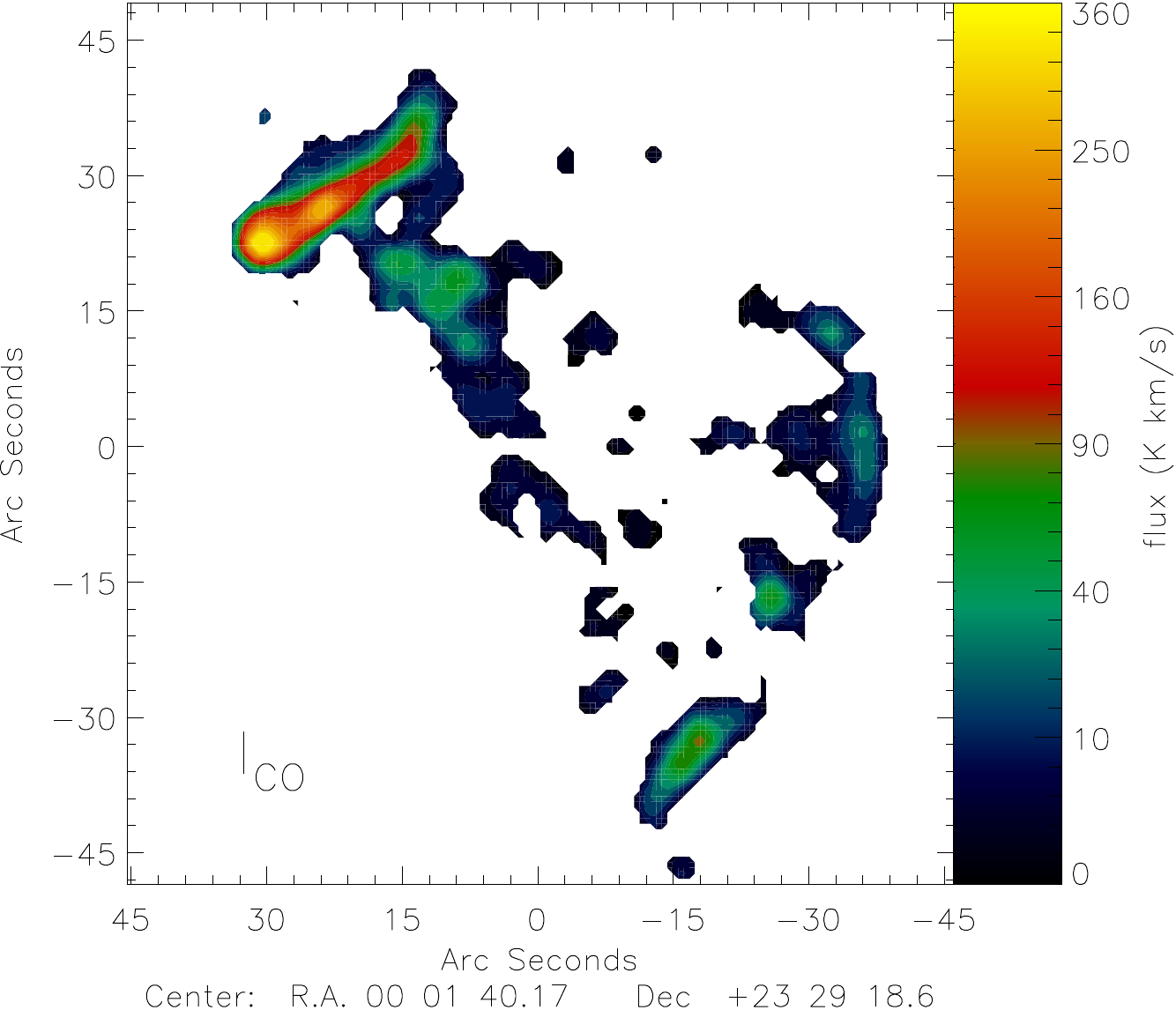}}
  \resizebox{\hsize}{!}{\includegraphics{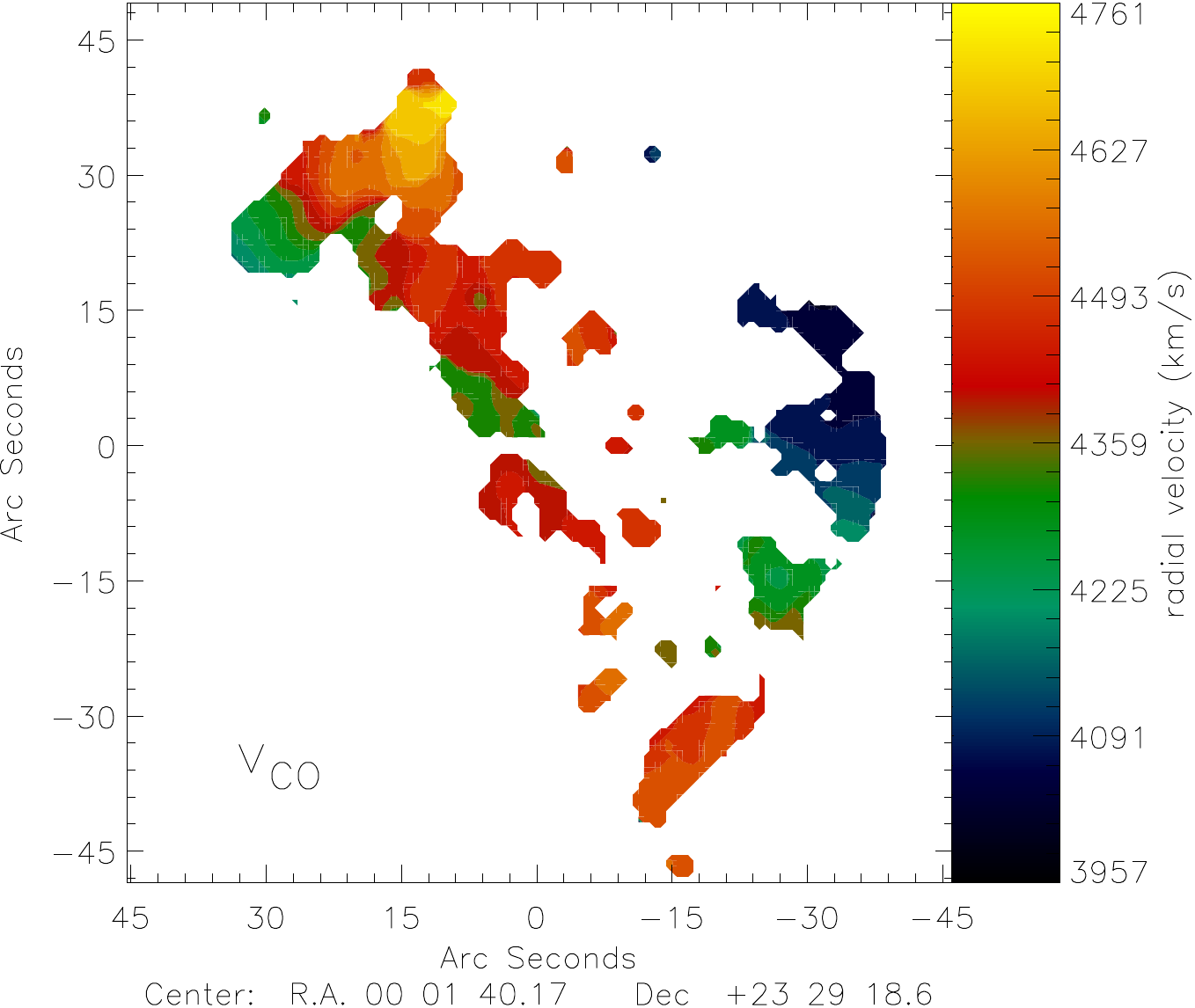}}
  \resizebox{\hsize}{!}{\includegraphics{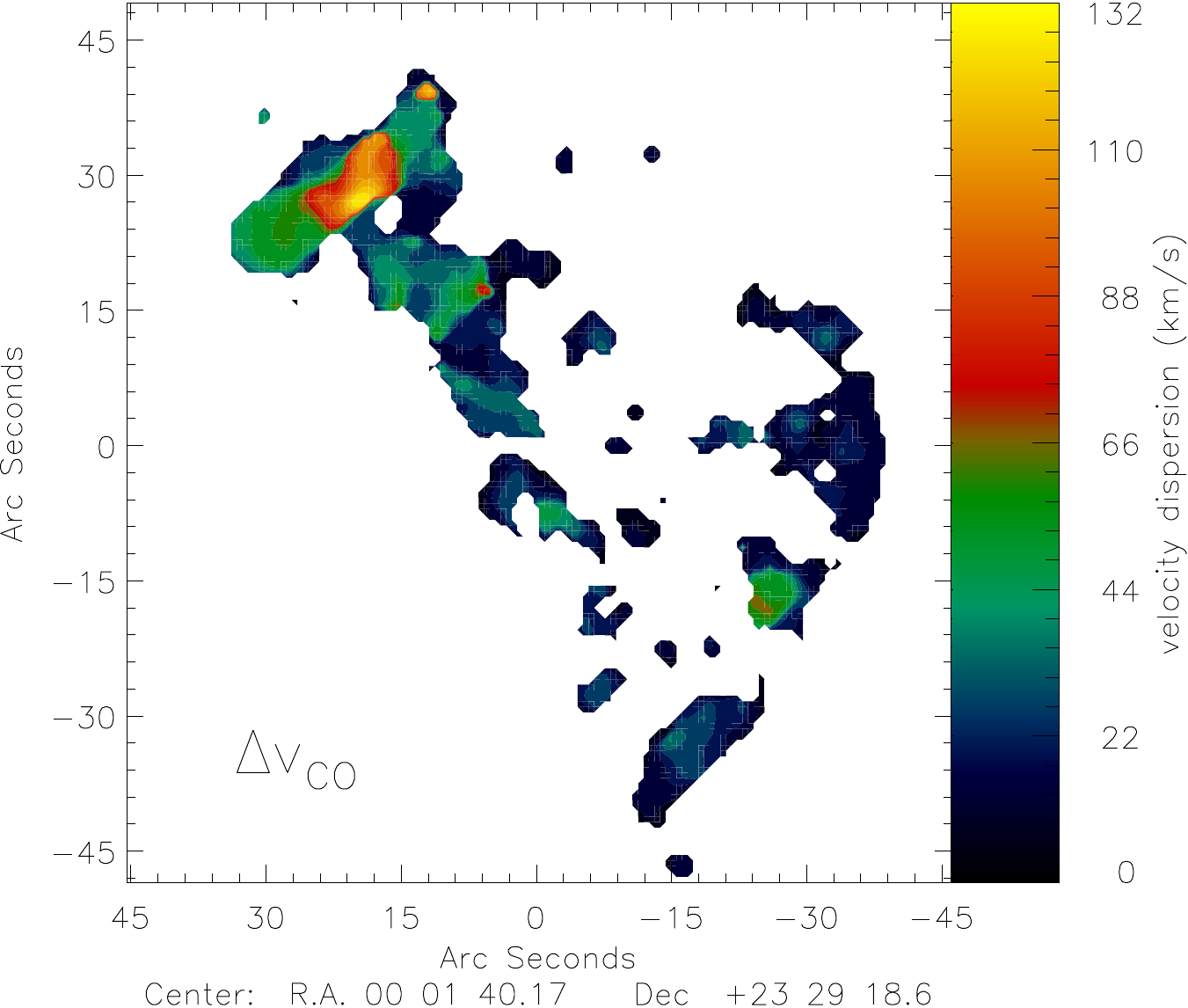}}
  \caption{Classical CO(1-0) moment maps.
  \label{fig:taffy_obs_mom_final_mom0c}}
\end{figure}

\begin{figure}[!ht]
  \centering
  \resizebox{\hsize}{!}{\includegraphics{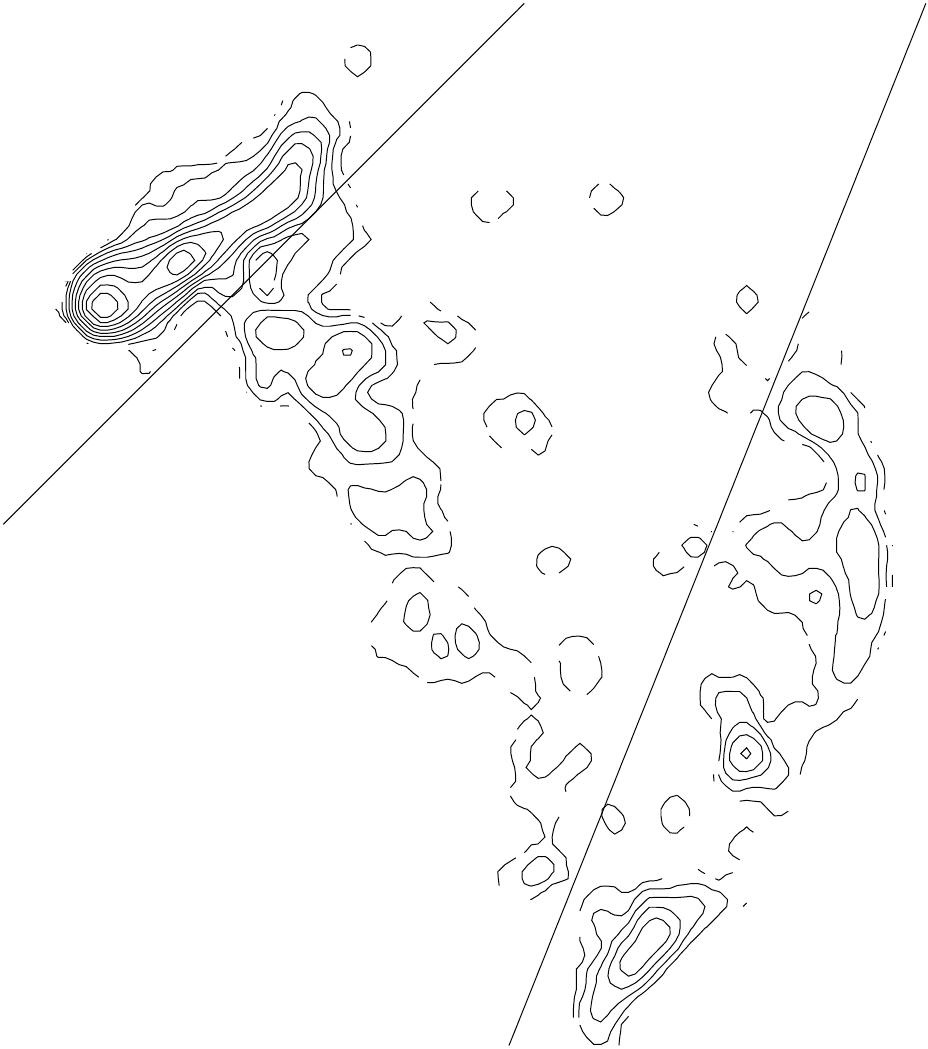}}
  \caption{Separation between the disk and bridge regions.
  \label{fig:bridge_separation}}
\end{figure}

\section{Additional moment maps and 3D views of the datacubes}

\begin{figure*}[!ht]
  \centering
  \resizebox{\hsize}{!}{\includegraphics{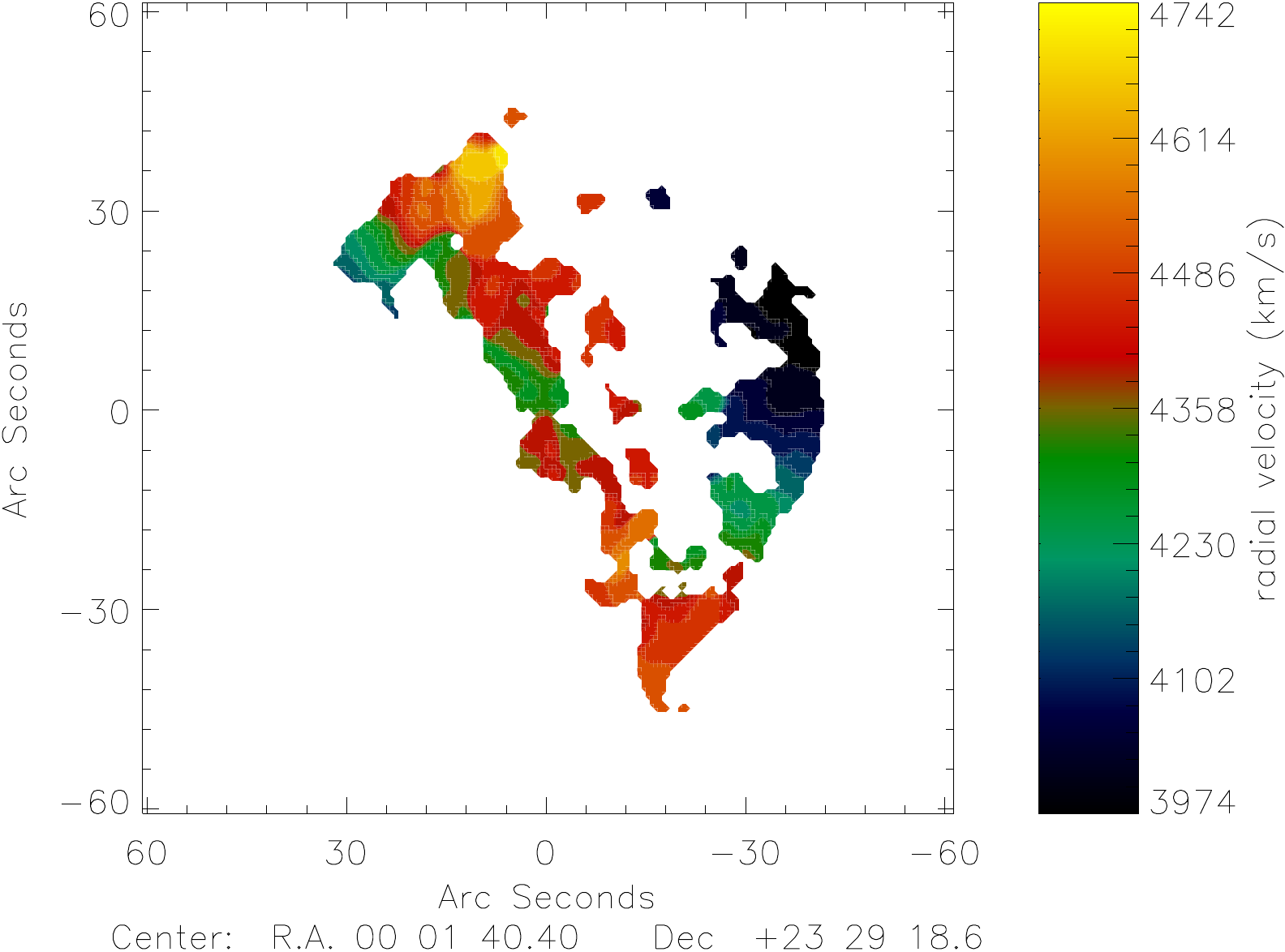}\includegraphics{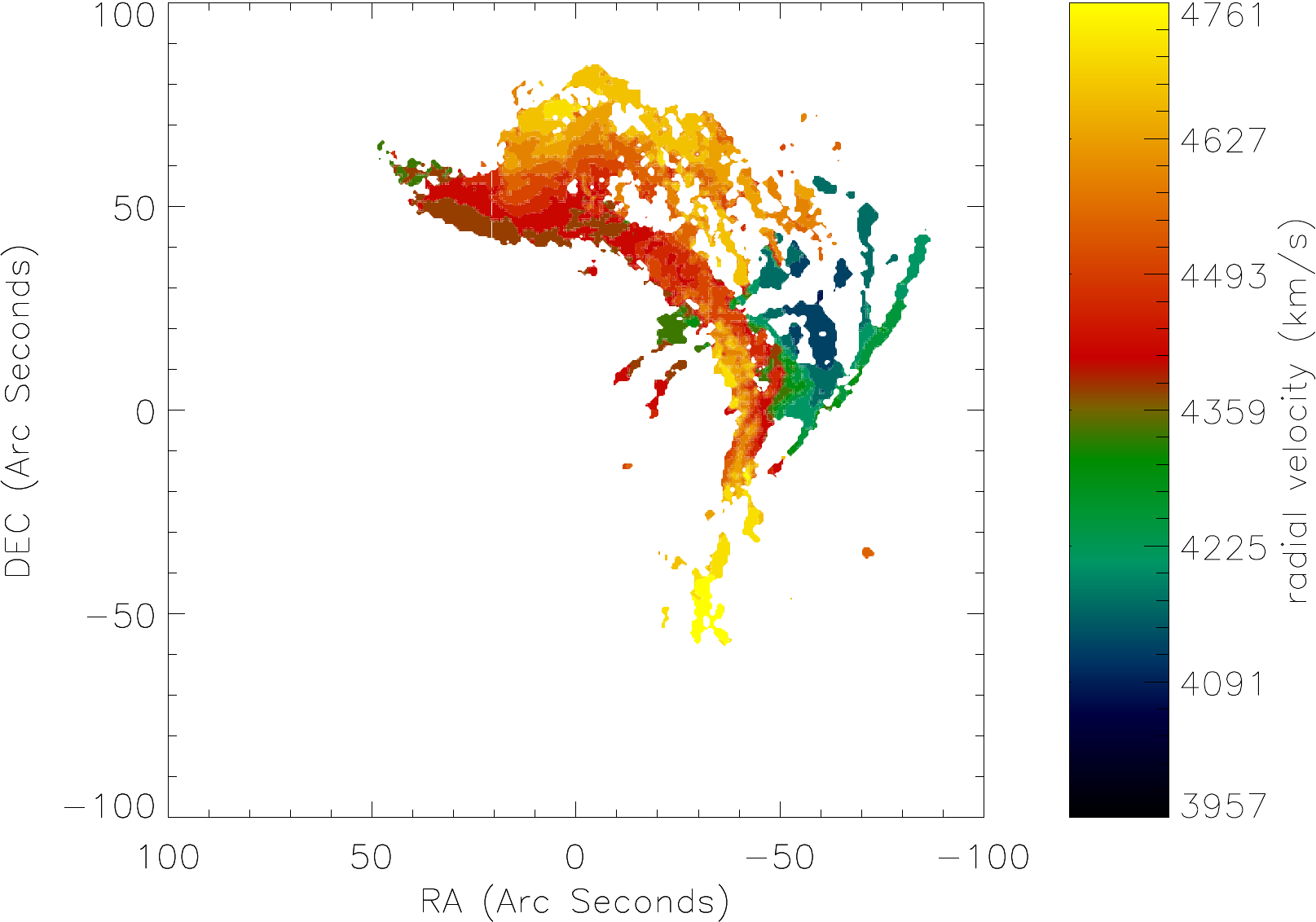}}
 \put(-480,35){\Large observations}
 \put(-220,35){\Large sim 19}
 \vspace{0cm}
  \resizebox{\hsize}{!}{\includegraphics{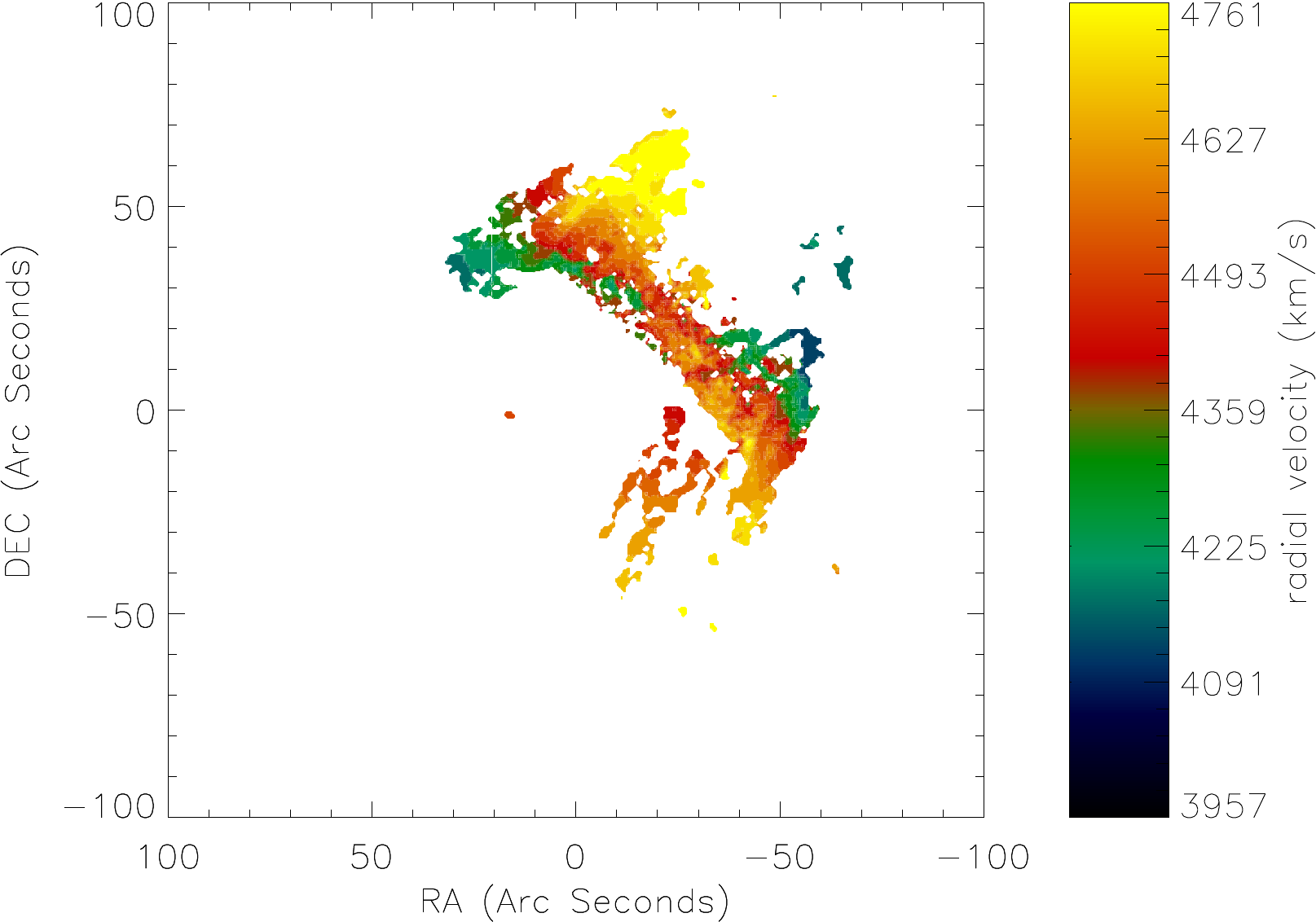}\includegraphics{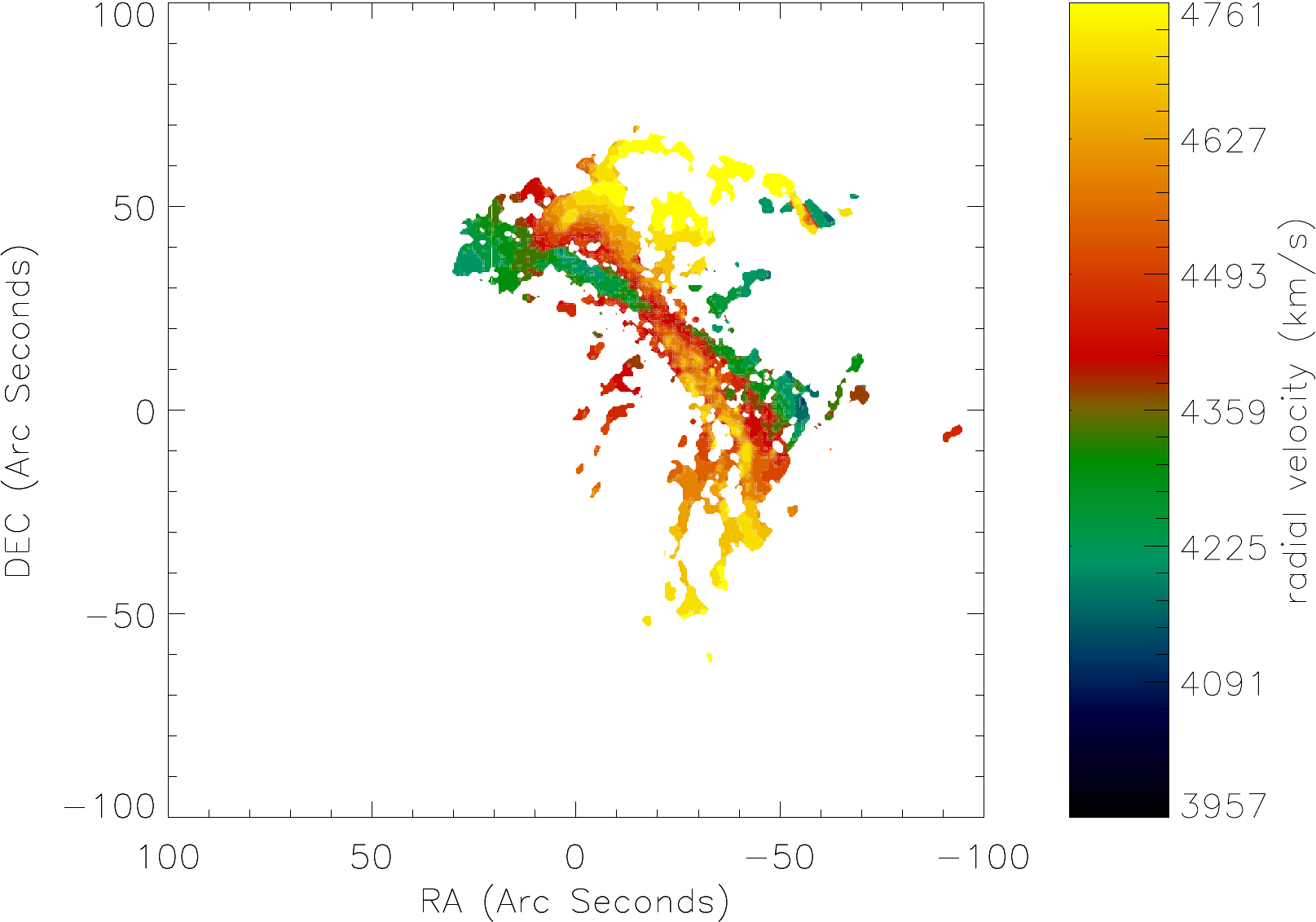}}
  \put(-480,35){\Large sim 19 fast}
  \put(-220,35){\Large sim 20}
  \caption{CO(1-0) moment~1 maps together with the model H$_2$ moment~0 maps. Upper left panel: PdBI observations, other panels: simulations.
  \label{fig:taffy_sm2_mom_final_mom1}}
\end{figure*}

\begin{figure*}[!ht]
  \centering
  \resizebox{\hsize}{!}{\includegraphics{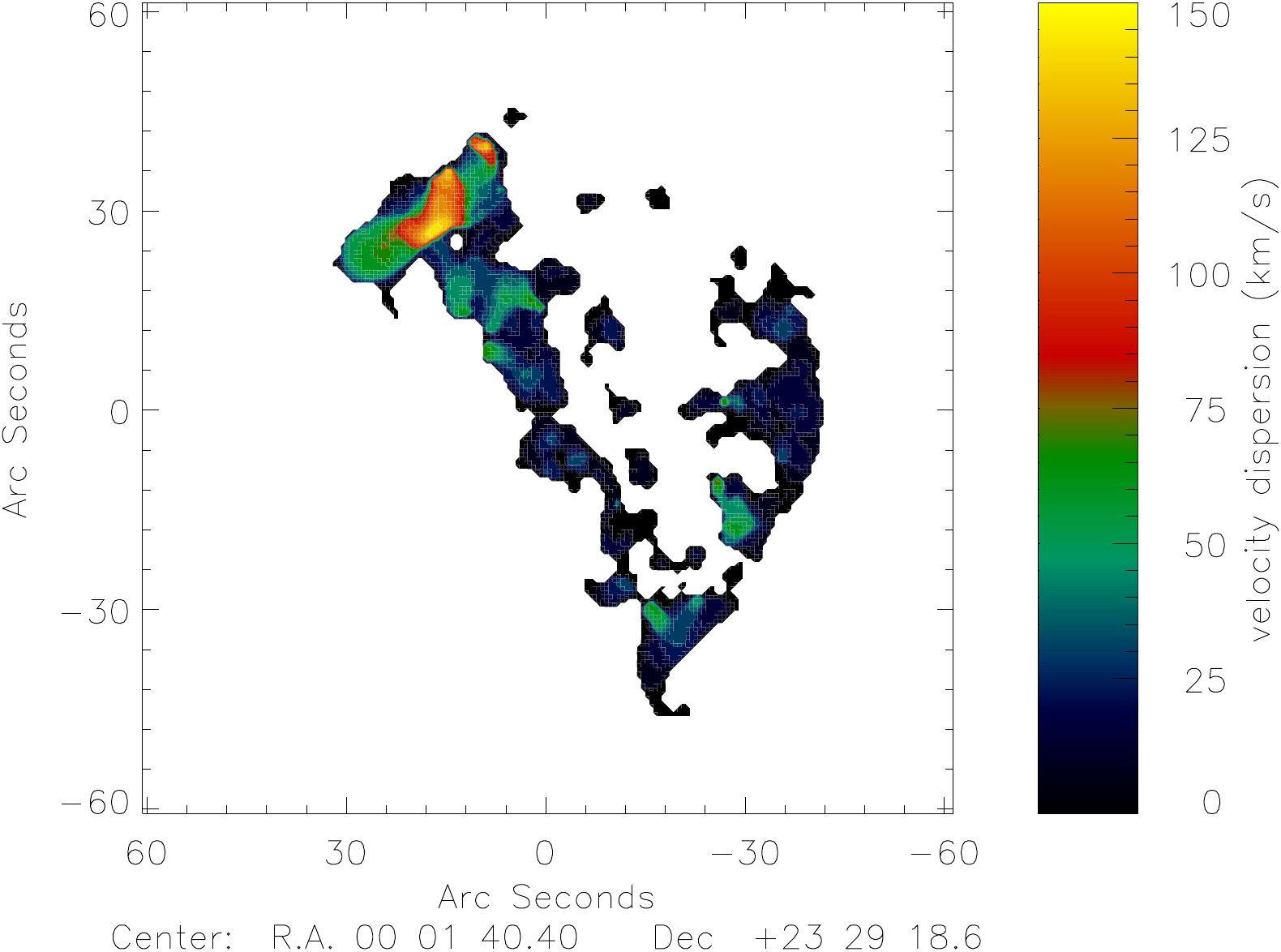}\includegraphics{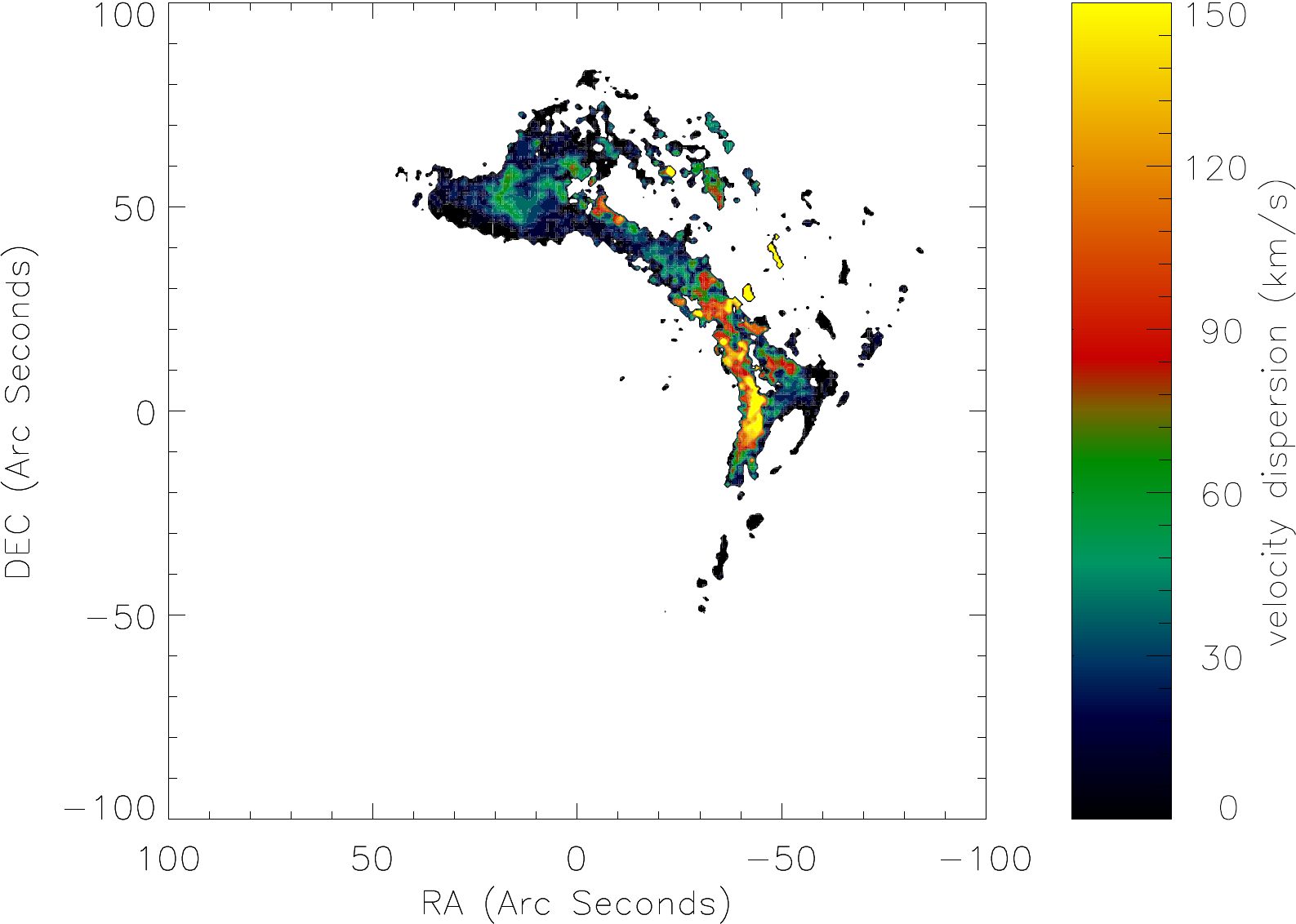}}
   \put(-480,35){\Large observations}
 \put(-220,35){\Large sim 19}
 \vspace{0cm}
  \resizebox{\hsize}{!}{\includegraphics{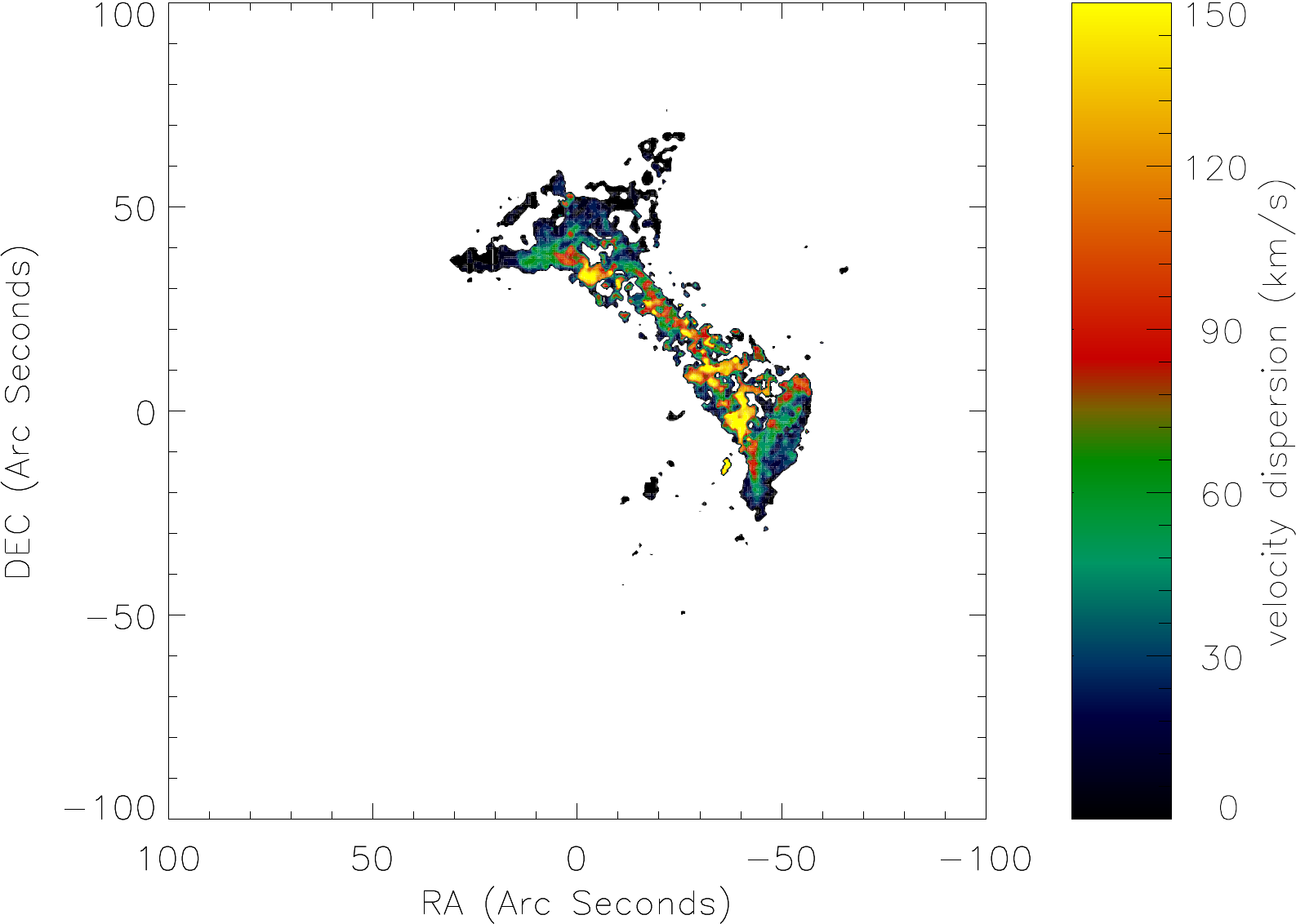}\includegraphics{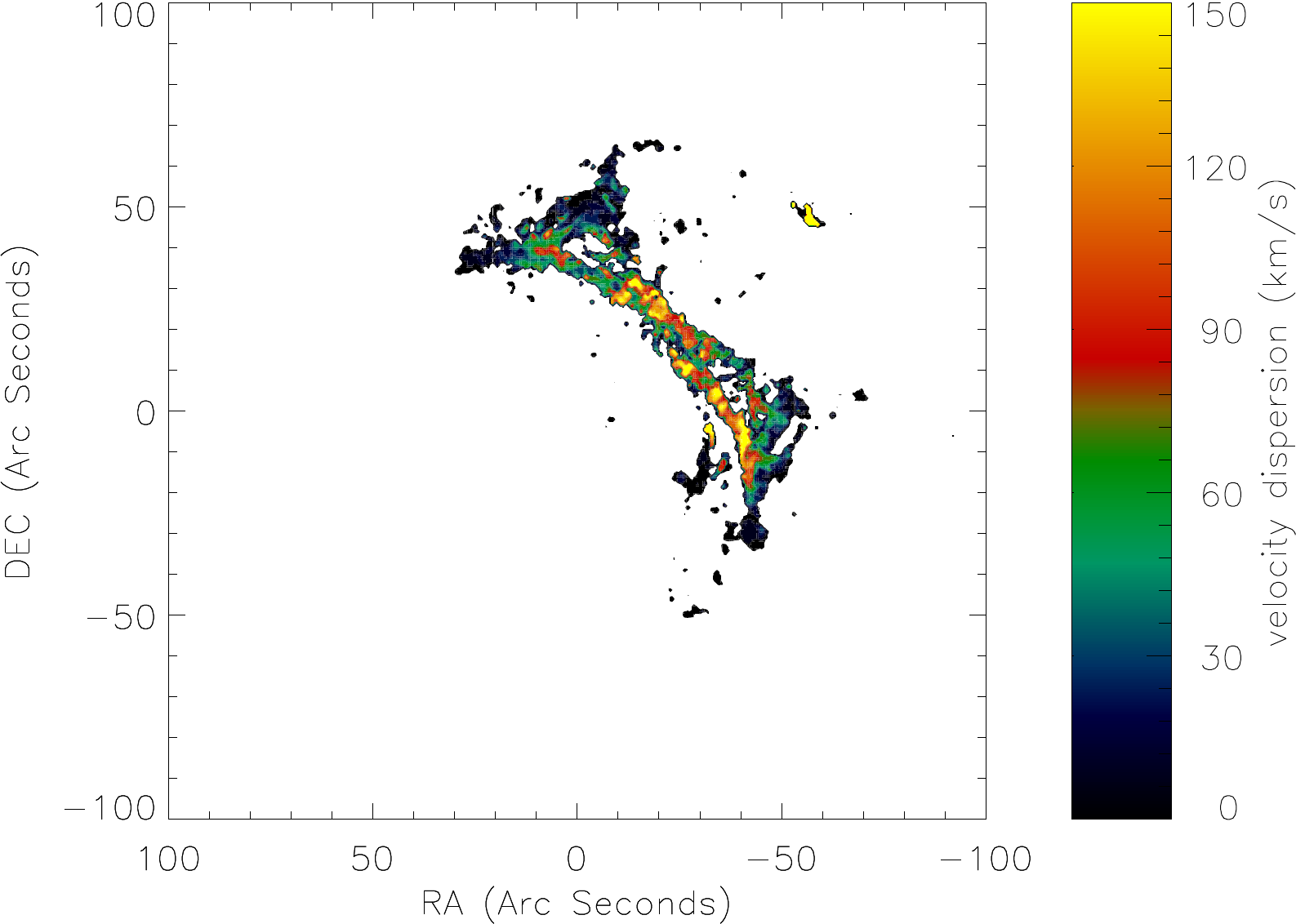}}
  \put(-480,35){\Large sim 19 fast}
  \put(-220,35){\Large sim 20}
  \caption{CO(1-0) moment~2 maps together with the model H$_2$ moment~0 maps. Upper left panel: PdBI observations, other panels: simulations.
  \label{fig:taffy_sm2_mom_final_mom2}}
\end{figure*}

\begin{figure*}[!ht]
  \centering
  \resizebox{\hsize}{!}{\includegraphics{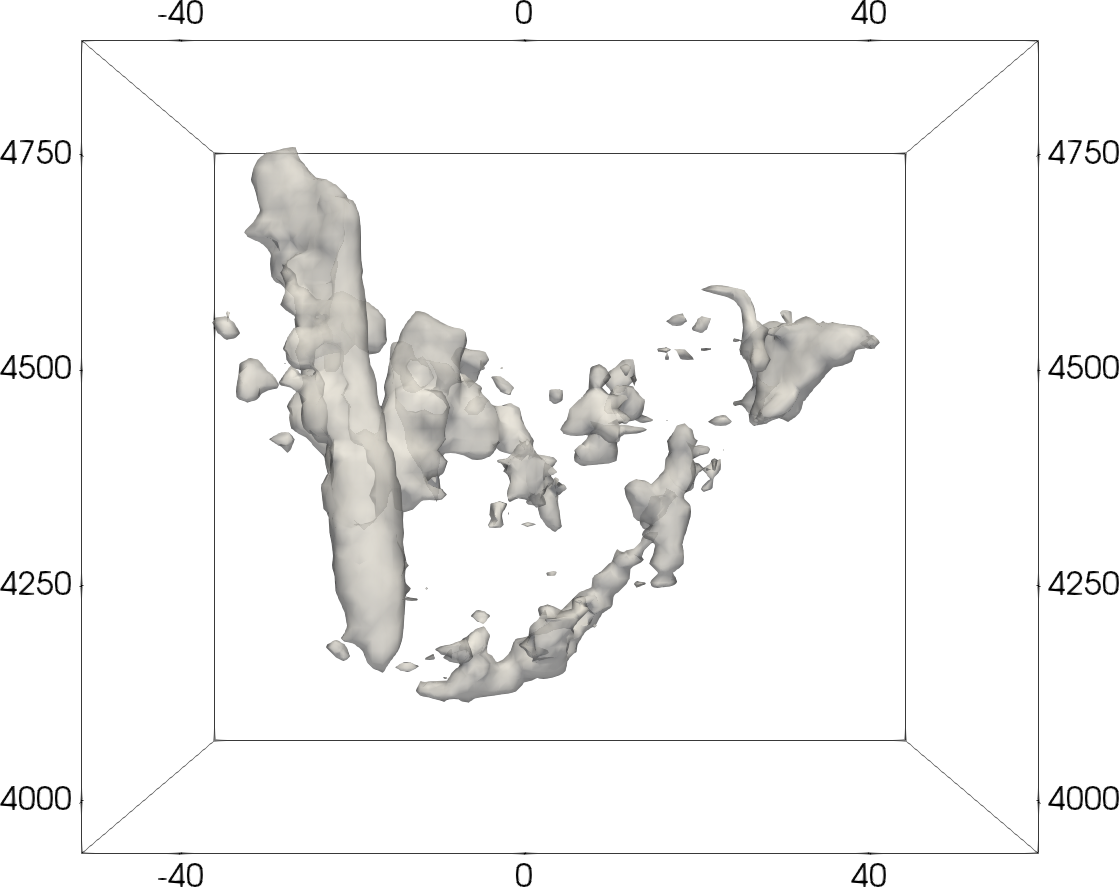}\includegraphics{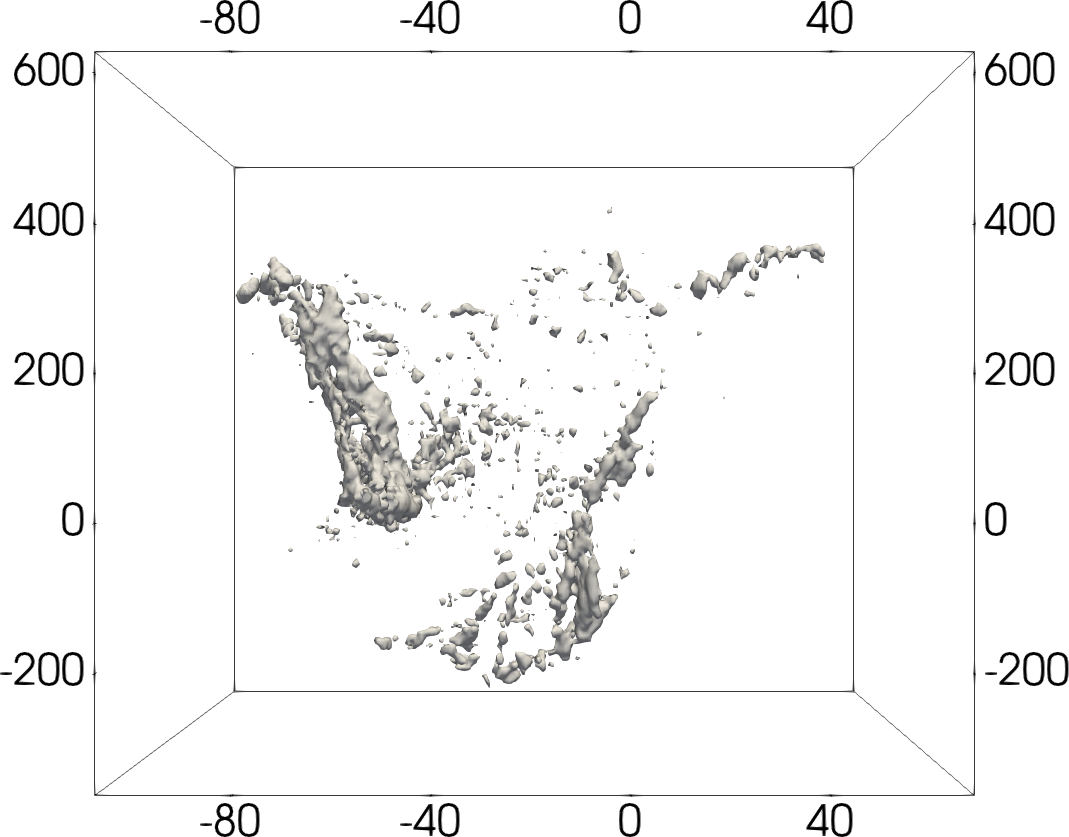}}
   \put(-460,190){\Large observations}
  \put(-200,170){\Large sim 19}
  \put(-530,100){\large $v_{\rm r}$ (km\,s$^{-1}$)}
  \put(-430,10){\large RA offset (arcsec)}
 \vspace{0cm}
  \resizebox{\hsize}{!}{\includegraphics{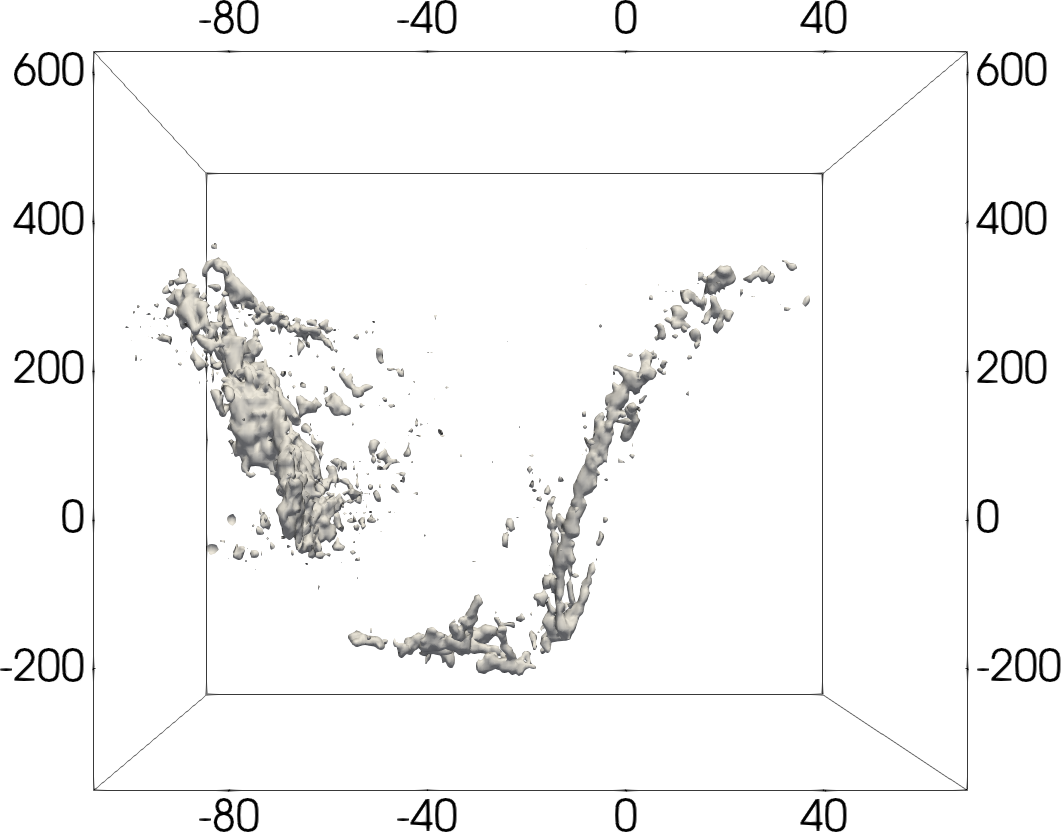}\includegraphics{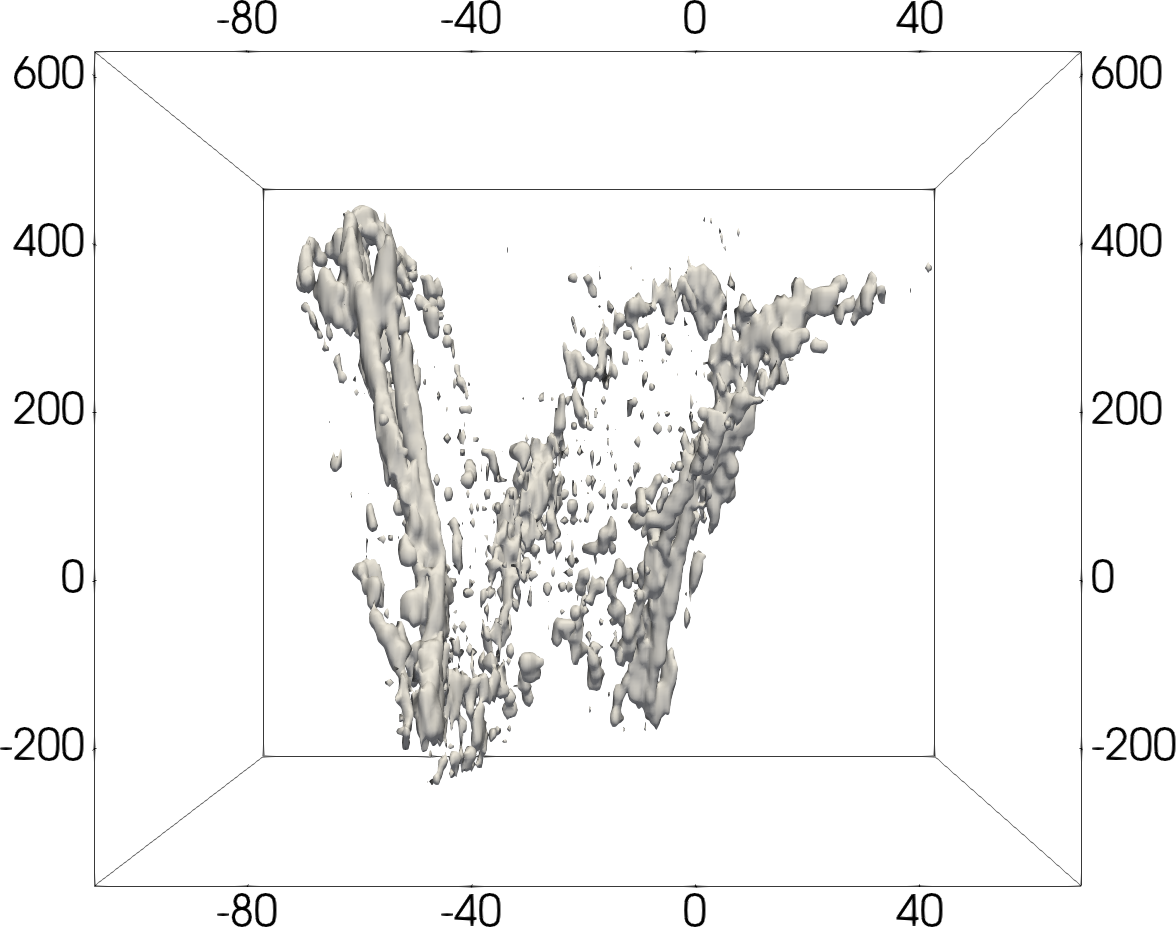}}
   \put(-460,170){\Large sim 19 fast}
  \put(-200,170){\Large sim 20}
  \caption{Second 3D view of the observed CO(1-0) datacube and the model H$_2$ data cubes. The axis labels are only shown for the observations. 
    For a better understanding of these views, three 3D animations of the rotating datacube are attached to this figure (\texttt{taffy\_cube3D\_z.gif}, 
      \texttt{taffy\_cube3D\_z1.gif}, and \texttt{taffy\_cube3D\_x.gif}).
  \label{fig:Taffy_sm2_3D_x_85d_eclump}}
\end{figure*}

\begin{figure*}[!ht]
  \centering
  \resizebox{\hsize}{!}{\includegraphics{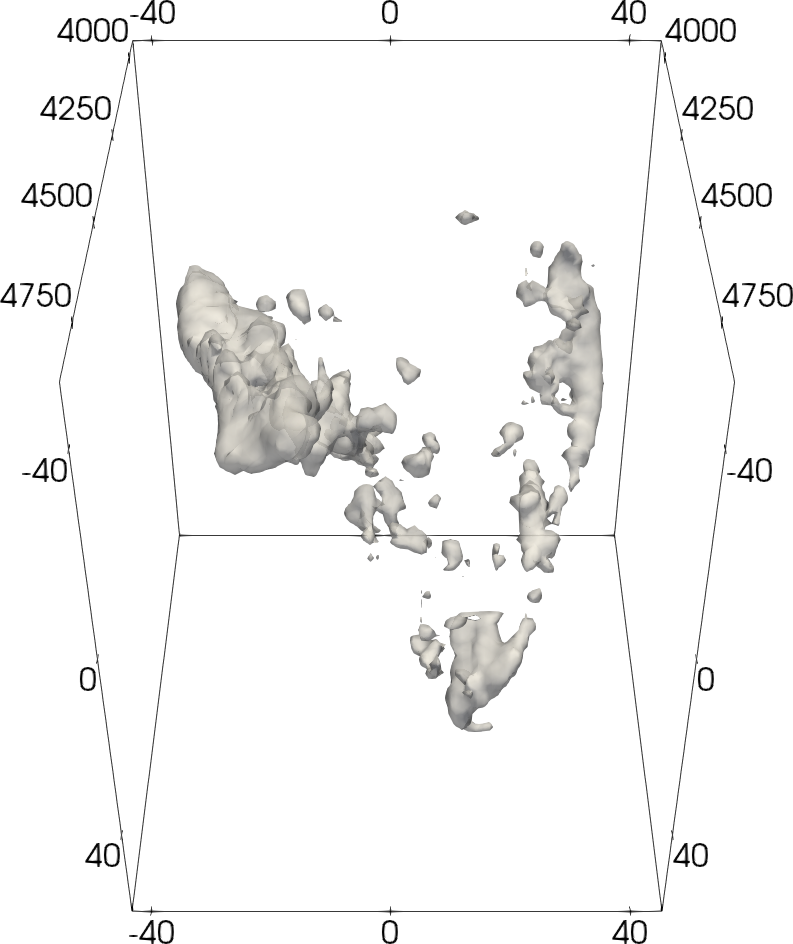}\includegraphics{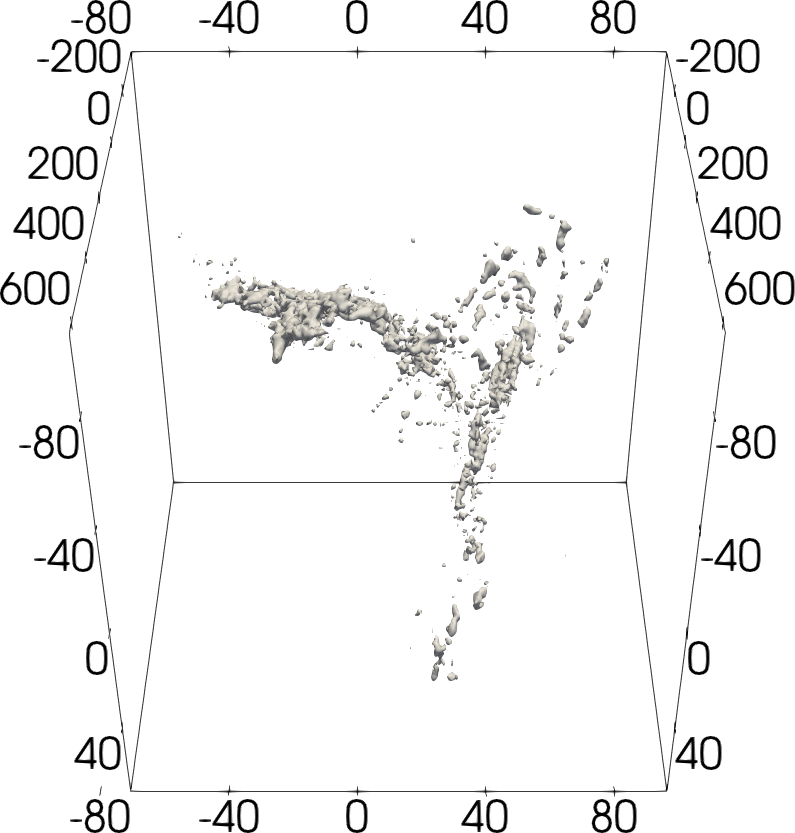}}
   \put(-460,270){\Large observations}
   \put(-200,230){\Large sim 19}
\put(-530,260){\large $v_{\rm r}$ (km\,s$^{-1}$)}
  \put(-430,10){\large DEC offset (arcsec)}
  \put(-530,110){\large RA offset (arcsec)}
 \vspace{0cm}
  \resizebox{\hsize}{!}{\includegraphics{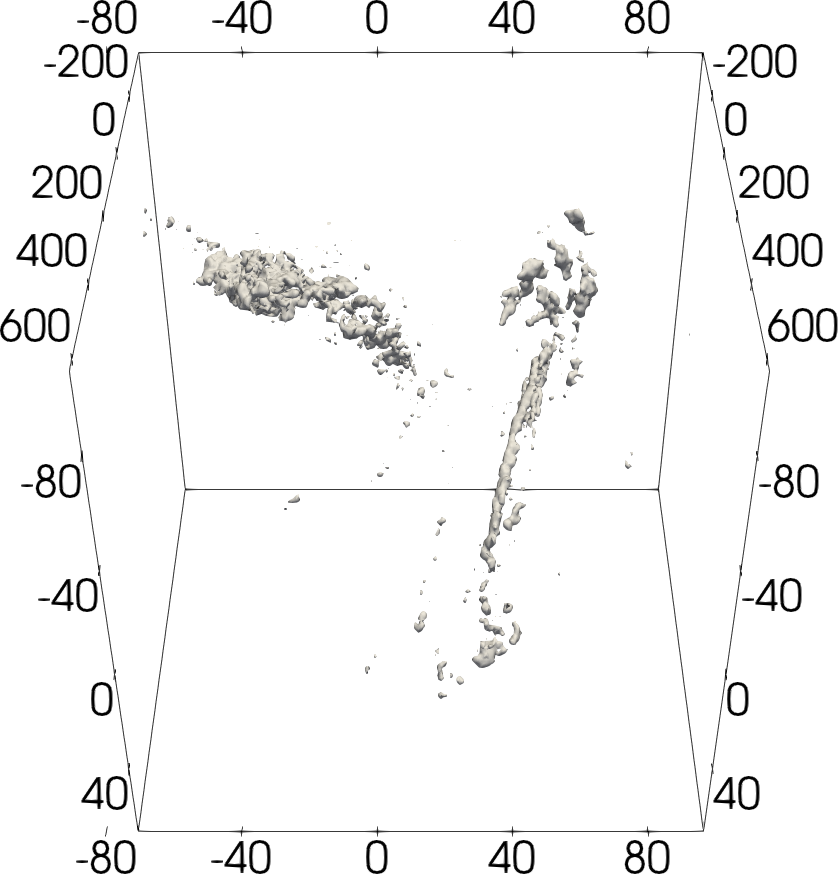}\includegraphics{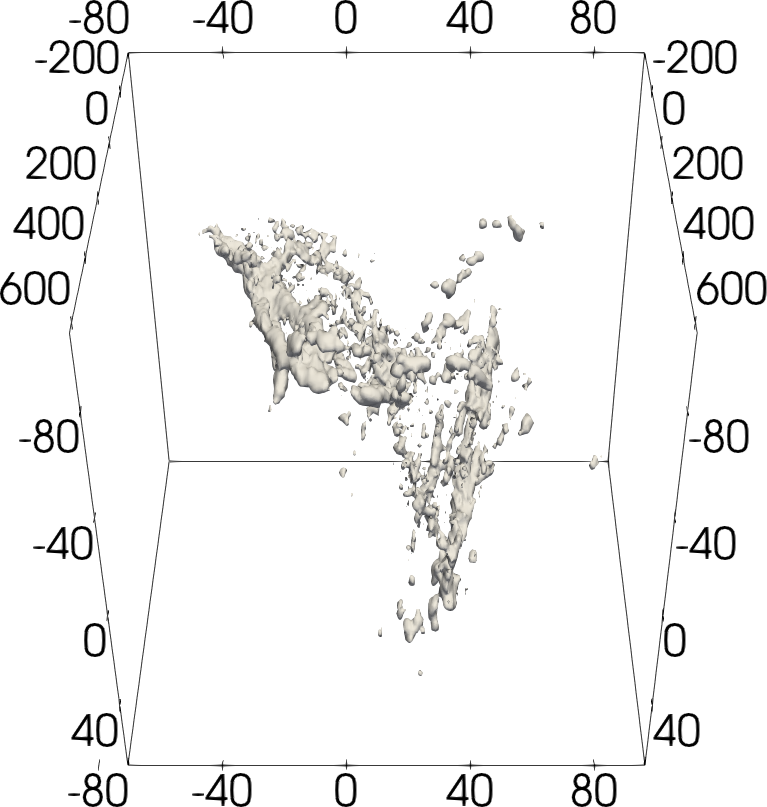}}
   \put(-460,230){\Large sim 19 fast}
  \put(-200,230){\Large sim 20}
  \caption{Third 3D view of the observed CO(1-0) datacube and the model H$_2$ data cubes. The axis labels are only shown for the observations.
    For a better understanding of these views, three 3D animations of the rotating datacube are attached to this figure (\texttt{taffy\_cube3D\_z.gif}, 
      \texttt{taffy\_cube3D\_z1.gif}, and \texttt{taffy\_cube3D\_x.gif}).
  \label{fig:Taffy_sm2_3D_z_130d_eclump}}
\end{figure*}

\begin{figure*}[!ht]
  \centering
  \resizebox{\hsize}{!}{\includegraphics{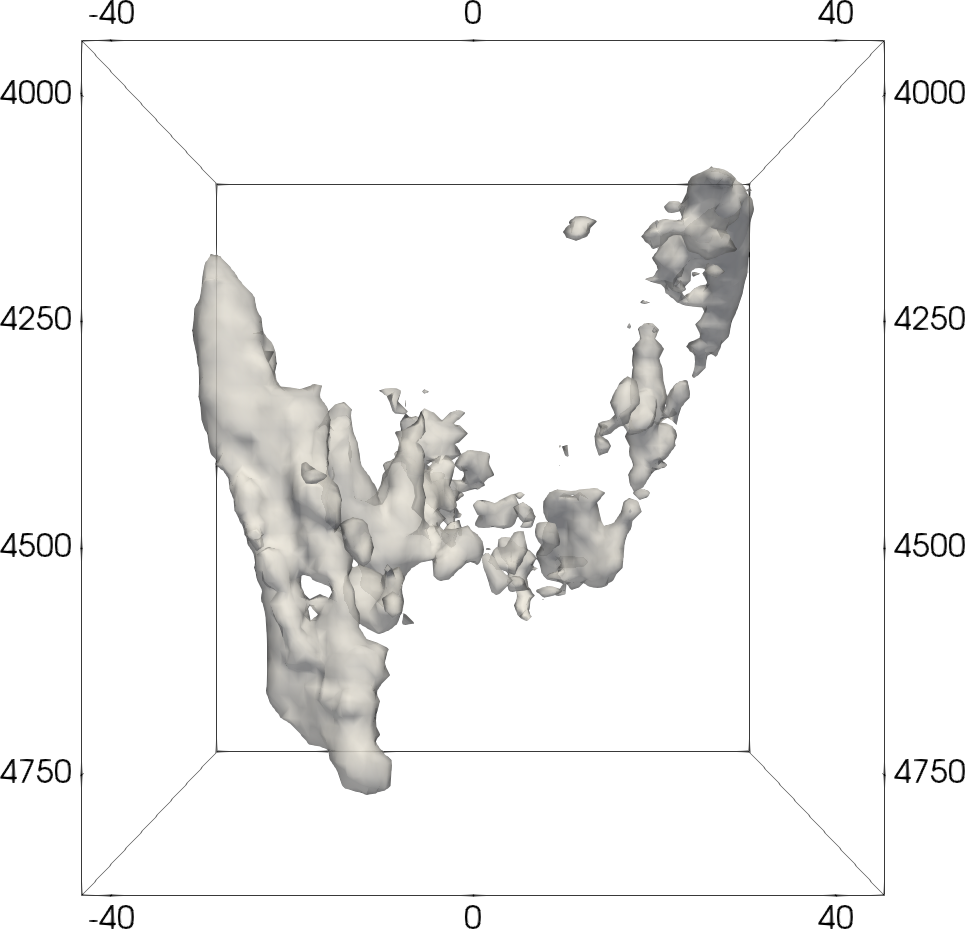}\includegraphics{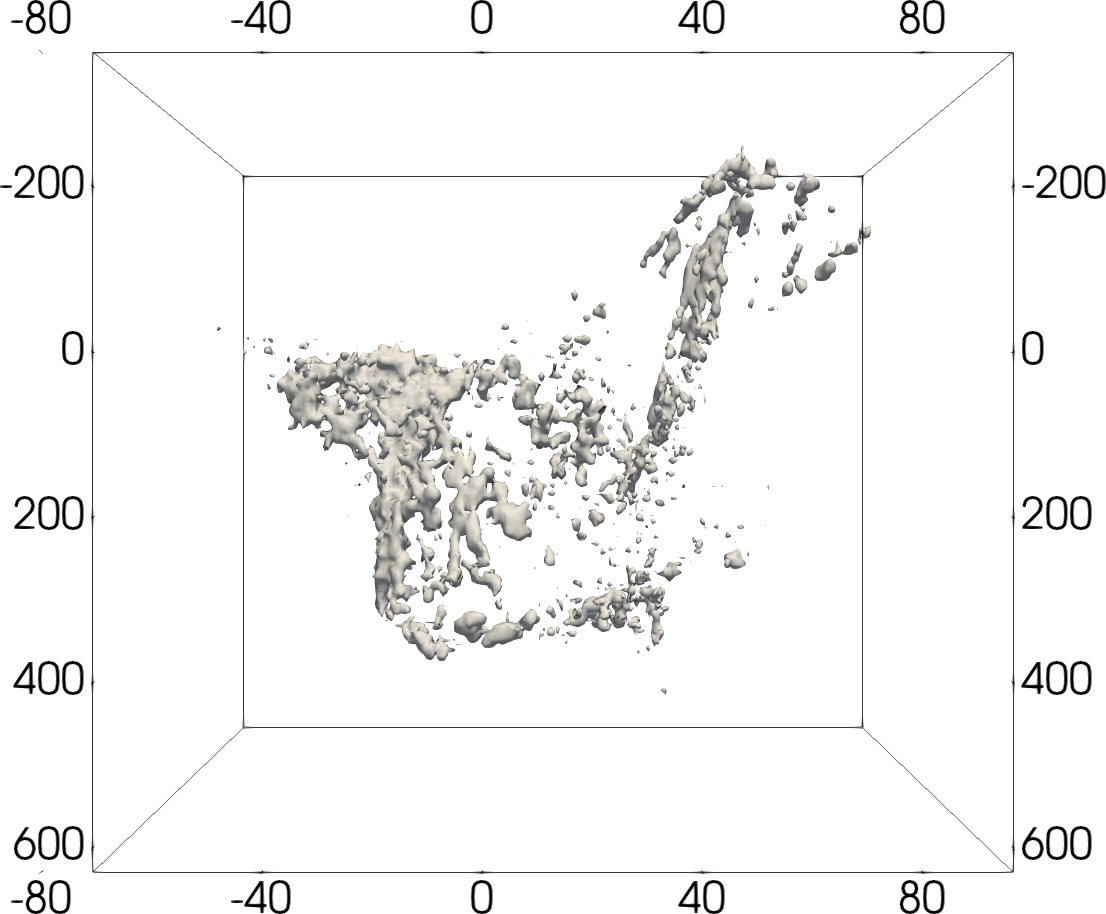}}
   \put(-440,200){\Large observations}
   \put(-180,200){\Large sim 19}
  \put(-530,130){\large $v_{\rm r}$ (km\,s$^{-1}$)}
  \put(-430,10){\large DEC offset (arcsec)}
 \vspace{0cm}
  \resizebox{\hsize}{!}{\includegraphics{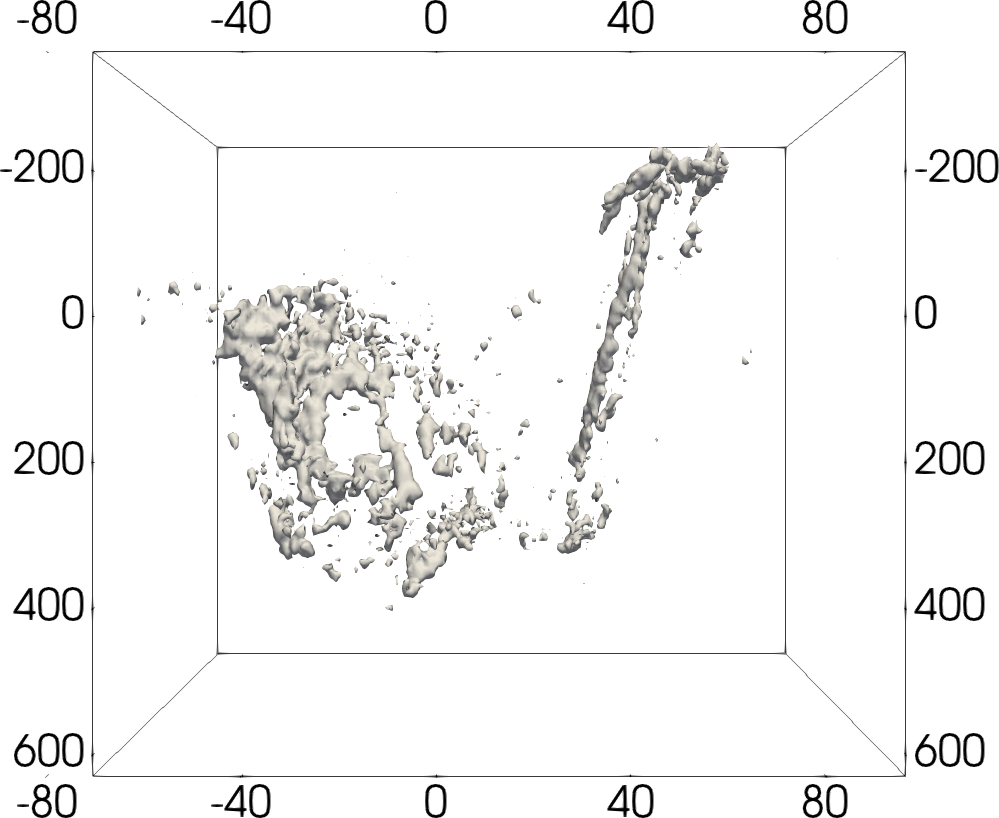}\includegraphics{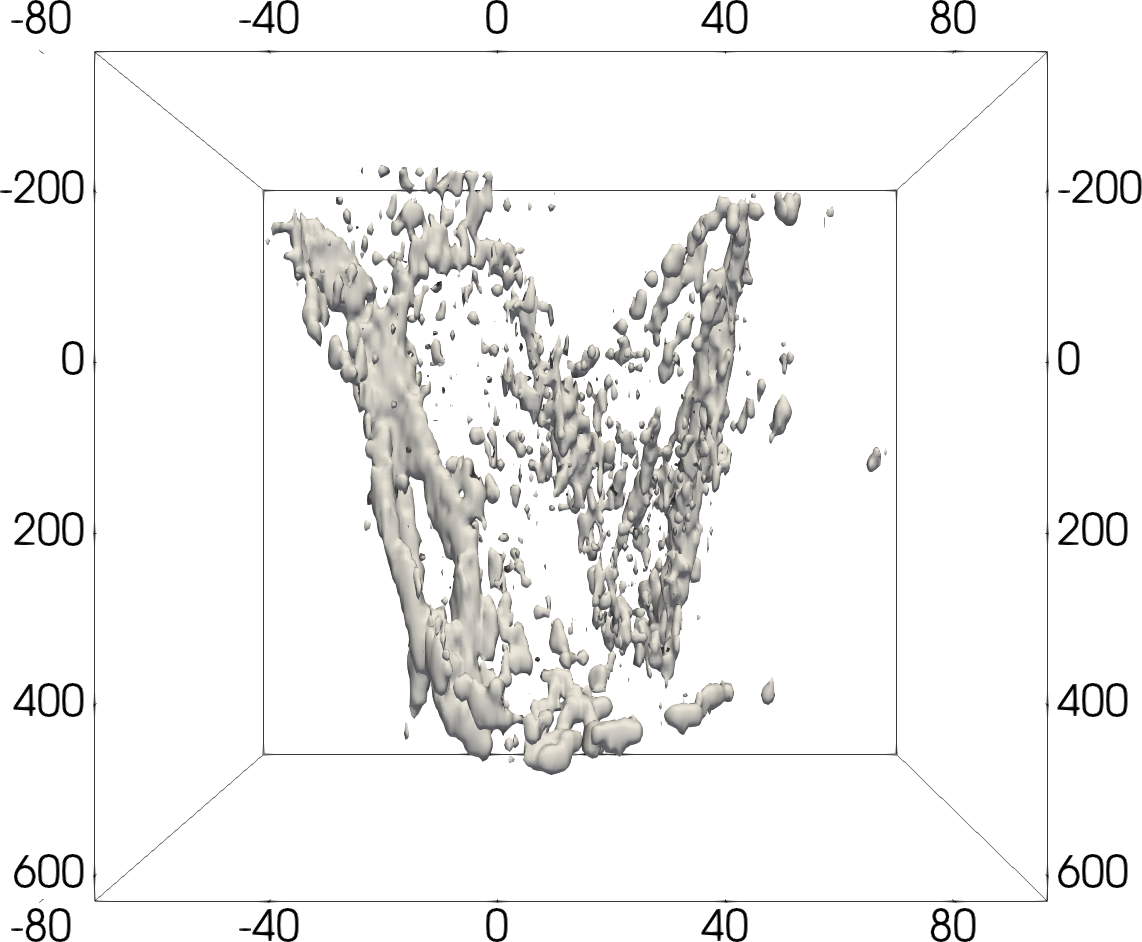}}
   \put(-440,210){\Large sim 19 fast}
  \put(-180,200){\Large sim 20}
  \caption{Fourth 3D view of the observed CO(1-0) datacube and the model H$_2$ data cubes. The axis labels are only shown for the observations.
    For a better understanding of these views, three 3D animations of the rotating datacube are attached to this figure (\texttt{taffy\_cube3D\_z.gif}, 
      \texttt{taffy\_cube3D\_z1.gif}, and \texttt{taffy\_cube3D\_x.gif}).
  \label{fig:Taffy_sm2_3D_z_180d_eclump}}
\end{figure*}

\section{Comparison with models \label{sec:compmodel1}}

Our modeling effort is based on the combination of a large-scale dynamical model (Sect.~\ref{sec:dynmodel})
together with a small-scale analytical model (Sect.~\ref{sec:anamodel}) to handle the properties of a
turbulent ISM in a simplified way. All cloud--cloud collisions conserve mass and momentum. 
Our method is akin to a sticky-particle scheme (e.g. Combes \& Gerin 1985)
where the cloud-cloud collisions are resolved due to the high time resolution.
The dynamical simulations follow Boltzmann's equation with a collisional term involving binary partially inelastic collisions.

The simulations do not include stellar feedback and do not follow the thermal 
evolution of the gas. For the thermal evolution of the gas in a galaxy-galaxy head-on 
collision we rely on the results of Yeager \& Struck (2019).
Our star formation recipe is based on cloud-cloud collisions (Sect.~\ref{sec:SFR}).
We verified that our SFR recipe based on cloud-cloud collisions leads to Schmidt-like star formation law
$\dot{\rho}_* \propto \rho^{1.5}$.

Following Robertson \& Goldreich (2012) and Mandal et al. (2020), we expect turbulent adiabatic heating to occur 
when the gas compression is faster than dissipation of turbulence $t_{\rm diss}$
(Sect.~\ref{sec:adcompmodel}). Since $t_{\rm diss}$ is not available from the dynamical model, we
compare $t_{\rm comp}$ to the $t_{\rm diss}$ the gas would have if it formed stars as in a galactic disk (i.e. following a Kennicutt-Schmidt
law). When compression energy exceeds that of stellar feedback, the velocity dispersion 
is expected to increase. In this case, we assume that the velocity dispersion of the clouds also increases such that star formation 
will be signicantly reduced (Sect.~\ref{sec:sfrsuppress1}).

\subsection{Large-scale dynamics - the dynamical model \label{sec:dynmodel}}

We used the dynamical simulations of Vollmer et al. (2012).
The ISM is simulated as a collisional component, i.e. as discrete particles that possess a mass and a 
radius and can have partially inelastic collisions. In contrast to smoothed 
particle hydrodynamics (SPH), which is a quasi-continuous approach where the particles cannot penetrate each other, our approach 
allows a finite penetration length, which is given by the mass-radius relation of the particles.
During the disk evolution, the cloud particles can have partially 
inelastic collisions, the outcome of which (coalescence, mass exchange, or fragmentation) is simplified following
the geometrical prescriptions of Wiegel (1994). 

The particle trajectories are integrated using an adaptive timestep for each particle. 
This method is described in Springel et al. (2001). 
The criterion for an individual timestep is $\Delta t_i=5~{\rm km\,s}^{-1}/a_i$, where $a_i$ is the acceleration of the particle
$i$. The minimum value of $\Delta t_i$ defines the global timestep used for the Burlisch-Stoer integrator that integrates the collisional component. 
The global timestep\footnote{In addition, the integrator divided this timestep at least into three sub-timesteps of about $3000$~yr.} is typically around $10^4$~yr.
For a velocity of $1000$~km\,s$^{-1}$ this corresponds to $\sim 10$~pc.

During each cloud-cloud collision the overlapping parts of the clouds are calculated. 
Let $b$ be impact parameter and $r_1$ and $r_2$ the radii of the larger and smaller clouds. 
If $r_1+r_2 > b > r_1-r_2$ the collision can result into fragmentation (high-speed encounter) or
mass exchange. If $b < r_1-r_2$ mass exchange or coalescence (low speed encounter) can occur.
If the maximum number of gas particles/cloud ($40000$) is reached, only coalescent or mass 
exchanging collisions are allowed. In this way a cloud mass distribution is naturally
produced. The cloud masses and velocities resulting from a cloud-cloud collision
are calculated by assuming mass and momentum conservation.
In Vollmer et al. (2012) we normalized the mass-size relation of the model clouds such that the gas mass of the bridge agrees
with that derived from CO observations of the Taffy system.
The cloud particle masses and radii range between $10^4$ and $10^6$~M$_{\odot}$ and 
$35$ and $145$~pc, respectively. 
The gas particles/clouds cannot be taken as the real clouds in the ISM of galactic disks, 
because the lifetime of giant molecular clouds (GMCs) of several $10$~Myr 
(e.g., Zamora-Aviles \& Vazquez-Semadeni 2014) does not permit frequent GMC-GMC collisions.
On the other hand, during an ISM-ISM collision as in the Taffy system, there will be a 
significant number of GMC-GMC collisions since the collision time is small $t \sim 1$~kpc$/(1000$~km\,s$^{-1}$)=$1$~Myr.
Following the direct cloud-cloud collision scenario of Harwit et al. (1987), the gas is heated 
to temperatures corresponding to a sizable fraction of the kinetic energy of the collision 
(millions of K). The shock-heated gas will then cool down with a rate that depends on its 
density. For a density of $10^3$~cm$^{-3}$ the cooling rate is about $10^4$~yr (Harwit et al. 1987). 
Note that there will also be collisions between the clouds and more diffuse gas
as simulated by Yeager \& Struck (2020).  
Since we are only interested in the dense molecular gas, our cloud particles can be 
identified with cool gas a few Myr after impact.

\subsection{Star formation \label{sec:SFR}}

In numerical simulations, the star formation recipe usually involves the gas density $\rho$ and the free-fall time 
$t_{\rm ff}=\sqrt{3\,\pi/(32\,G \rho)}$: $\dot{\rho}_* \propto \rho\, t_{\rm ff}^{-1} \propto \rho^{1.5}$.
In our dynamical model the star formation rate (SFR) is proportional to the cloud-cloud collision rate and stars are formed in cloud-cloud collisions. 

The newly created star particles have zero mass
(they are test particles) and the positions and velocities of the colliding clouds after the collision. 
These particles then move passively with the whole system. 
The information about the time of creation is attached to each newly created star particle.
The UV emission of a star particle in the two GALEX bands
is modeled by the UV flux from single stellar population models from STARBURST99 (Leitherer et al. 1999).
The age of the stellar population equals the time since the creation of the star particle.
The total UV distribution is then the extinction-free distribution of the UV emission of the newly created
star particles. 
 
\begin{figure}[!ht]
  \centering
  \resizebox{\hsize}{!}{\includegraphics{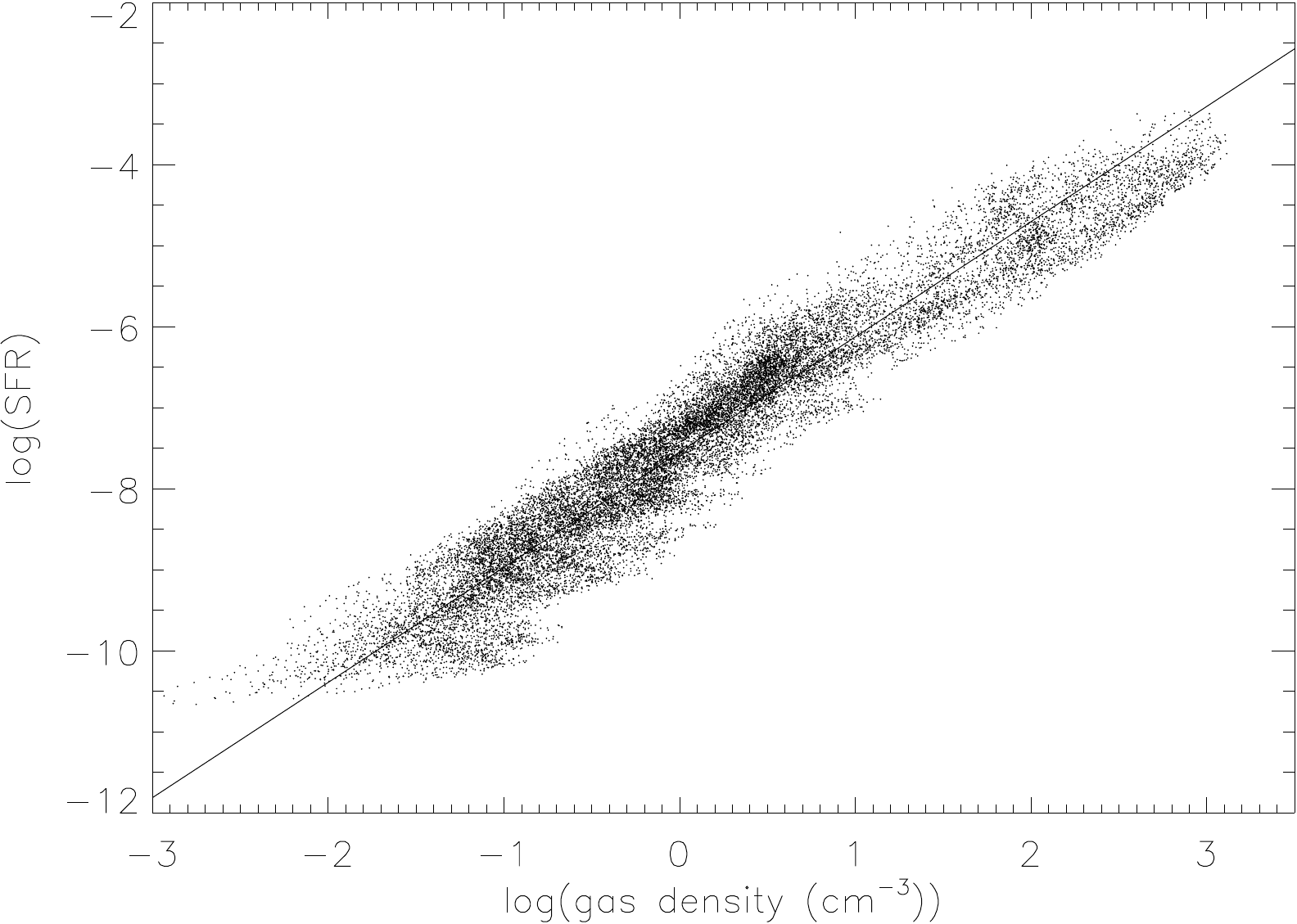}}
  \resizebox{\hsize}{!}{\includegraphics{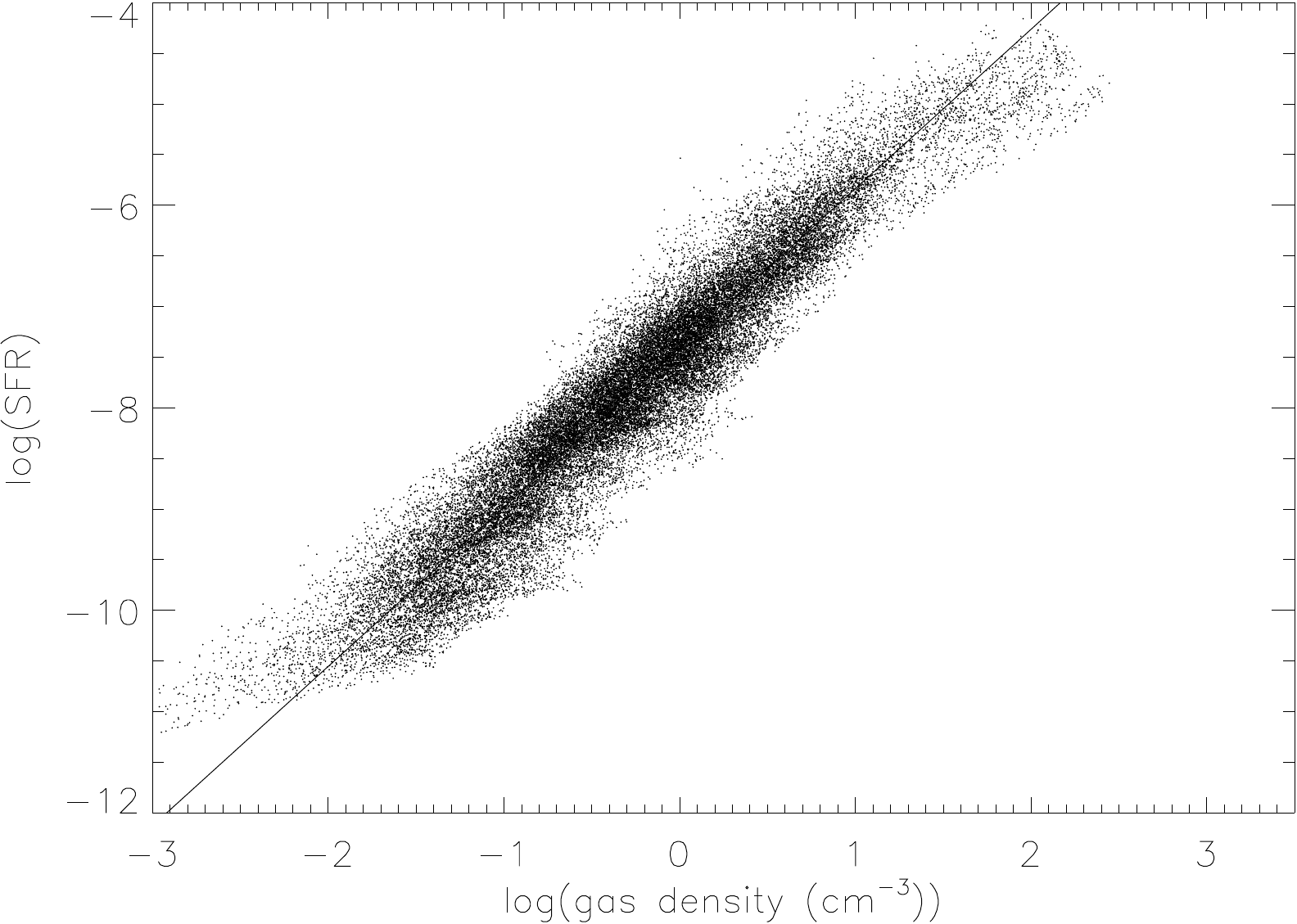}}
  \resizebox{\hsize}{!}{\includegraphics{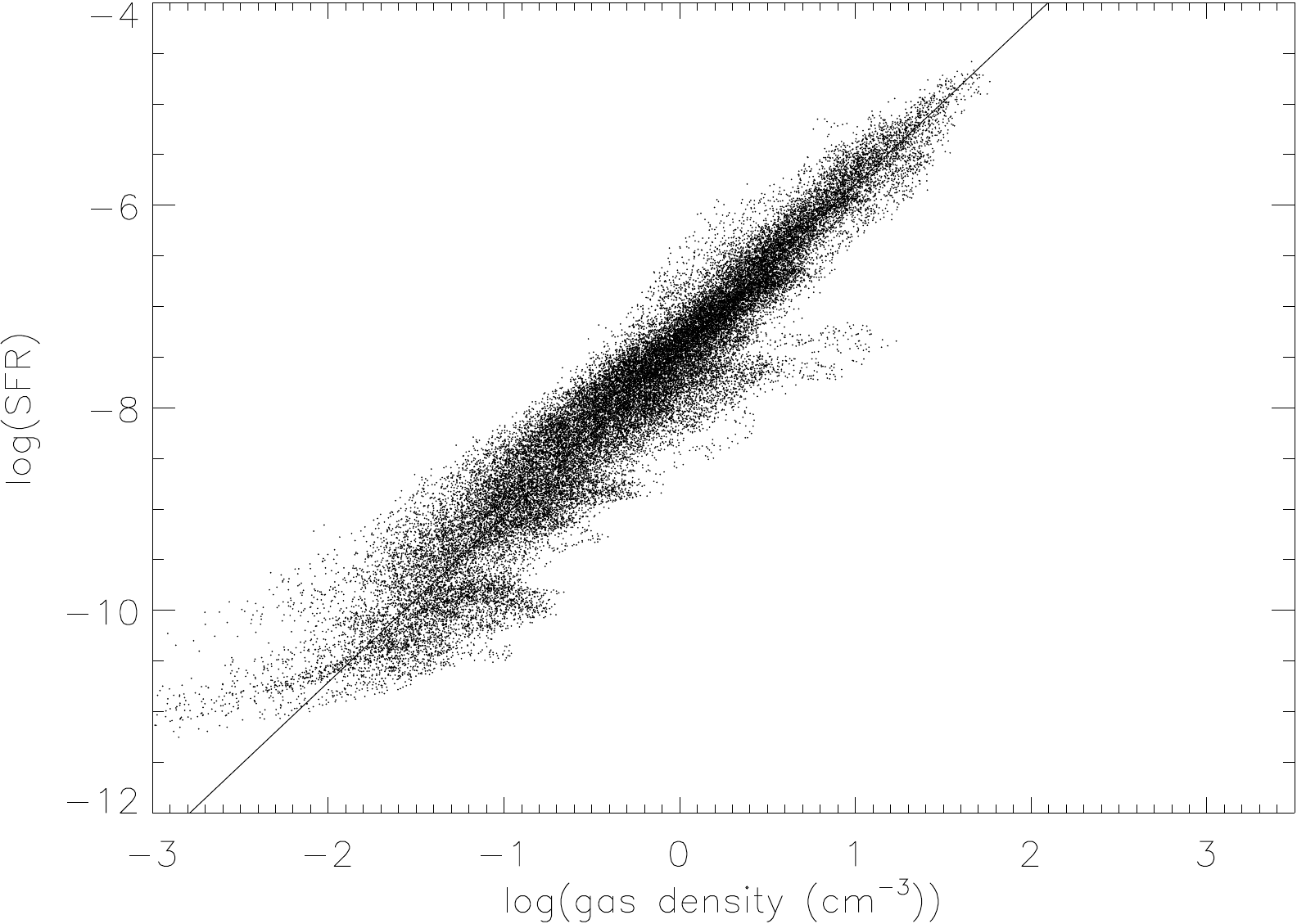}}
\caption{The local star formation rate $\dot{\rho}_*$ (in arbitrary units) as a function of the volume density $\rho$. 
Upper panel: unperturbed simulation after $0.5$~Gyr.
Middle panel: simulation~19 at impact. Lower panel: simulation~19 $20$~Myr after impact. Solid lines: linear regressions.
The slope of the correlation $\dot{\rho}_* \propto \rho^n$ for the unperturbed galaxy simulation is $n=1.4$, whereas it is 
$n=1.6$ for the Taffy simulation.
  \label{fig:KSlawvolume}}
\end{figure}

\begin{figure}[!ht]
  \centering
  \resizebox{8cm}{!}{\includegraphics{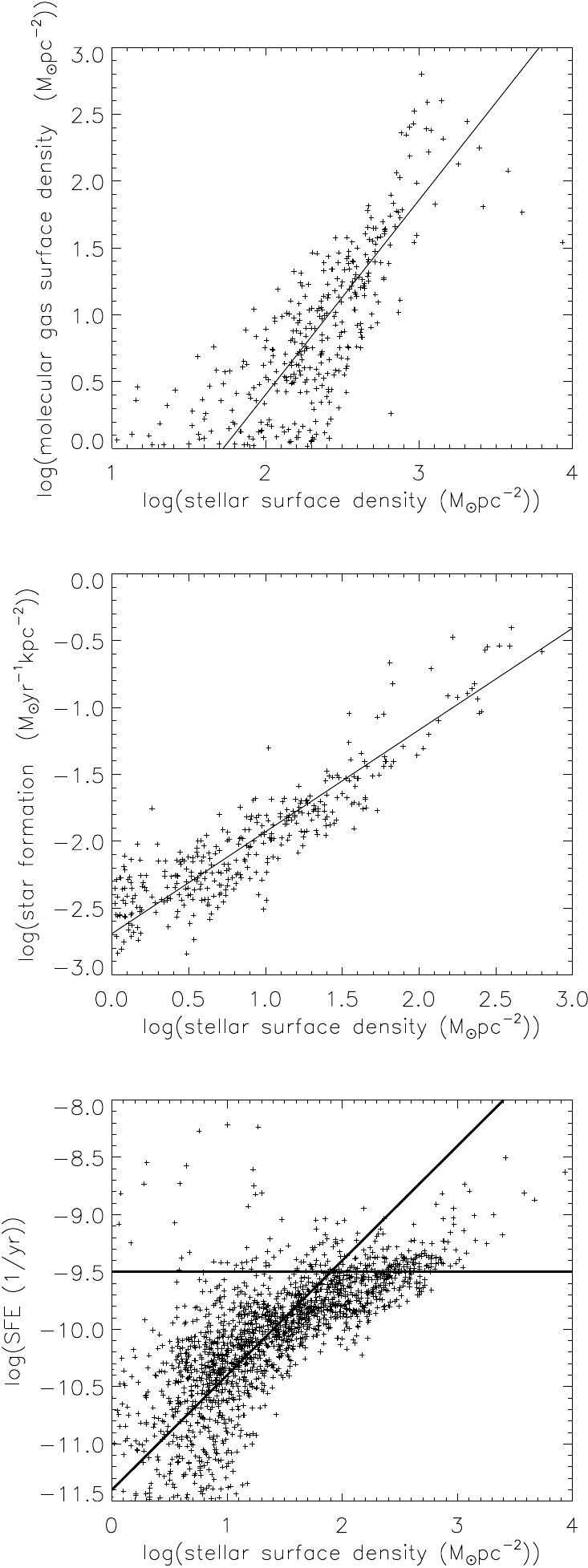}}
\caption{Model of an unperturbed spiral galaxy. Upper panel: star formation rate as a function of the molecular
gas surface density. Middle panel:  star formation rate as a function of the stellar surface density.
Lower panel: star formation efficiency with respect to the molecular gas as a function of the stellar surface density.
The solid lines mark the observed relations found by Leroy et al. (2008).
  \label{fig:compleroy}}
\end{figure}

We verified that our SFR recipe based on cloud-cloud collisions leads to the same exponent (1.4-1.6; Fig.~\ref{fig:KSlawvolume})
of the gas density in a simulation of an isolated spiral galaxy and for the Taffy system at impact
and $\sim 20$~Myr after impact. As a consequence, our code reproduces the observed SFR-total gas surface 
density, SFR-molecular gas surface density, and SFR-stellar surface density relations (Vollmer et al. 2012a). 
To go a step further we show the comparison of our model results with observed
scaling relations for the molecular gas surface density, star formation rate, and star formation efficiency in Fig.~\ref{fig:compleroy}.
The model relations agree quite well with the observed relations.

\subsection{Small-scale ISM properties - the analytical model:  \label{sec:anamodel}}

The model of Vollmer \& Beckert (2003) and Vollmer \& Leroy (2011) considers the warm, cold, and molecular phases of the ISM as a 
single turbulent gas. The gas is taken to be clumpy, so that the local density can be enhanced relative to the average density of the disk.
From the local density, the free-fall time of an individual self-gravitating gas clump is used as the timescale governing star formation.
The  star formation rate is used to calculate the rate of energy injection by supernovae. 
Turbulence is driven by this energy injection into turbulent eddies that have a characteristic length scale $l_{\rm driv}$ 
and a characteristic velocity $v_{\rm turb}$; $l_{\rm driv}$ and $v_{\rm turb}$ are linked to the volume filling factor of self-gravitating GMCs $\Phi_{\rm V}$.
All model parameters are described in Table~\ref{tab:parameters}.
The Vollmer \& Beckert (2003) model does not address the spatial inhomogeneity of the turbulent driving nor the mechanics of turbulent driving 
and dissipation. It is assumed that the energy input rate into the ISM due to supernovae is cascaded to smaller scales without loss. 
The energy of self-gravitating clouds is dissipated via cloud contraction and star formation.
The smallest scale investigated by the analytical model is the scale where the gas clouds become self-gravitating.
The size, density, and turbulent crossing time of these clouds are $l_{\rm cl}=l_{\rm driv}/\delta$, $\rho_{\rm cl}=\langle \rho \rangle/\Phi_{\rm V}$, 
and $t_{\rm turb,cl}=l_{\rm cl}/v_{\rm turb,cl}=\delta^{-0.5} l_{\rm driv}/v_{\rm turb}$, where $\langle \rho \rangle$ is the large-scale gas density.  

\begin{table*}
\begin{center}
\caption{Model Parameters.\label{tab:parameters}}
\begin{tabular}{lll}
\hline
Parameter & Unit & Explanation \\
\hline
$G=5 \times 10^{-15}$ & pc$^{3}$yr$^{-1}$M$_{\odot} ^{-1}$ & gravitation constant \\
$l_{\rm driv}$ & pc & turbulent driving length scale \\
$v_{\rm turb}$ & pc\,yr$^{-1}$ & gas turbulent 3D velocity dispersion at $l_{\rm driv}$ \\
$l_{\rm cl}$ & pc & cloud size \\
$v_{\rm turb,cl}$ & pc\,yr$^{-1}$ & cloud 3D velocity dispersion \\
$\sigma_{\rm cl}$ & pc\,yr$^{-1}$ & cloud 1D velocity dispersion \\
$\delta=l_{\rm driv}/l_{\rm cl}$ & & scaling between driving length scale and cloud size \\
$\Phi_{\rm V}$ & & volume filling factor of self-gravitating clouds \\
$\langle \rho \rangle$ & M$_{\odot}$pc$^{-3}$ &  mean gas density \\
$t_{\rm ff}=\sqrt{3\,\pi/(32\,G\, \langle \rho \rangle)}$ & yr & \\
$\rho_{\rm cl}=\langle \rho \rangle /\Phi_{\rm V}$ & M$_{\odot}$pc$^{-3}$ & cloud density \\
$t_{\rm ff,cl}$ & yr & cloud free fall timescale at size $l_{\rm cl}$ \\
$t_{\rm turb,cl}$ & yr & cloud turbulent timescale at size $l_{\rm cl}$ \\
$t_{\rm life,cl}$ & yr & cloud lifetime \\
$t_{\rm dep}$ & yr & gas depletion timescale \\
$\dot{\rho}_{*}$ & M$_{\odot}$pc$^{-3}$yr$^{-1}$ & star formation rate per unit volume \\
$\xi=4.6 \times 10^{-8}$ & pc$^2$yr$^{-2}$ & constant relating SN energy input to SF \\
$\epsilon_{\rm ff}$ & & star formation efficiency per free fall time \\
$f_{\rm SF}$ & & fraction of the star-forming molecular gas mass \\
$\epsilon_*$ & & cloud mass fraction converted into stars \\
$\epsilon_{\rm life}=t_{\rm ff,cl}/t_{\rm life,cl}$ & & cloud free-fall time divided by the lifetime \\
$t_{\rm diss}=l_{\rm driv}/v_{\rm turb}$ & yr & turbulent dissipation timescale \\
$t_{\rm comp}=\rho/(d\rho/dt)$ & yr & gas compression timescale \\ 
\hline
\end{tabular}
\end{center}
\end{table*}

Following Vollmer \& Leroy (2011) the star formation rate per unit volume is given by 
\begin{equation}
\label{eq:rhostar}
\dot{\rho}_*=\Phi_{\rm V} \rho\,t_{\rm ff,cl}^{-1}= \sqrt{\Phi_{\rm V}} \rho\,t_{\rm ff}^{-1}=\epsilon_{\rm ff} \rho\,t_{\rm ff}^{-1} \ ,
\end{equation}
where $\Phi_{\rm V}^{-1}=\rho_{\rm cl}/\rho$ is the overdensity of self-gravitating clouds, $\rho$ the gas density, $t_{\rm ff,cl}$ the free-fall 
time of a self-gravitating gas cloud, $t_{\rm ff}=\sqrt{3\,\pi/(32\,G\,\rho)}$, and $\epsilon_{\rm ff} = \sqrt{\Phi_{\rm V}} \propto t_{\rm turb}/t_{\rm ff}$ 
the star formation efficiency per free-fall time.
Vollmer et al. (2017) found that for star formation rates comparable to those of nearby spiral galaxies and gas velocity dispersions around 
$10$~km\,s$^{-1}$, $\Phi_{\rm V}$ is about constant and has values of a few times $0.001$, consistent with the findings of Leroy et al. (2017) in M~51. 
In the following we will show that $\epsilon_{\rm ff}=\sqrt{\Phi_{\rm V}} \propto v_{\rm turb}$,
which is consistent with the predictions of feedback-regulated star formation in turbulent, self-gravitating, strongly star-forming galactic gas disks
(Ostriker \& Shetty 2011, Faucher-Gigu\`ere et al. 2013; however, see Krumholz et al. 2018 for a different point of view).

For self-gravitating clouds with a Virial parameter of unity the turbulent crossing time equals twice the free-fall time:
\begin{equation}
2\,t_{\rm ff,cl} = 2\, \sqrt{\frac{3\,\pi \Phi_{\rm V}}{32\,G \langle \rho \rangle}} = \frac{\sqrt{3}\,l_{\rm cl}}{2\,v_{\rm turb,cl}} \ ,
\end{equation}
where $l_{\rm cl}$ and $v_{\rm turb,cl}$ are the size and turbulent 3D velocity dispersion of the cloud.
Using Larson's law ($l_{\rm cl}/v_{\rm turb,cl} = l_{\rm driv}/v_{\rm turb}/\sqrt{\delta}$),
the star formation rate per unit volume is
\begin{equation}
\dot{\rho}_*=\frac{4\,\sqrt{\delta}}{\sqrt{3}} \Phi_{\rm V} \langle \rho \rangle v_{\rm turb}/l_{\rm driv}\ .
\end{equation}
We can connect the energy input into the ISM by SNe directly to the star formation rate. With the assumption of a
constant initial mass function independent of environment one can write
\begin{equation}
\label{eq:energyflux}
\frac{1}{2} \langle \rho \rangle \frac{v_{\rm turb}^3}{l_{\rm driv}}=\xi \dot{\rho}_*\ .
\end{equation}
This leads to the following expression for the volume filling factor: 
\begin{equation}
\Phi_{\rm V}=\frac{\sqrt{3}\,v_{\rm turb}^2}{8 \sqrt{\delta} \xi}\ ,
\label{eq:vturb}
\end{equation}
and the star formation law becomes
\begin{equation}
\label{eq:sfr123}
\dot{\rho}_* = \sqrt{\frac{\sqrt{3}}{8 \sqrt{\delta} \xi}} v_{\rm turb} \rho\,t_{\rm ff}^{-1} \ .
\end{equation}
We thus find $\epsilon_{\rm ff} \propto v_{\rm turb}$, which is equivalent to Eq.~22 of Ostriker \& Shetty (2011), Eq.~37 of Faucher-Gigu\`ere et al. (2013),
and Eq.~54 of Krumholz et al. (2018).

Using Eq.~\ref{eq:rhostar} and Eq.~\ref{eq:energyflux}, the large-scale turbulent crossing time, which equals the turbulent dissipation timescale, is 
\begin{equation}
t_{\rm turb}=t_{\rm diss}=\frac{l_{\rm driv}}{v_{\rm turb}}=\frac{v_{\rm turb}^2}{2\,\xi\,\sqrt{\Phi_{\rm V}}} \sqrt{\frac{3\,\pi}{32\,G \langle \rho \rangle}}\ .
\label{eq:tdiss}
\end{equation}
Inserting Eq.~\ref{eq:vturb} into Eq.~\ref{eq:tdiss} leads to the final expression for the turbulent
dissipation timescale:
\begin{equation}
\label{eq:tdiss1}
t_{\rm diss}=v_{\rm turb} \sqrt{\frac{6\,\pi \sqrt{\delta}}{\sqrt{3}\,32\,G \langle \rho \rangle \xi}}\ .
\end{equation}

Alternatively, we can assume a constant $\epsilon_{\rm ff}$ (Krumholz \& McKee 2005, Krumholz et al. 2012).
In this case the equation for the energy injection and dissipation becomes 
\begin{equation}
\frac{1}{2} \langle \rho \rangle \frac{v_{\rm turb}^3}{l_{\rm driv}}=\xi \epsilon_{\rm ff} \langle \rho \rangle \, t_{\rm ff}^{-1}
\end{equation}
and the dissipation timescale is
\begin{equation}
t_{{\rm diss},\epsilon}=\frac{v_{\rm turb}^2}{2\,\xi\,\epsilon_{\rm ff}} \sqrt{\frac{3\,\pi}{32\,G\,\langle \rho \rangle}}\ .
\end{equation}
This timescale equals $t_{\rm diss}$ (Eq.~\ref{eq:tdiss}) for $v_{\rm turb}=\epsilon_{\rm ff} \sqrt{\frac{8\,\xi \sqrt{\delta}}{\sqrt{3}}}=6.5$~km\,s$^{-1}$.
For higher velocity dispersions $t_{{\rm diss},\epsilon} > t_{\rm diss}$. 

Within the framework of Vollmer et al. (2017) the dependence of $\epsilon_{\rm ff}$ on the turbulent velocity dispersion is
$\epsilon_{\rm ff} \propto \sqrt{v_{\rm turb}}$ leading to $t_{\rm diss} \propto v_{\rm turb}^{1.5}$.
For $v_{\rm turb} > 10$~km\,s$^{-1}$, Eq.~\ref{eq:tdiss1} represents the lower limit for the dissipation timescale.
Since we require $t_{\rm comp} < t_{\rm diss}$ for turbulent adiabatic compression, this lower limit of $t_{\rm diss}$ is an appropriate,
conservative choice.

\begin{figure}[!ht]
  \centering
  \resizebox{\hsize}{!}{\includegraphics{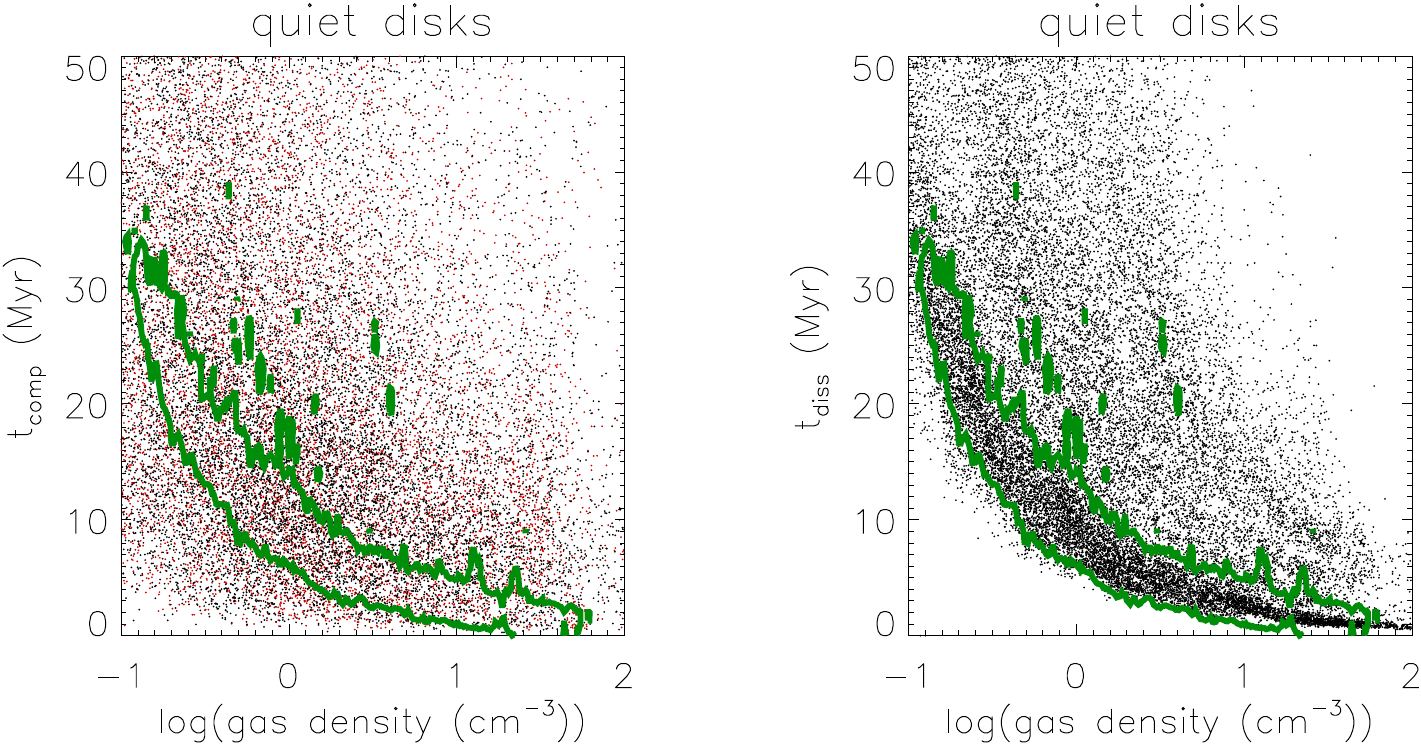}}
  \caption{Compression (Eq.~\ref{eq:tdiss1}, left panel) and dissipation (Eq.~\ref{eq:tdiss1}, right panel) timescales
    as a function of the mean gas density $\langle \rho \rangle$ for the quite disks before the interaction. 
    Negative compression timescales, i.e. gas expansion, are marked
    as red points in the left panel. The green contours mark the regions of highest particle density in the  
    $\langle \rho \rangle$-$t_{\rm diss}$ relation.
  \label{fig:tcomptdissquiet1}}
\end{figure}

\begin{figure}[!ht]
  \centering
  \resizebox{\hsize}{!}{\includegraphics{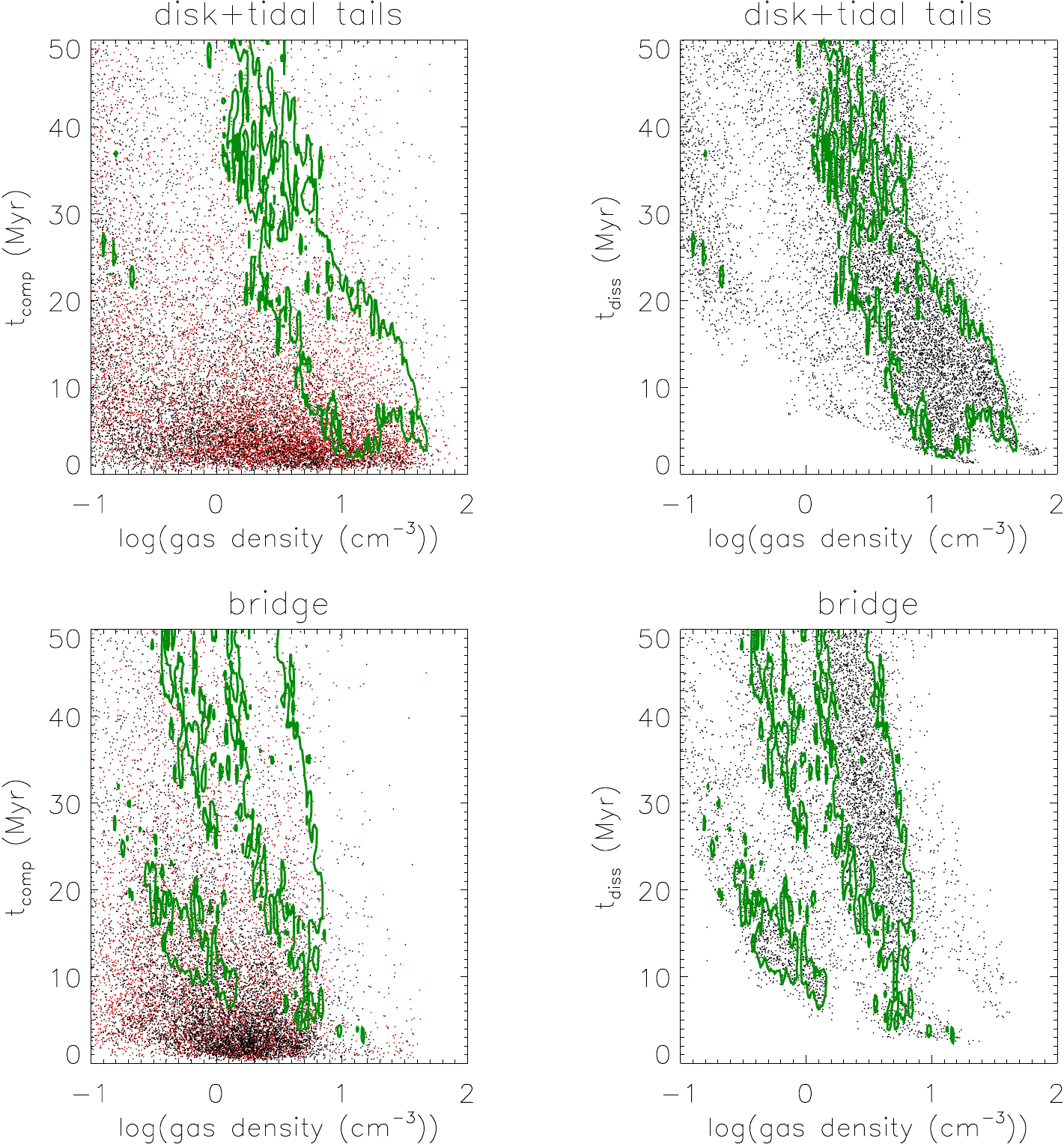}}
  \caption{Compression (Eq.~\ref{eq:tdiss1}, left panel) and dissipation (Eq.~\ref{eq:tdiss1}, right panel) timescales
    as a function of the mean gas density $\langle \rho \rangle$ for the timestep of interest of sim19.
    The meaning of the colors is the same as in Fig.~\ref{fig:tcomptdissquiet1}. Upper panels: all gas particles within the
    geometrically defined disk and tidal tail regions. Lower panels: all gas particles within the geometrically defined bridge region.
  \label{fig:tcomptdiss_TAFFY22new_10new1}}
\end{figure}

The dissipation timescale $t_{\rm diss}$ is compared to the compression timescale $t_{\rm comp}$ for the quiet disks before the
interaction in Fig.~\ref{fig:tcomptdissquiet1} and for the system $\sim 20$~Myr after impact in
Fig.~\ref{fig:tcomptdiss_TAFFY22new_10new1}. The dissipation timescale of the quiet disks
(right panel of Fig.~\ref{fig:tcomptdissquiet1}) shows the $1/\sqrt{\langle \rho \rangle}$-dependence of Eq.~\ref{eq:tdiss1}.
Roughly half of the particles have a 1D velocity dispersion of about $10$~km\,s$^{-1}$ (green contours), 
and about $25$\,\% have twice that velocity dispersion. Three quarters of all particles
have $t_{\rm comp} > t_{\rm diss}$ (left panel of Fig.~\ref{fig:tcomptdissquiet1}).

The picture changes for the system at the time of interest where we geometrically divided the system into a bridge and 
disk+tidal tail regions. The majority of the gas particles of the system show significantly higher velocity dispersions 
and thus higher $t_{\rm diss}$ (right panels of Fig.~\ref{fig:tcomptdiss_TAFFY22new_10new1}).
At the same time the compression timescale of the majority of particles is significantly shorter than those
of the quiet disks (left panels of Fig.~\ref{fig:tcomptdiss_TAFFY22new_10new1}). 
About half of the particles have $t_{\rm comp} < t_{\rm diss}$.
The gas densities in the bridge do not exceed $\langle \rho \rangle \sim 10$~cm$^{-1}$ 
(lower panels of Fig.~\ref{fig:tcomptdiss_TAFFY22new_10new1}) which is due to the coarse spatial resolution of our simulations.
The gas particles located within the bridge region almost exclusively have high velocity dispersions (lower right panel of 
Fig.~\ref{fig:tcomptdiss_TAFFY22new_10new1}) and show $t_{\rm comp} < t_{\rm diss}$ (lower left panel of 
Fig.~\ref{fig:tcomptdiss_TAFFY22new_10new1}).

\subsection{Turbulent adiabatic compression \label{sec:adcompmodel}}

In our simulation of an isolated spiral galaxy the 1D velocity dispersion of the model
clouds is constant, $v_{\rm disp} \sim 10$~km\,s$^{-1}$ , during 1 Gyr.
Since there is no stellar feedback, the cloud velocity dispersion is increased when the gas is compressed.
In kinetic theory, particles move with random motions around the sound speed and over a length scale given by the collision mean free path. 
In the eddy-viscosity model (Boussinesq approximation), eddies also move with random motions, at a typical speed given by the
turbulent velocity dispersion and over a typical length scale called the mixing length. Since these time scales are well-resolved in our 
simulations, we can identify the particle/cloud velocity dispersion with the velocity dispersion of the largest turbulent eddies.

Since the dissipation timescale is not part of the dynamical model, we compare the gas compression timescale to the turbulent dissipation
timescale $t_{\rm diss}=l_{\rm driv}/v_{\rm turb}$ following Eq.~\ref{eq:tdiss1}, i.e. in the absence of adiabatic compression (Eq.~\ref{eq:energyflux}).
The large-scale velocity dispersion and density are taken from the dynamical model.
Eq.~\ref{eq:tdiss1} implies that these quantities approximately correspond to their values at the turbulent driving lengthscale.
Within an unperturbed galactic disk the driving lengthscale is $l_{\rm driv}=v_{\rm turb} t_{\rm turb} \sim 100$~pc and $30$~pc at 
densities of $n \sim 1$~cm$^{-3}$ and $10$~cm$^{-3}$, respectively. These values are broadly consistent with (i) the
length scale at which Elmegreen et al. (2003) observed a break in the Fourier transform power spectrum of azimuthal optical 
and H{\sc i} intensity scans and (ii) the vertical thickness of the Galactic cold neutral medium (Wolfire et al. 2003).
The driving length in the bridge is estimated in Sect.~\ref{sec:sfrsuppress}.
Using Eq.~\ref{eq:tdiss} instead of Eq.~\ref{eq:tdiss1} leads to equivalent numbers
of bridge clouds affected by adiabatic compression at the time of interest (today).

The timescales $t_{\rm diss}$ (Eq.~\ref{eq:tdiss1}) and $t_{\rm comp}$ (Eq.~\ref{eq:tcomp1}) are important to identify the primary source of energy loss.
If $t_{\rm diss}$ is shorter than $t_{\rm comp}$, then the dominant energy injection
mechanism is star formation and cloud-scale dissipation is more important than adiabatic compression.  
This is true for galactic disks (Vollmer \& Beckert 2003). 

The compression timescale was calculated using the continuity equation
\begin{equation}
\label{eq:tcomp}
\frac{{\rm d}\rho}{{\rm d}t} + \nabla \cdot (\rho \vec{v}) = 0\ .
\end{equation}
All quantities which are needed to derive $t_{\rm comp}$ and $t_{\rm diss}$ are calculated from the dynamical model 
via a Smoothed-Particle Hydrodynamics (SPH)-type algorithm involving the $50$ nearest neighbouring particles.

\subsection{Star formation suppression caused by turbulent adiabatic compression \label{sec:sfrsuppress1}}

It is generally assumed that within the disks of isolated galaxies turbulence is driven
by energy injection through stellar feedback (SN explosions). In an equilibrium state a balance 
between turbulent pressure and gravity is reached leading to a global virial equilibrium state of the GMCs (Heyer et al. 2009). 
If the energy injection through large-scale gas compression exceeds that of stellar feedback deduced via the star formation rate,
the velocity dispersion of the largest eddies is expected to increase.
In this case, we presume that the velocity dispersion of the turbulent 
substructures/clouds also increases (Fig.~2 of Mandal et al. 2020). In our toy model, we decided to
suppress star formation during a cloud-cloud collision if the energy injection by large-scale gas
compression exceeds that from stellar feedback expected from an ISM that forms stars according to
a Kennicutt-Schmidt law. For the latter case, the turbulent energy dissipation timescale $t_{diss}$ can 
be calculated via our analytical model. 

We included the effect of star formation suppression by turbulent adiabatic compression
in the following way: if for a cloud-cloud collision $t_{\rm comp} > 0$ and $t_{\rm comp} < t_{\rm diss}$, no stellar particle is created.
In addition, rapid expansion also suppresses star formation ($|t_{\rm comp}| < t_{\rm diss}/5$).
The factor $1/5$ was derived heuristically. It allowed us to clearly separate the bridge from the disk regions.
This second criterion does not play a dominant role for the outcome of the star formation suppression.
We verified that in a simulation of an isolated spiral galaxy this is only the case for a negligible 
fraction of the gas clouds. Until a few Myr after impact a significant portion of the shocked gas will be hot (Yeager \& Struck 2019)
and will not be able to form stars. Since the compression timescale is extremely short, our
star formation suppression recipe ensures the absence of star formation in the hot gas.

\subsection{Suppressed star formation in the Taffy bridge \label{sec:sfrtaffy}}

\begin{figure}[!ht]
  \centering
  \resizebox{\hsize}{!}{\includegraphics{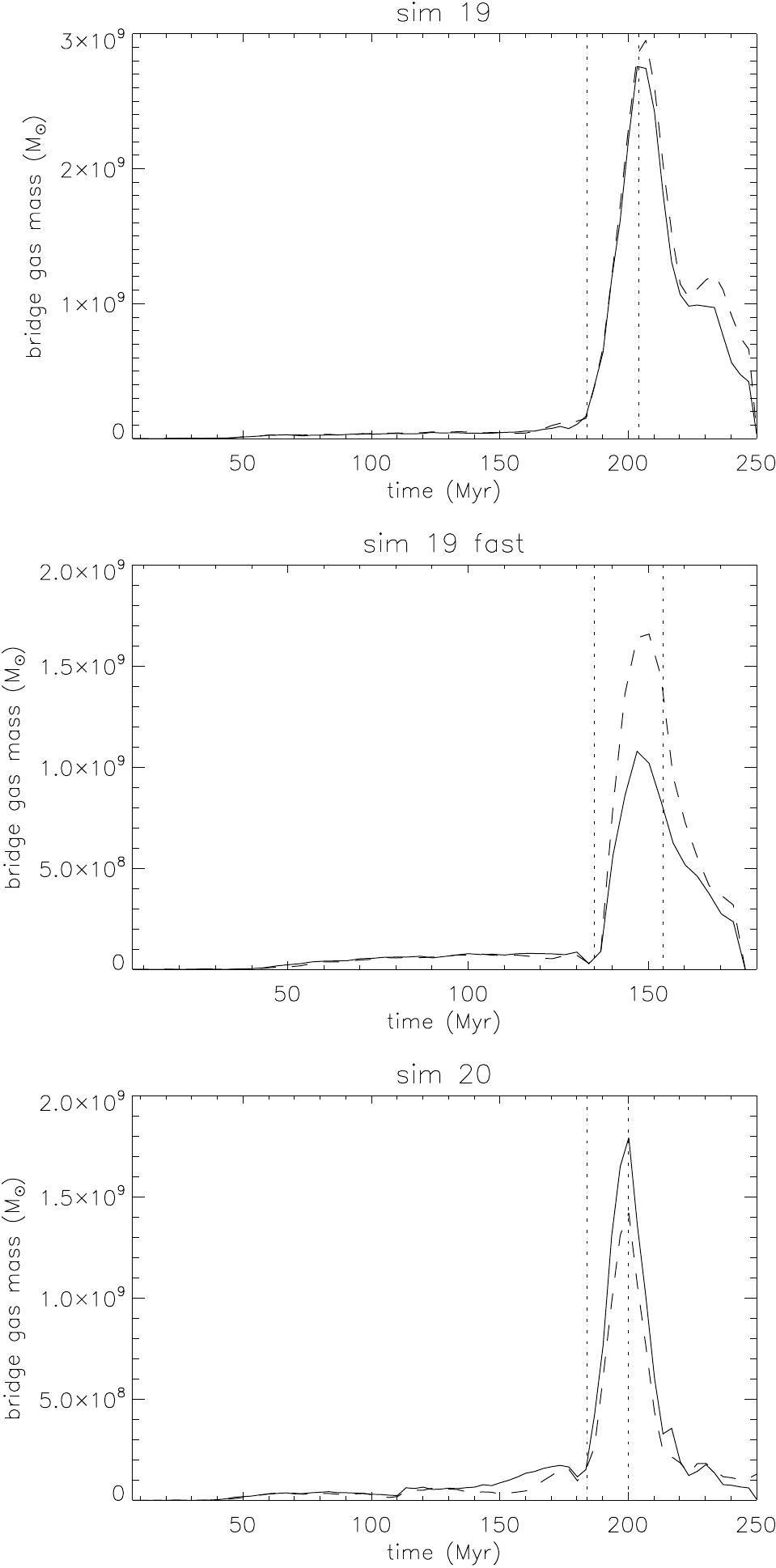}}
\caption{Evolution of the total gas mass in the bridge. The dotted vertical lines mark the impact time and the
time of interest (today).
  \label{fig:taffybridge_2}}
\end{figure}

We calculated the star formation rate within our simulations using the cloud-cloud collisions as described in Sect.~\ref{sec:sfrsuppress}.
In the following, we separate the bridge region from the disk regions based on geometry and the gas
density. These conditions appear appropriate based on examining the separation in three dimensions.
Fig.~\ref{fig:taffybridge_2} shows the gas mass in the model bridges. The total gas masses range between $10^9$~M$_{\odot}$ for
sim19fast to almost $3 \times 10^9$~M$_{\odot}$ for sim19.

\begin{figure}[!ht]
  \centering
  \resizebox{\hsize}{!}{\includegraphics{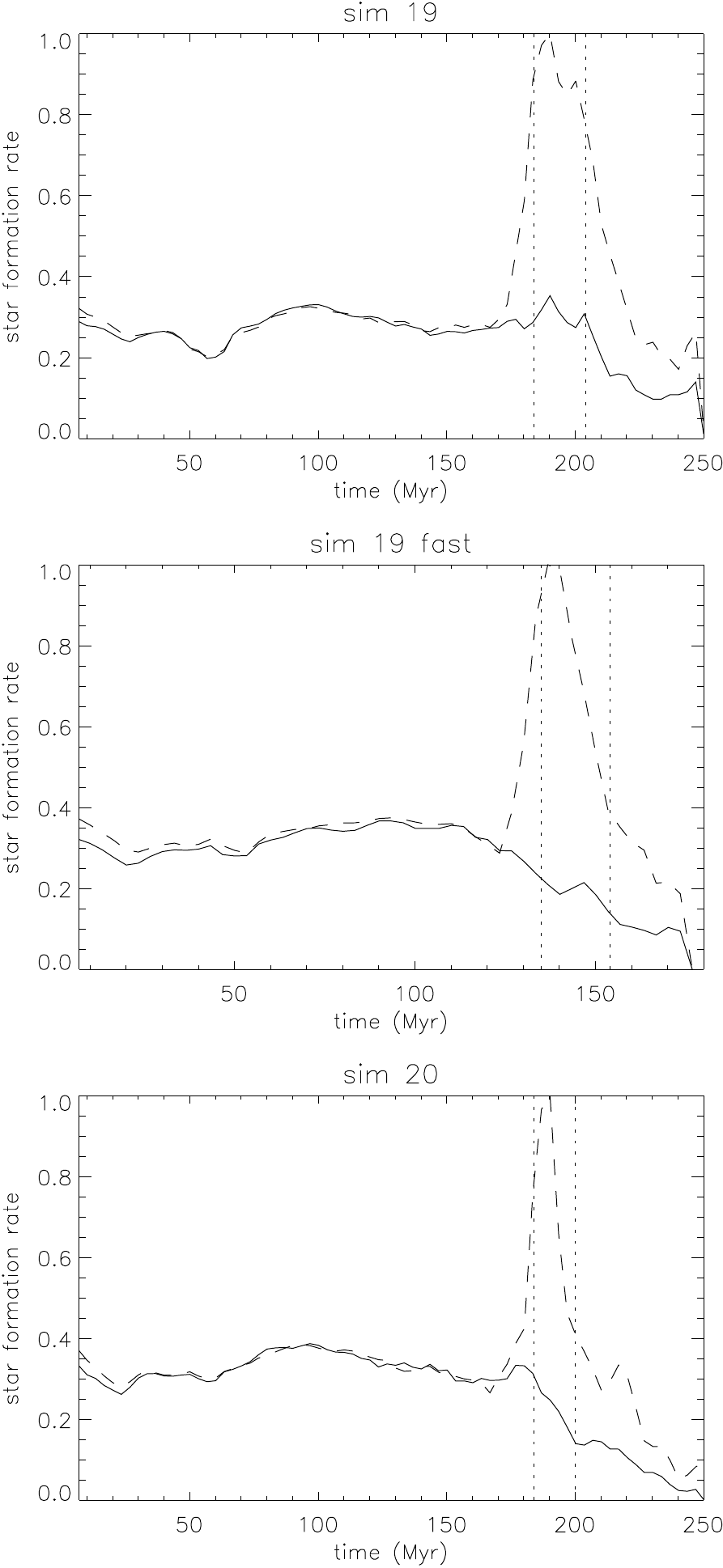}}
\caption{Evolution of the normalized total star formation rate. Solid line: with turbulent adiabatic compression. 
  Dashed line: without turbulent adiabatic compression. The dotted vertical lines mark the impact time and the
time of interest (today).
  \label{fig:taffybridge_1}}
\end{figure}

\begin{figure}[!ht]
  \centering
  \resizebox{\hsize}{!}{\includegraphics{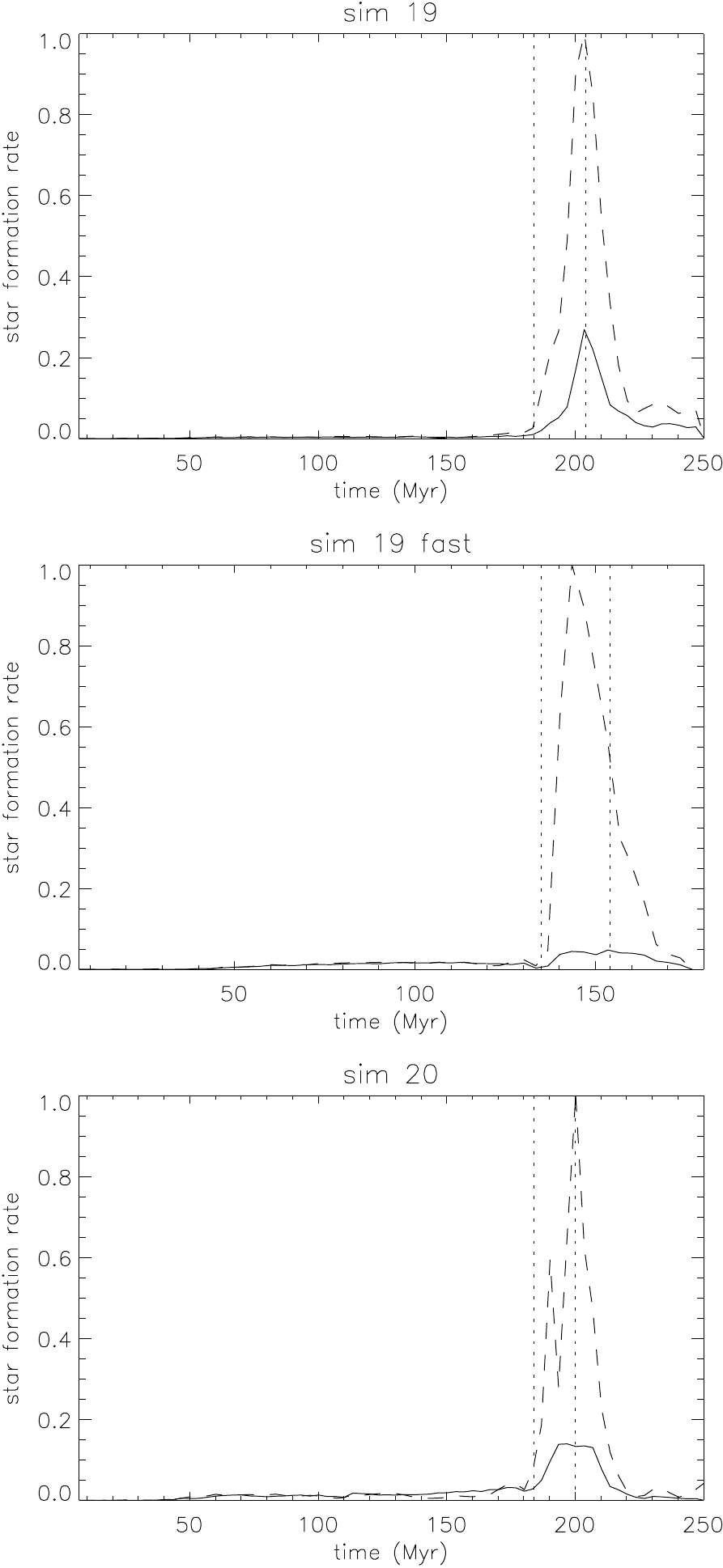}}
\caption{Evolution of the normalized total star formation rate in the bridge. Solid line: with turbulent adiabatic compression. 
  Dashed line: without turbulent adiabatic compression. The dotted vertical lines mark the impact time and the
time of interest (today).
  \label{fig:taffybridge_3}}
\end{figure}

The total (disk and bridge) star formation rate is shown in Fig.~\ref{fig:taffybridge_1} for all three models.
It is constant during about $2/3$ of the evolution of the system. The evolution of the total star formation rate of the models without 
adiabatic compression is much different from that of the models with adiabatic compression. 
The star formation rate within the bridge region is shown in Fig.~\ref{fig:taffybridge_3}.
The comparison between Fig.~\ref{fig:taffybridge_1} and Fig.~\ref{fig:taffybridge_3} shows that the strong increase of the total star formation rate 
is caused by the star formation in the bridge region.
Without adiabatic compression
the star formation rate rapidly increases by a factor of three, whereas the star formation rate in the models with adiabatic 
compression stays constant or slowly declines. Our conditions for turbulent adiabatic compression
therefore efficiently suppresses star formation in the bridge gas by a factor of $3$ to $5$.

\begin{figure*}[!ht]
  \centering
  \resizebox{16cm}{!}{\includegraphics{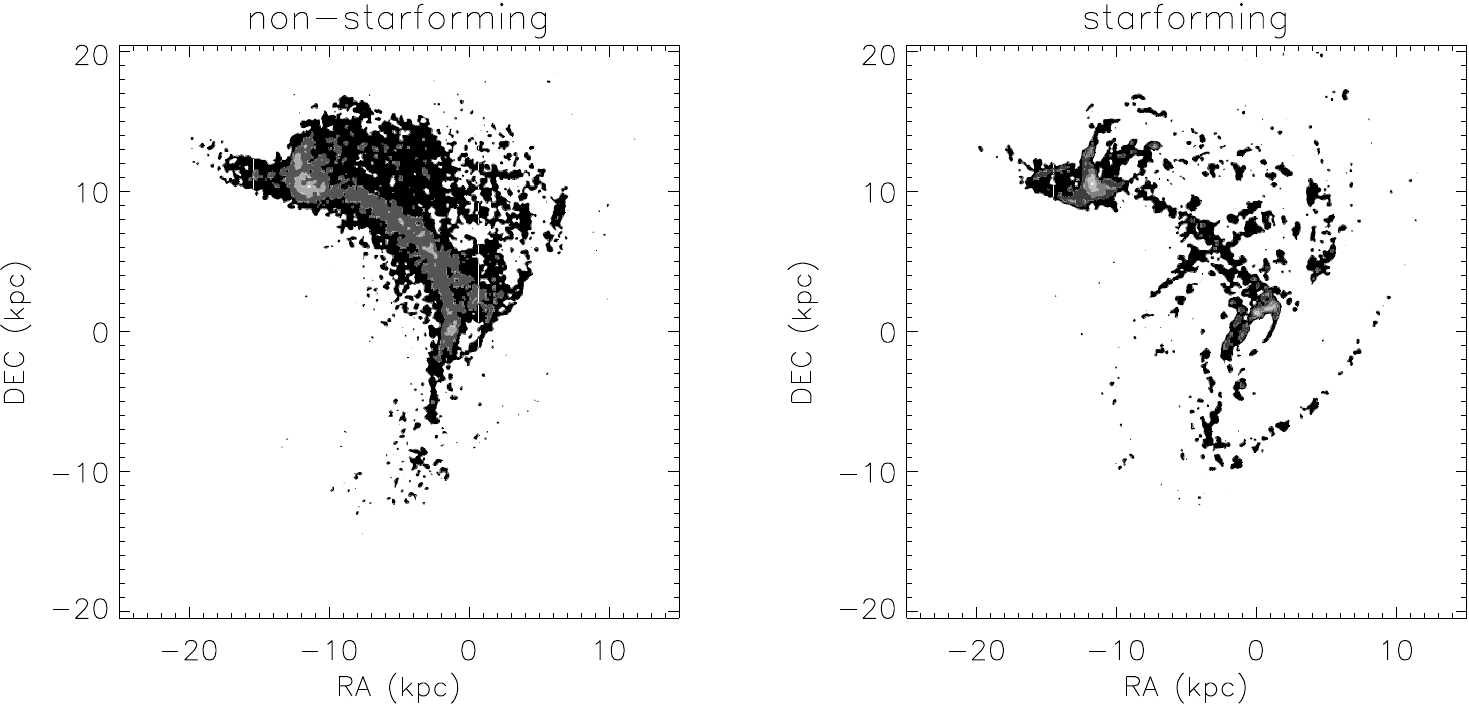}}
  \put(-400,35){\Large sim 19}
  \vspace{0cm}
  \resizebox{16cm}{!}{\includegraphics{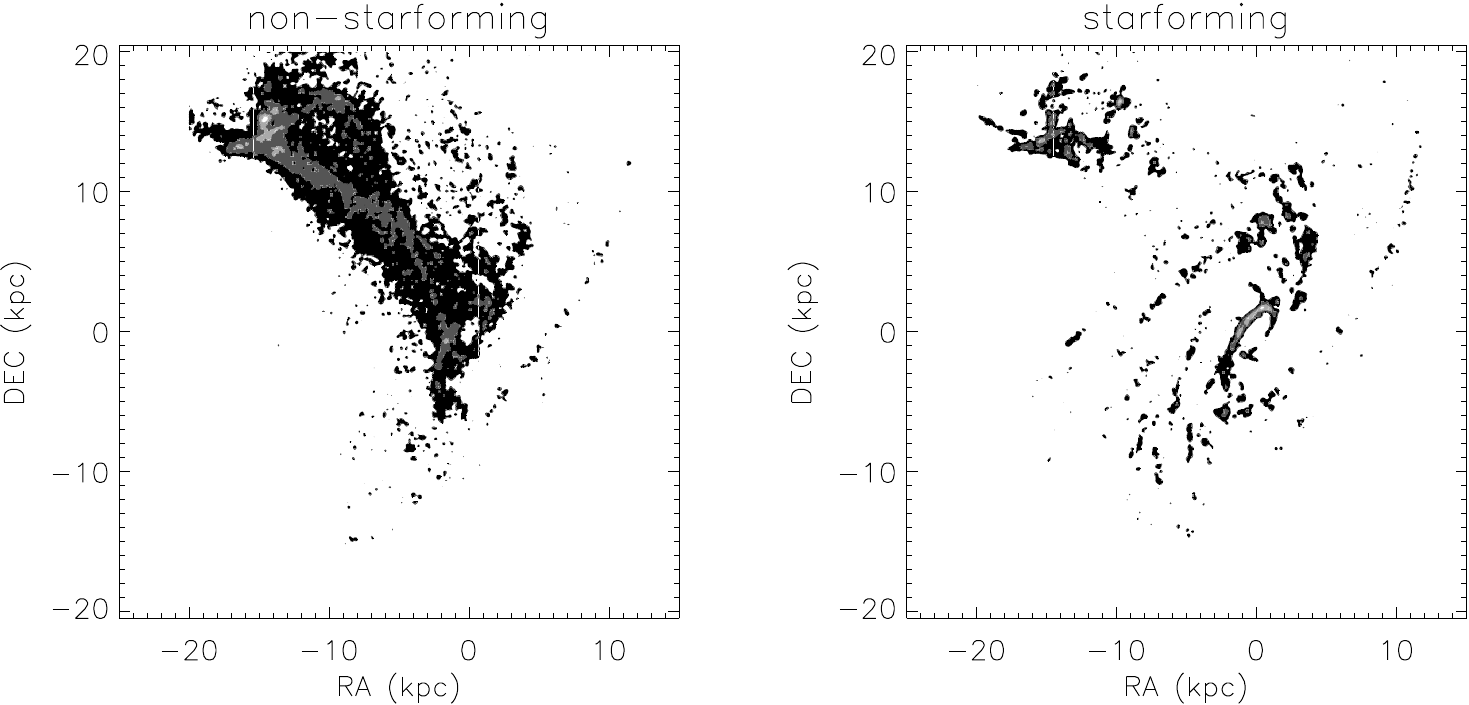}}
  \put(-400,35){\Large sim 19 fast}
  \vspace{0cm}
  \resizebox{16cm}{!}{\includegraphics{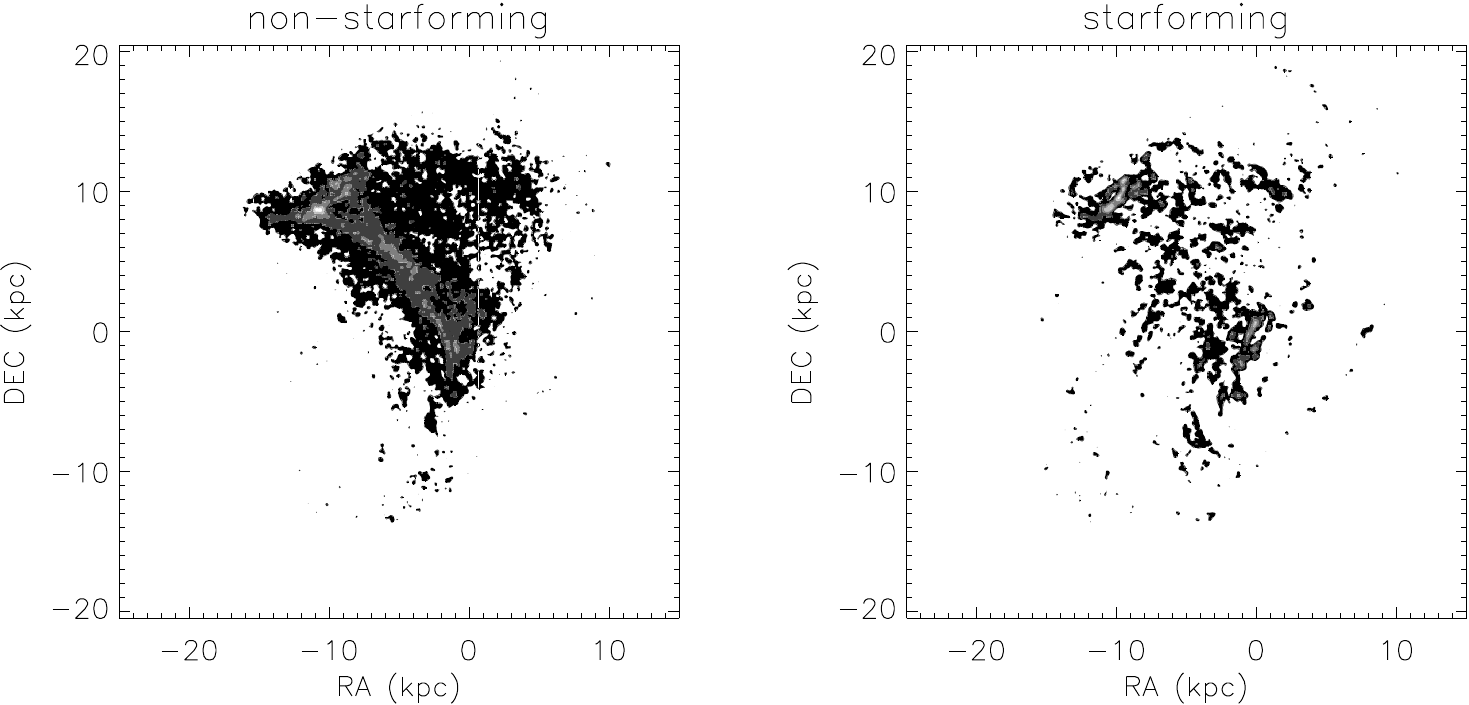}}
   \put(-400,35){\Large sim 20}
  \caption{Model maps of non-starforming and starforming gas.
  \label{fig:taffy22_gr}}
\end{figure*}

\begin{figure*}[!ht]
  \centering
  \resizebox{16cm}{!}{\includegraphics{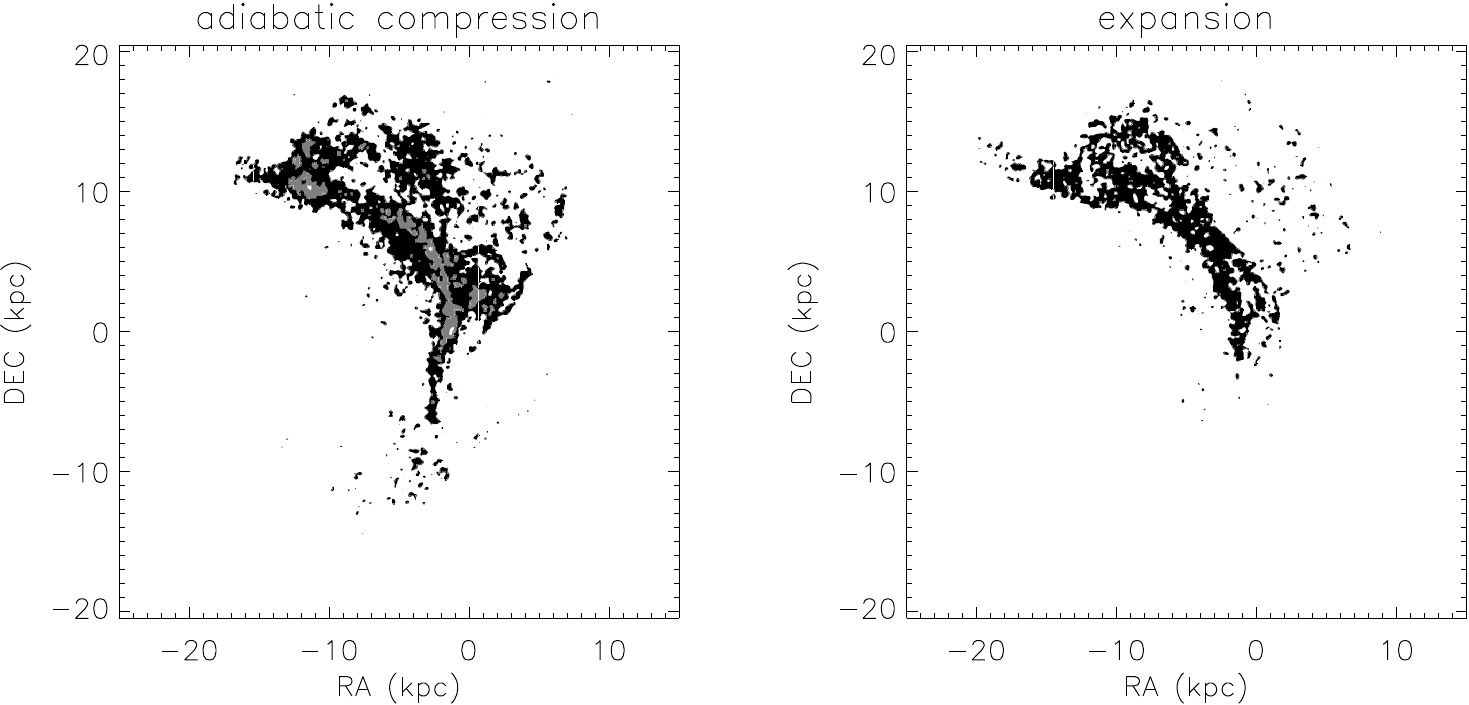}}
  \put(-400,35){\Large sim 19}
  \vspace{0cm}
  \resizebox{16cm}{!}{\includegraphics{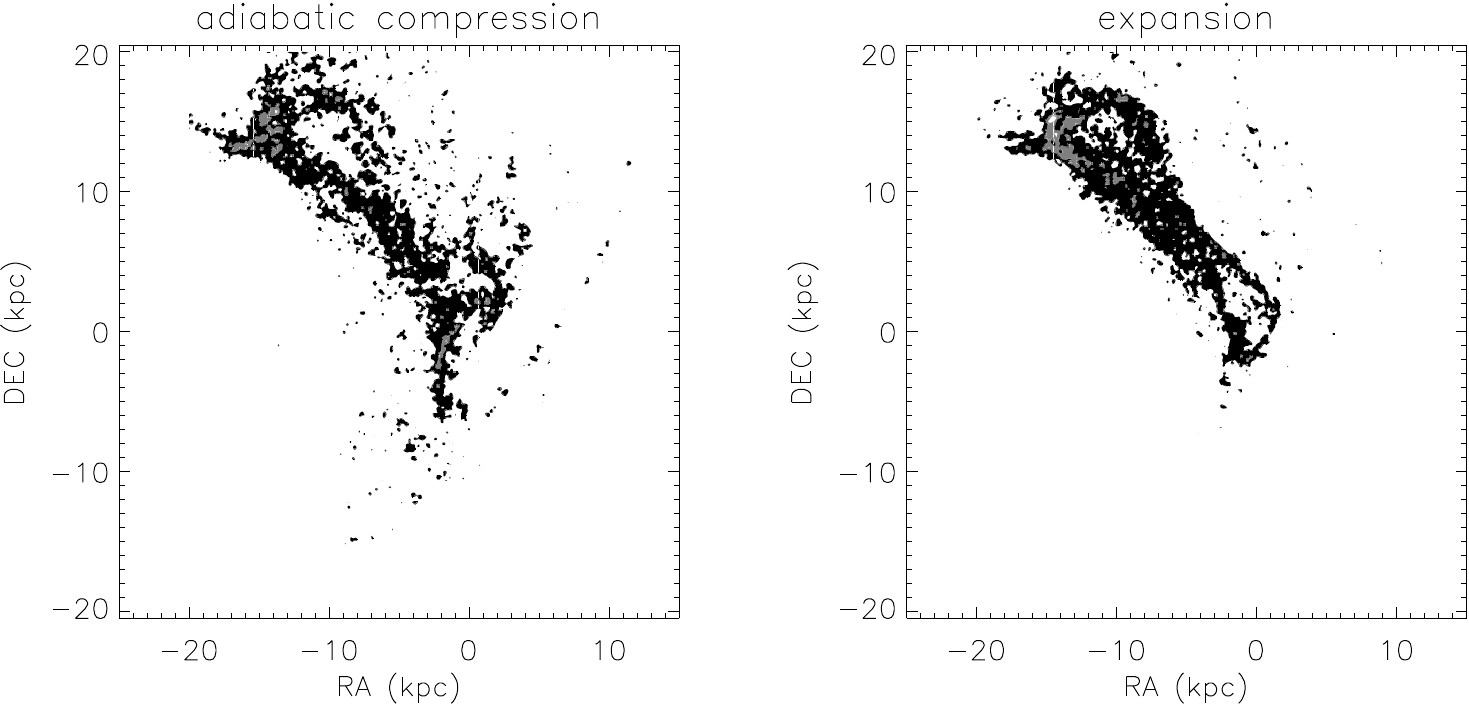}}
  \put(-400,35){\Large sim 19 fast}
  \vspace{0cm}
  \resizebox{16cm}{!}{\includegraphics{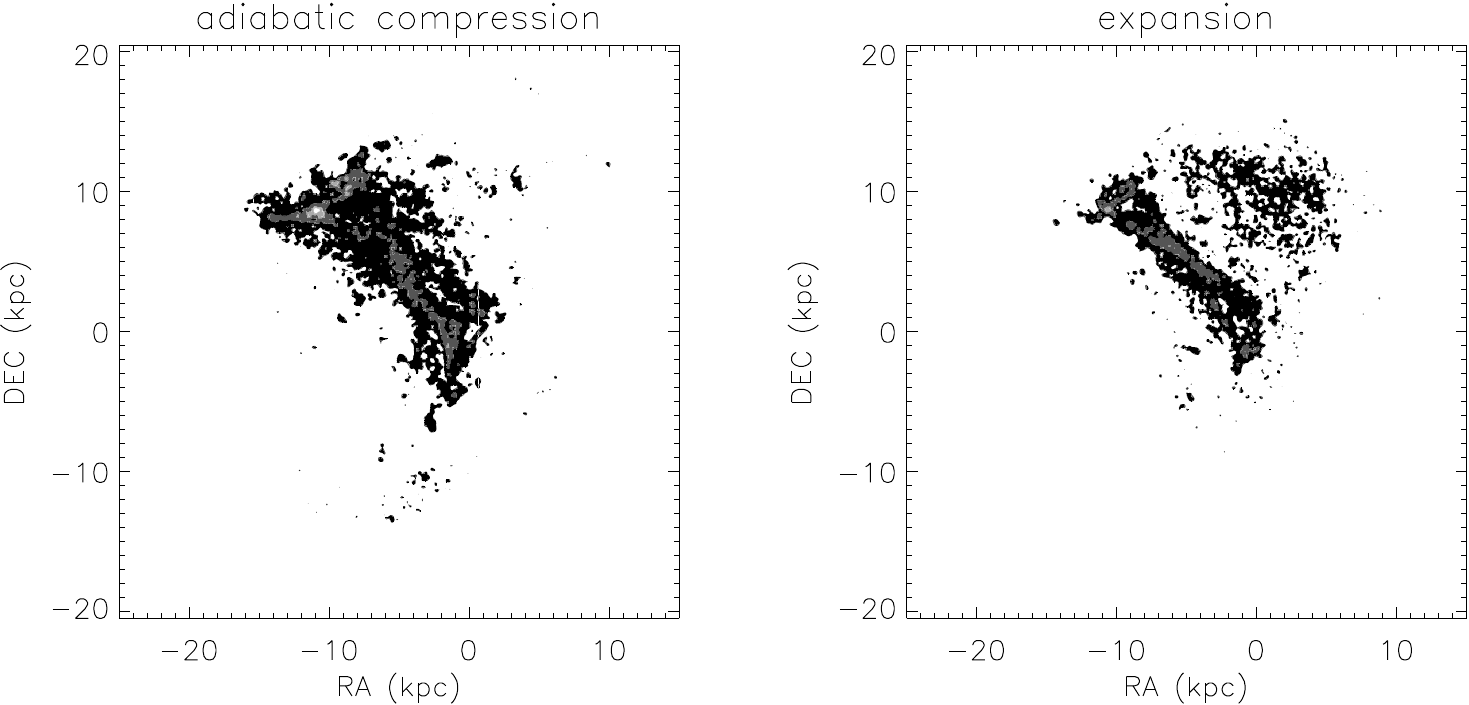}}
   \put(-400,35){\Large sim 20}
  \caption{Model maps of gas affected by turbulent adiabatic compression and rapid expansion.
  \label{fig:taffy22_gr1}}
\end{figure*}

To illustrate the effect of turbulent adiabatic compression we separated the star-forming and non-starforming gas particles
for the times of interest of the three simulations (Fig.~\ref{fig:taffy22_gr}). This shows that our conditions cleanly separate the
clouds in the disk and bridge regions. It is worth noting that turbulent adiabatic compression affects gas particles of all
volume and column densities.
In all models there are about $30$--$60$\,\% more particles affected by turbulent adiabatic compression than by rapid expansion.
The distributions of these particles for the models are shown in Fig.~\ref{fig:taffy22_gr1}.

\begin{figure*}[!ht]
  \centering
  \resizebox{\hsize}{!}{\includegraphics{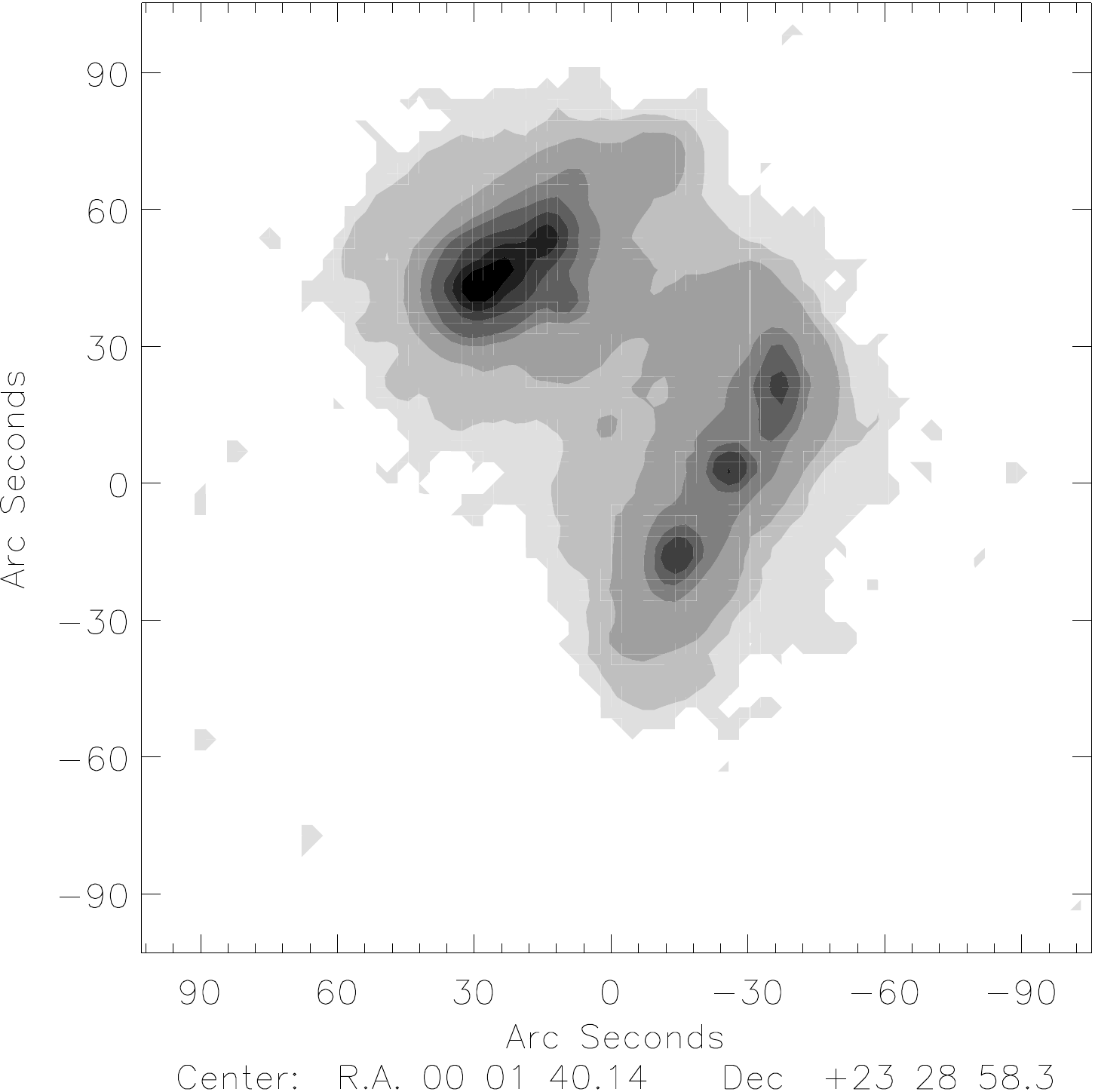}\includegraphics{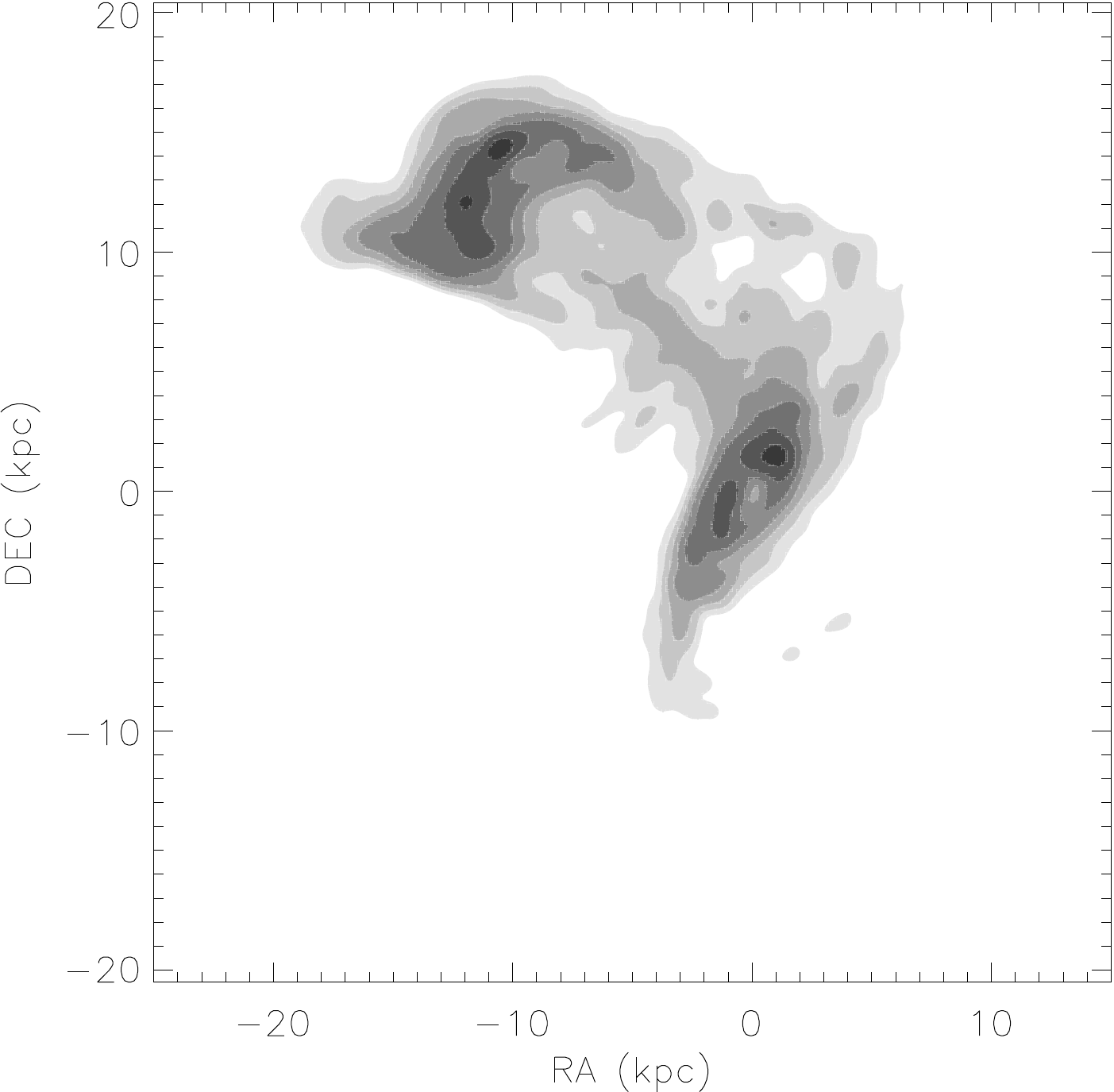}}
  \put(-480,45){\Large observations}
 \put(-215,35){\Large sim 19}
 \vspace{0cm}
  \resizebox{\hsize}{!}{\includegraphics{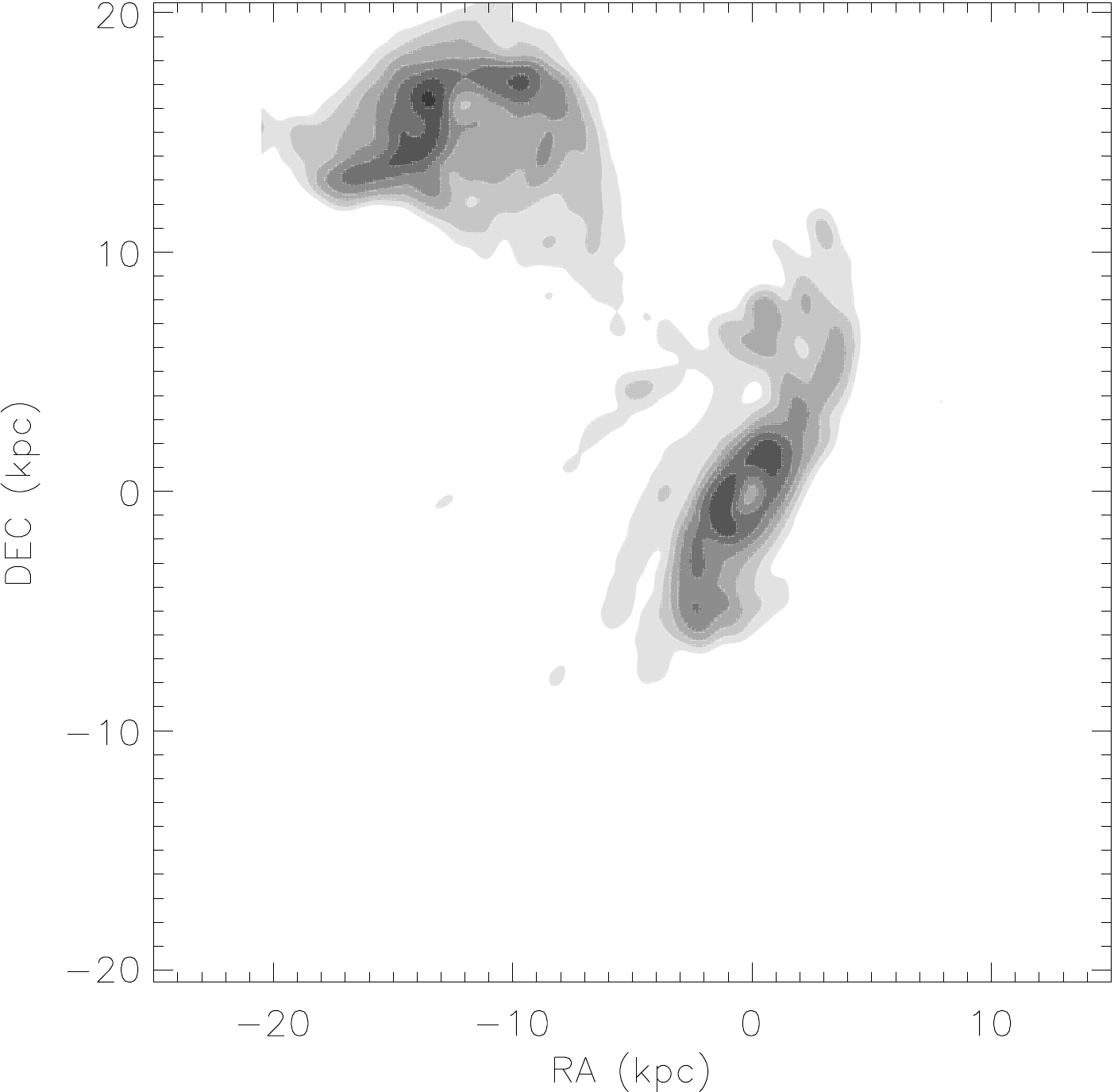}\includegraphics{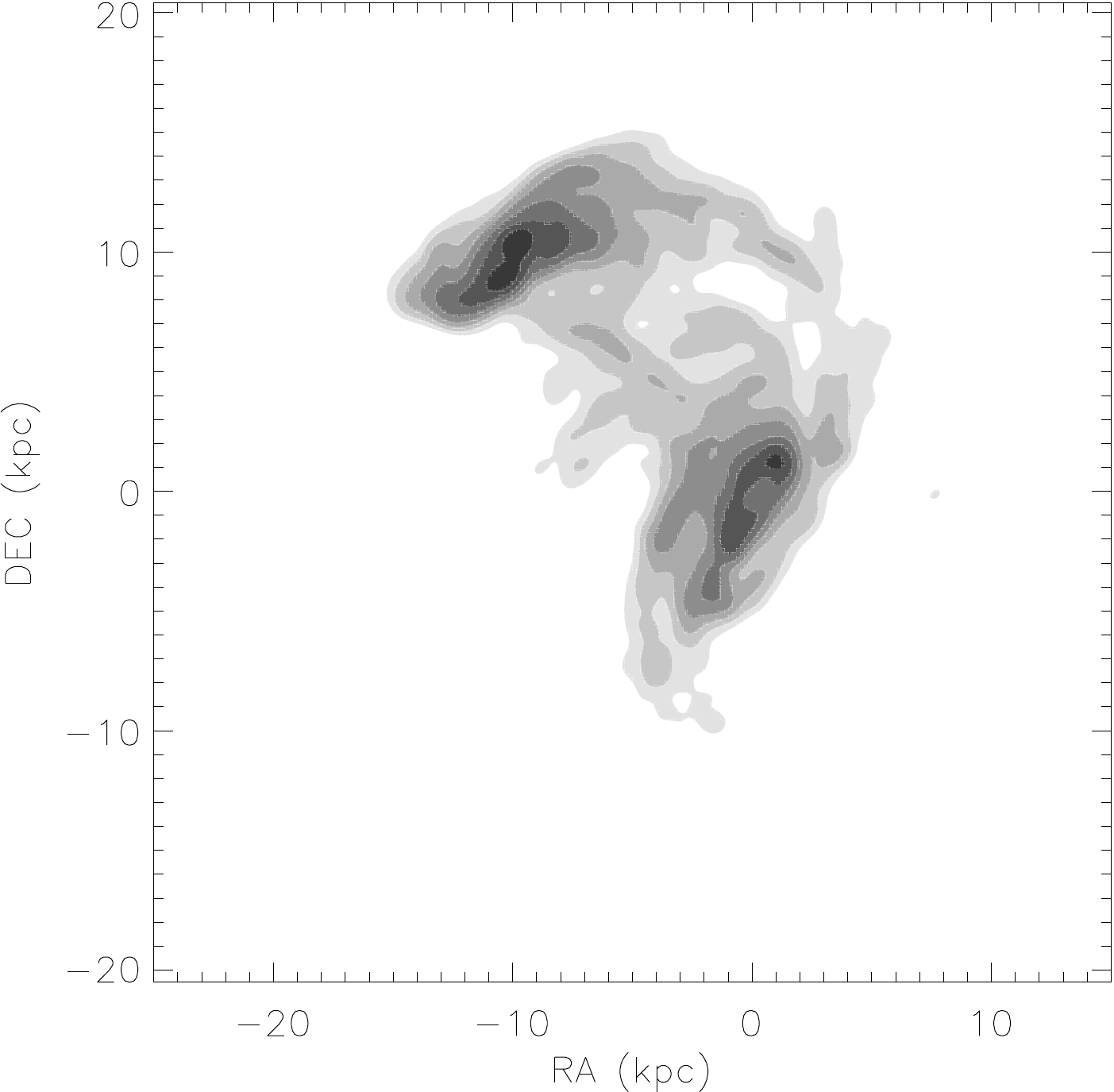}}
  \put(-480,35){\Large sim 19 fast}
  \put(-215,35){\Large sim 20}
  \caption{Observed star formation map based on Spitzer 24~$\mu$m and GALEX FUV maps together with the model star formation maps.
  \label{fig:taffy_sfr}}
\end{figure*}

\begin{figure*}[!ht]
  \centering
  \resizebox{\hsize}{!}{\includegraphics{taffy_sfr.pdf}\includegraphics{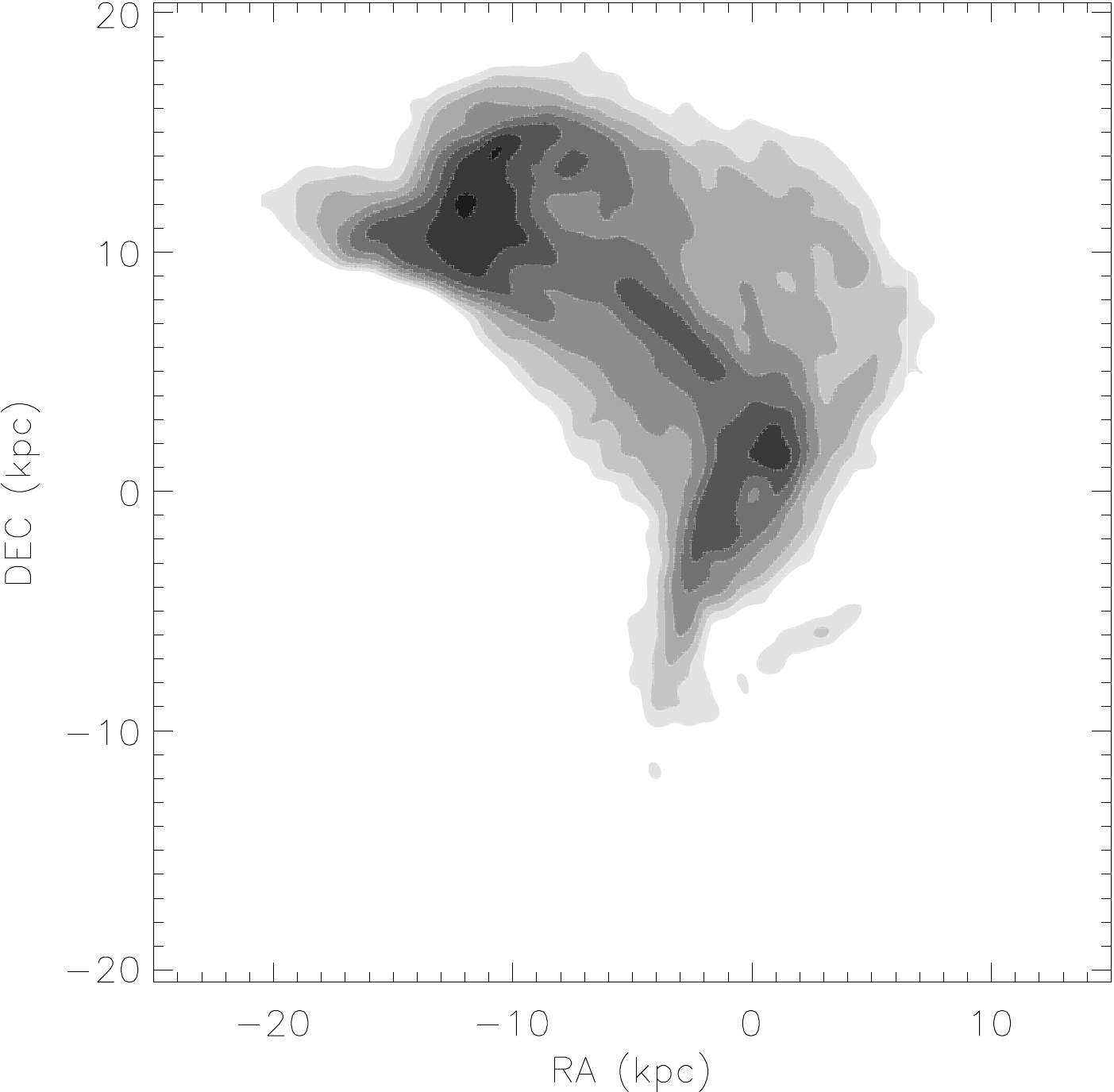}}
  \put(-480,45){\Large observations}
 \put(-215,35){\Large sim 19}
 \vspace{0cm}
  \resizebox{\hsize}{!}{\includegraphics{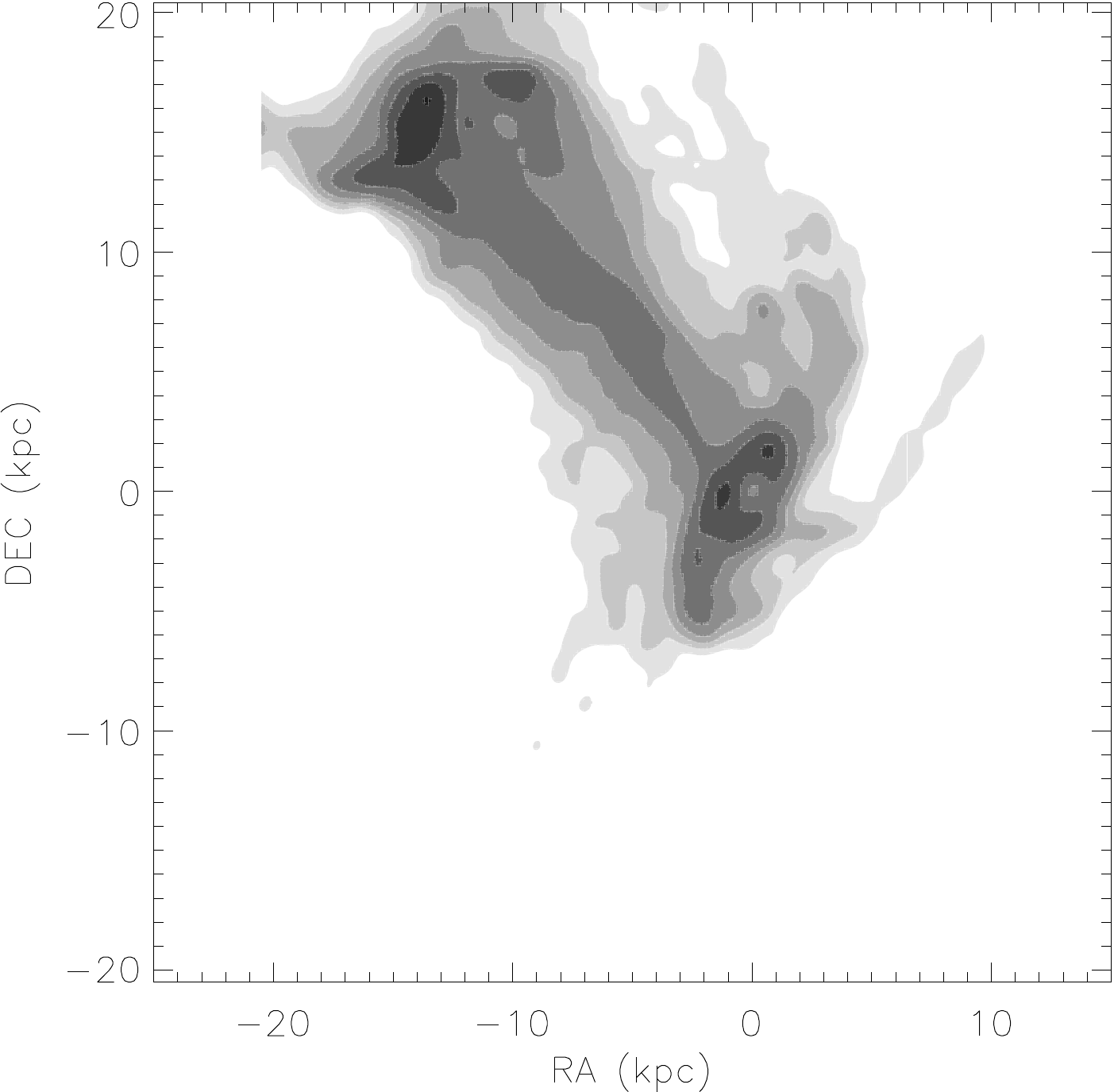}\includegraphics{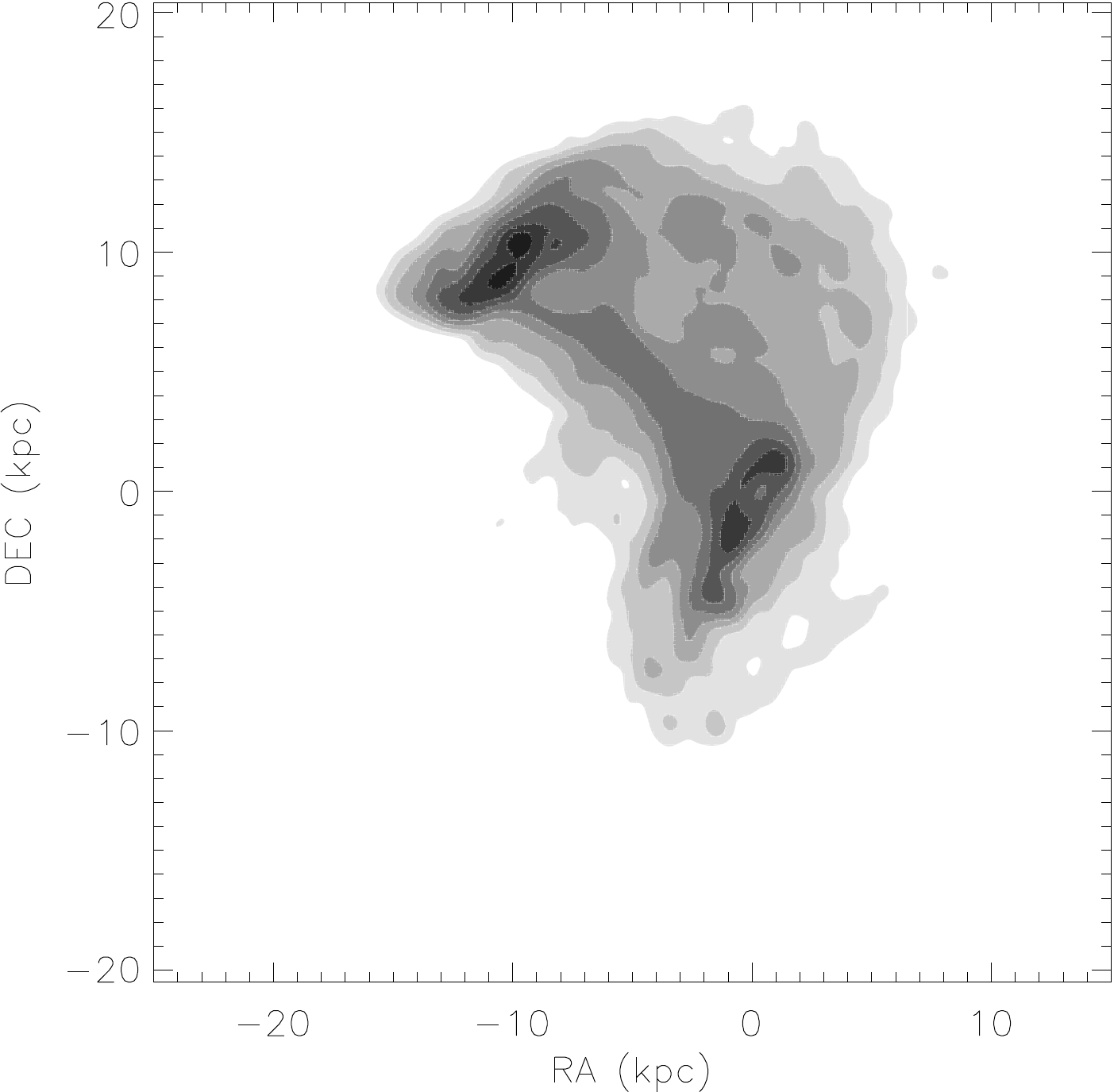}}
  \put(-480,35){\Large sim 19 fast}
  \put(-215,35){\Large sim 20}
  \caption{Observed star formation map based on Spitzer 24~$\mu$m and GALEX FUV maps together with the model star 
    formation maps without the suppression of star formation by turbulent adiabatic compression.
  \label{fig:taffy_sfr1}}
\end{figure*}

The resulting FUV emission maps based on the models including turbulent adiabatic compression together with the observed GALEX FUV map are 
presented in Fig.~\ref{fig:taffy_sfr}. As expected, the morphology of the FUV emission is very similar to that of the Spitzer $8$~$\mu$m emission
(Fig.~\ref{fig:taffy_spitzer}).
The corresponding maps from the models without turbulent adiabatic compression are shown in 
Fig.~\ref{fig:taffy_sfr1}. The GALEX FUV image does not show structures whose morphology resembles that of the CO emission with the
exception of the compact star formation region close to UGC~12915. The FUV images of sim19 and sim20 still show some trace of the dense
bridge gas. Overall, sim19fast most resembles the GALEX UV image: the emission UGC~12914 and the bridge region are well-reproduced.
However, as for the gas distribution, the model northern bridge filament is not present in the observations.

\clearpage

\begin{figure}[!ht]
  \centering
  \resizebox{\hsize}{!}{\includegraphics{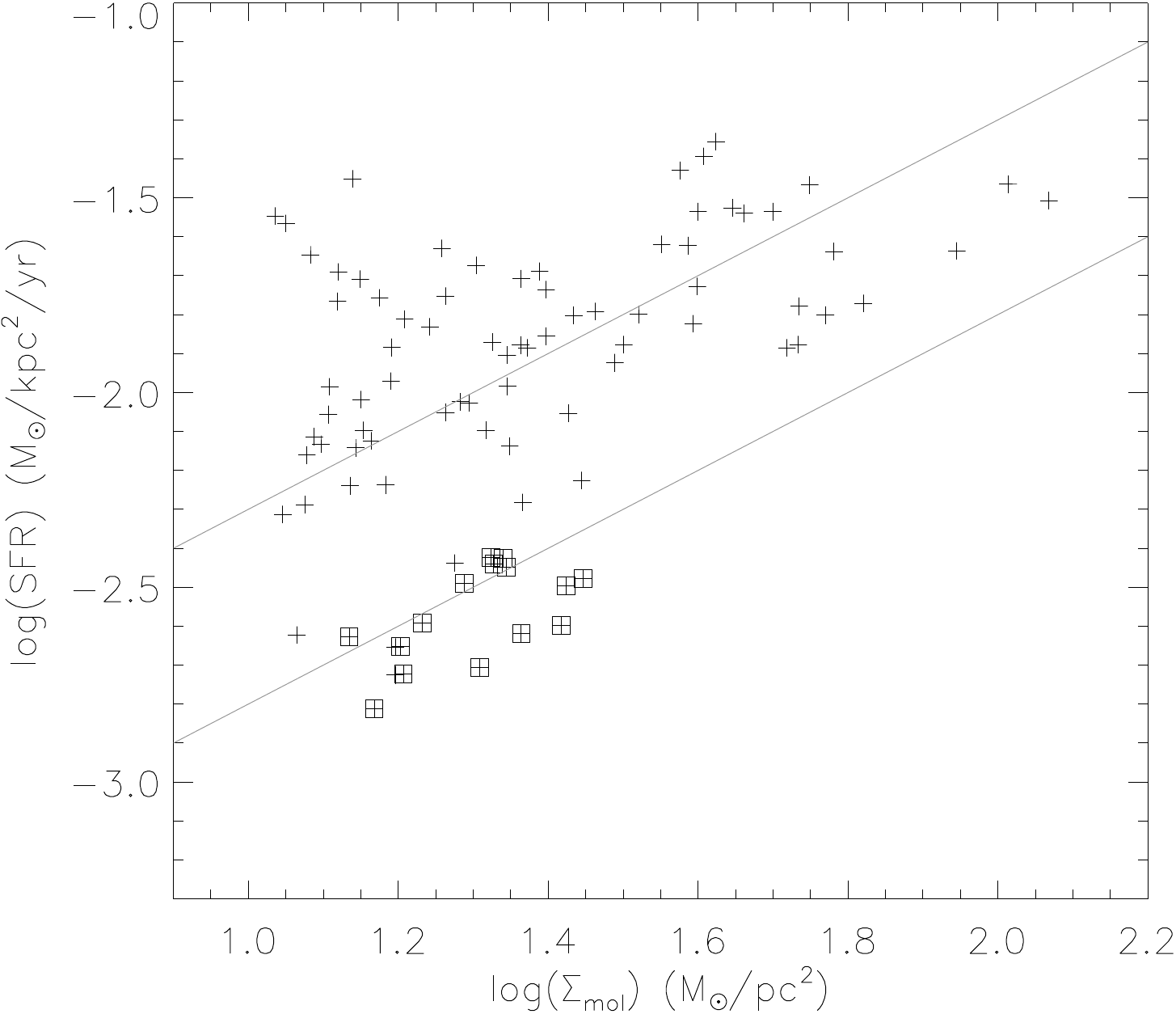}}
 \put(-80,35){\Large sim 19}
 \vspace{0cm}
  \resizebox{\hsize}{!}{\includegraphics{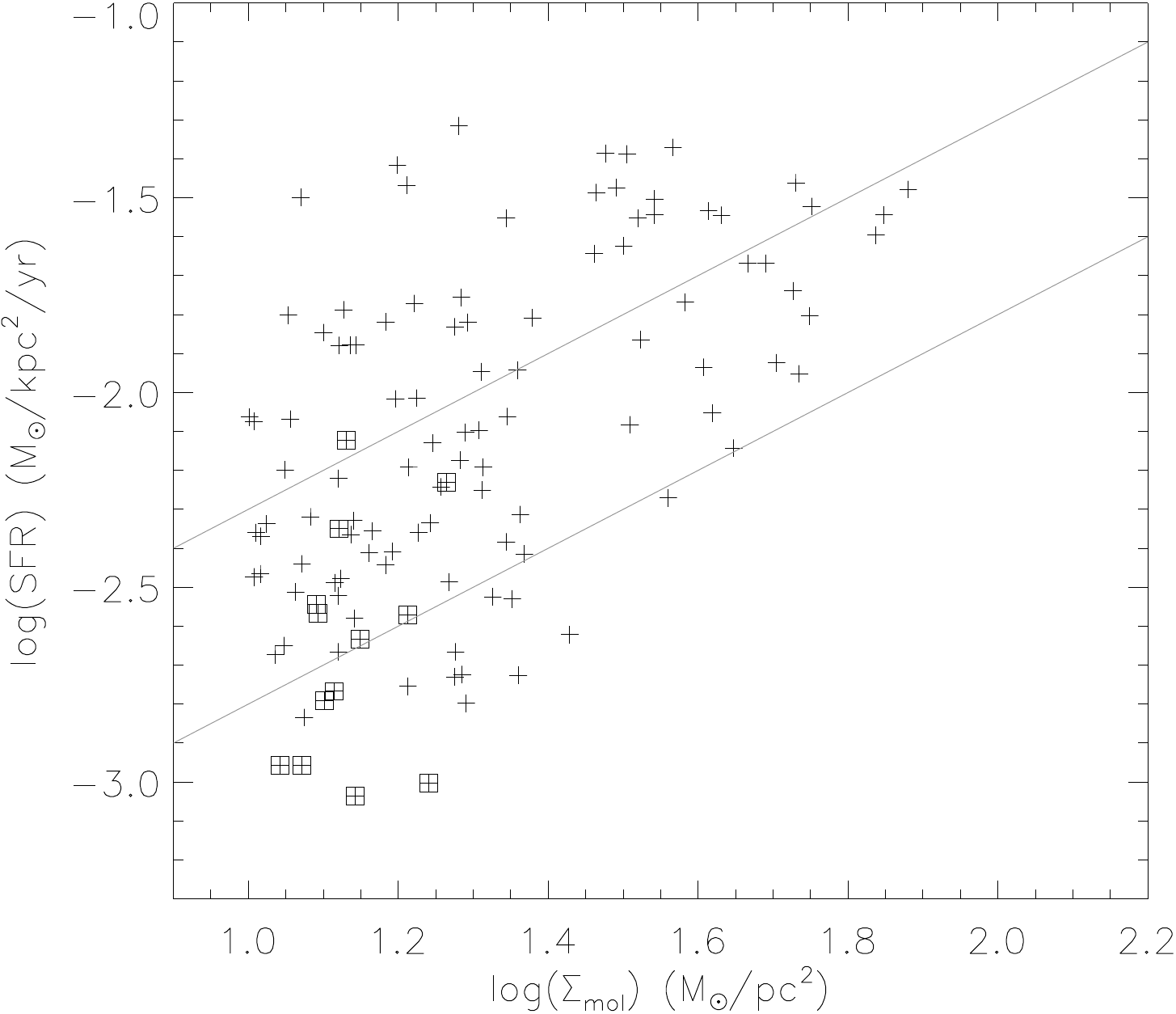}}
  \put(-80,35){\Large sim 19 fast}
  \vspace{0cm}
  \resizebox{\hsize}{!}{\includegraphics{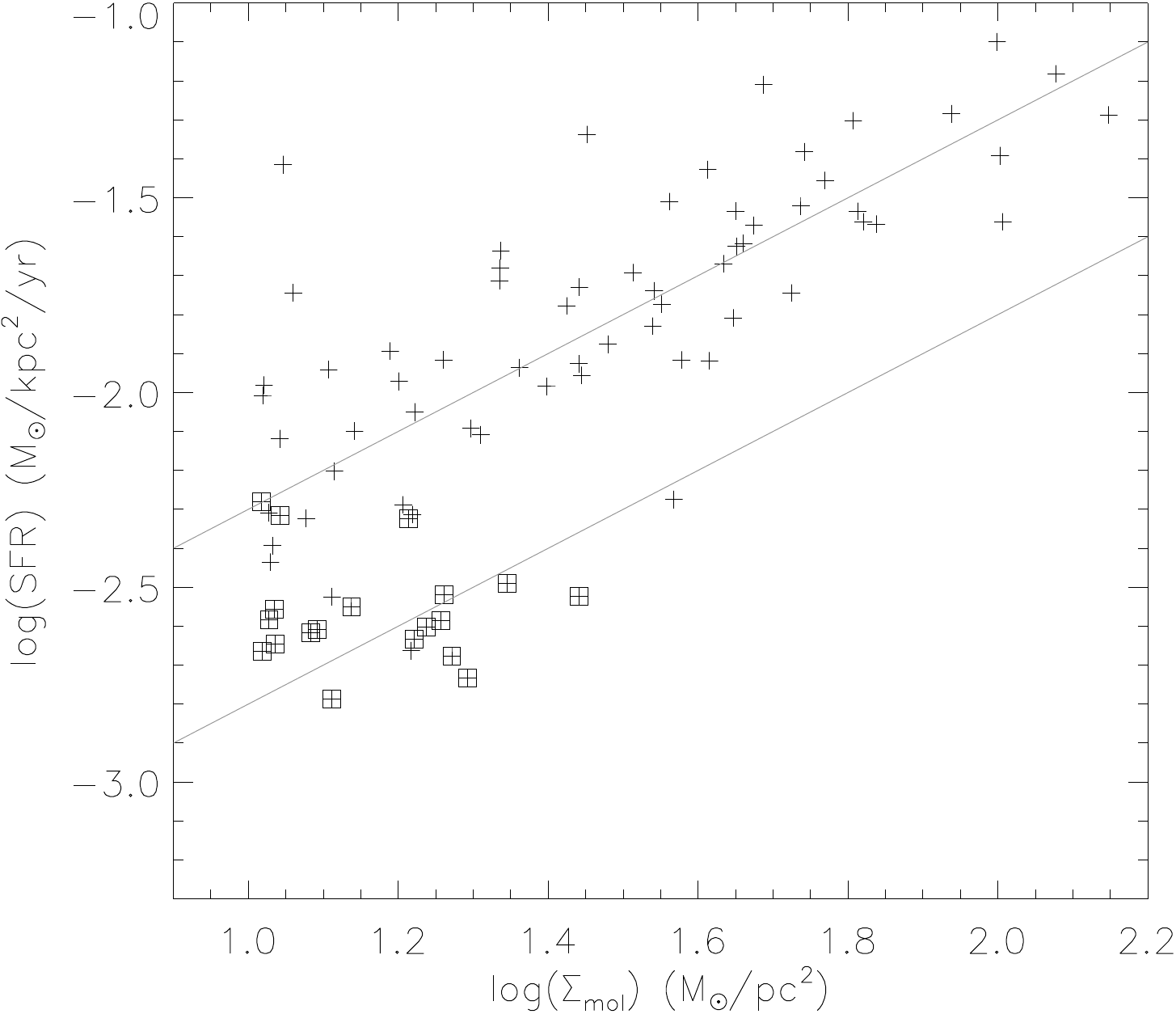}}
  \put(-80,35){\Large sim 20}
  \caption{Star formation as a function of the molecular gas surface density at a spatial resolution of $10''$.
  The resolution elements of the bridge region are marked with boxes. The solid blue lines mark molecular depletion times of
  $1.6$ and $5$~Gyr.
  \label{fig:KSlaw}}
\end{figure}

\begin{figure}[!ht]
  \centering
  \resizebox{\hsize}{!}{\includegraphics{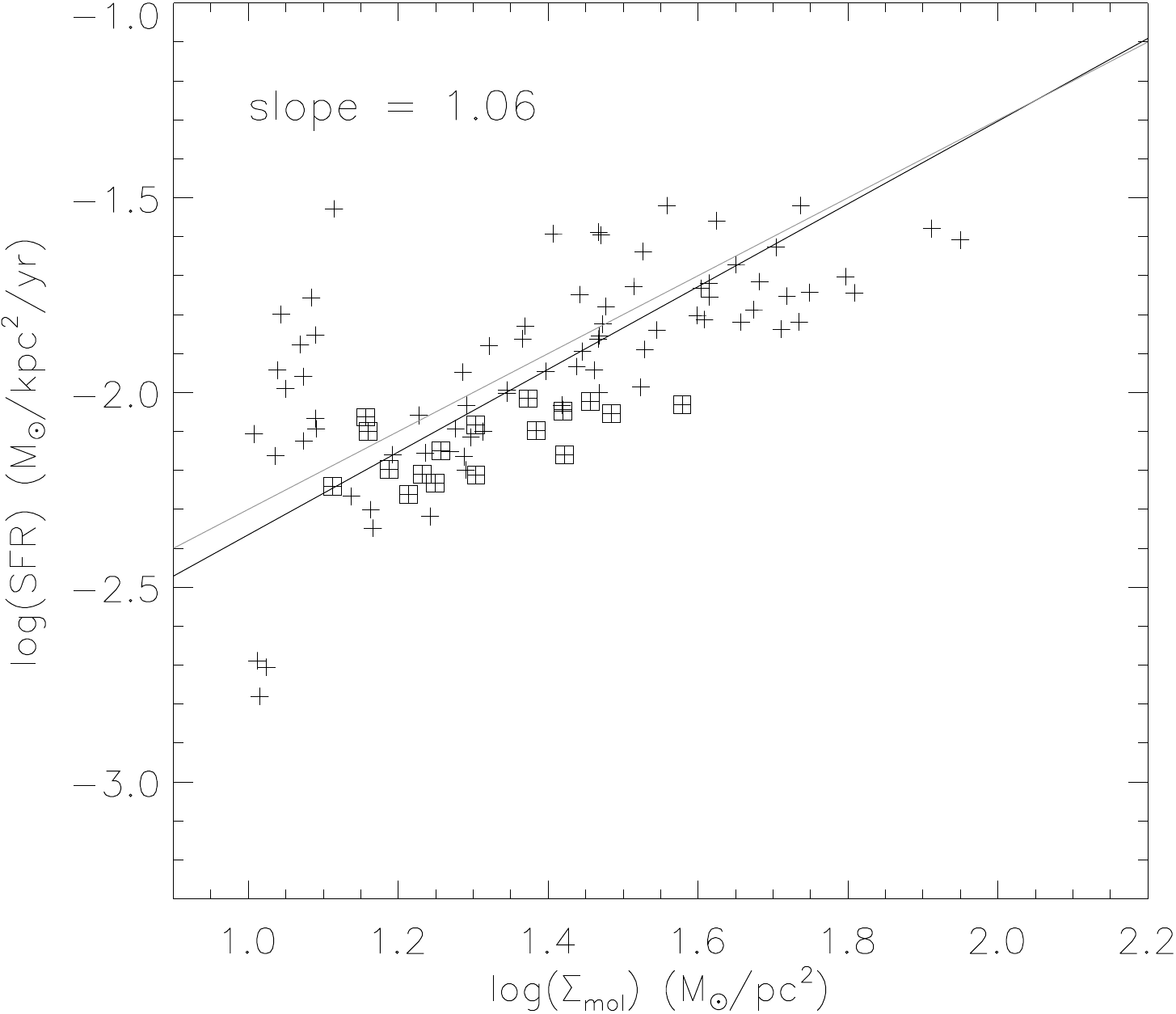}}
 \put(-80,35){\Large sim 19}
 \vspace{0cm}
  \resizebox{\hsize}{!}{\includegraphics{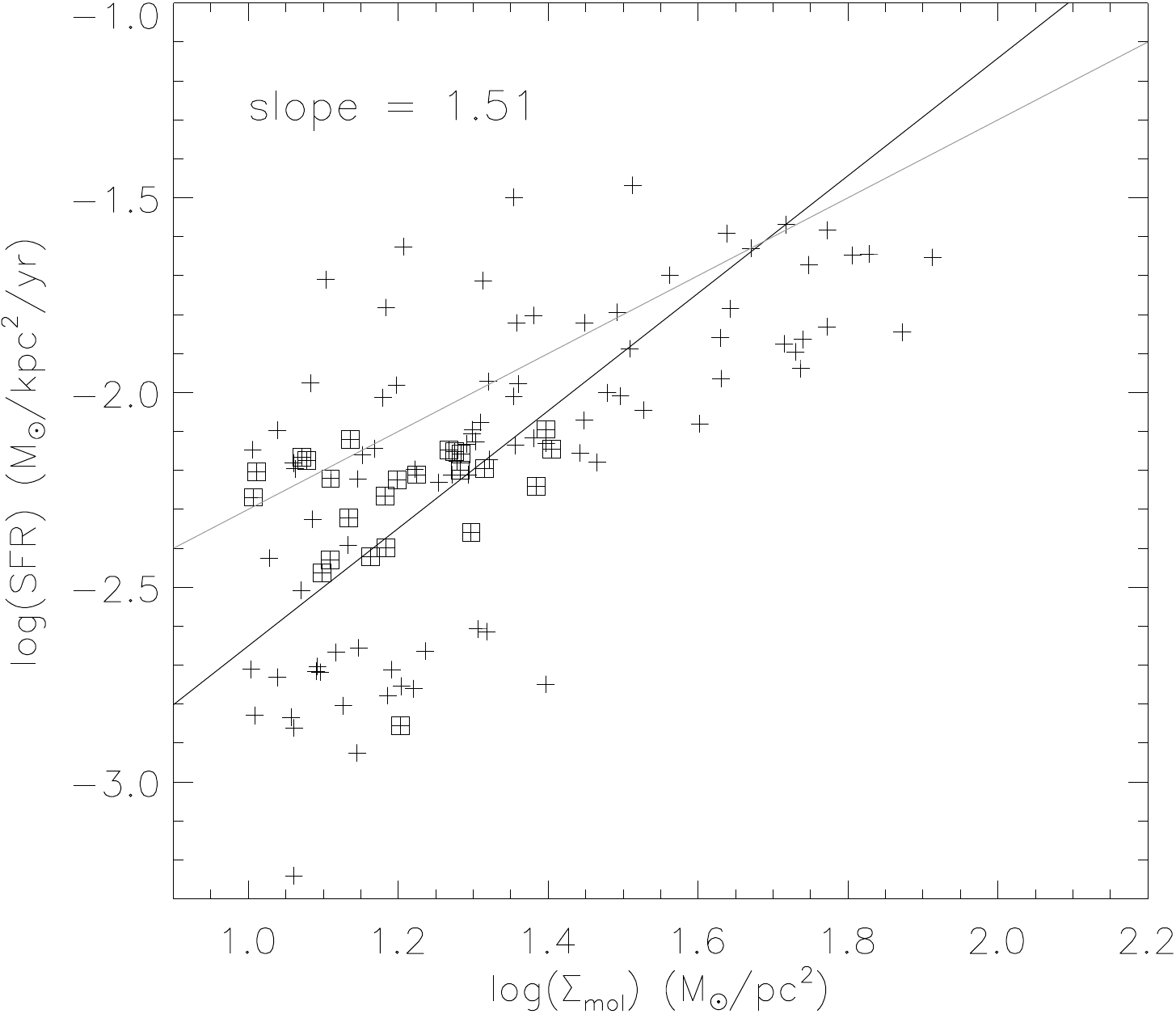}}
  \put(-80,35){\Large sim 19 fast}
  \vspace{0cm}
  \resizebox{\hsize}{!}{\includegraphics{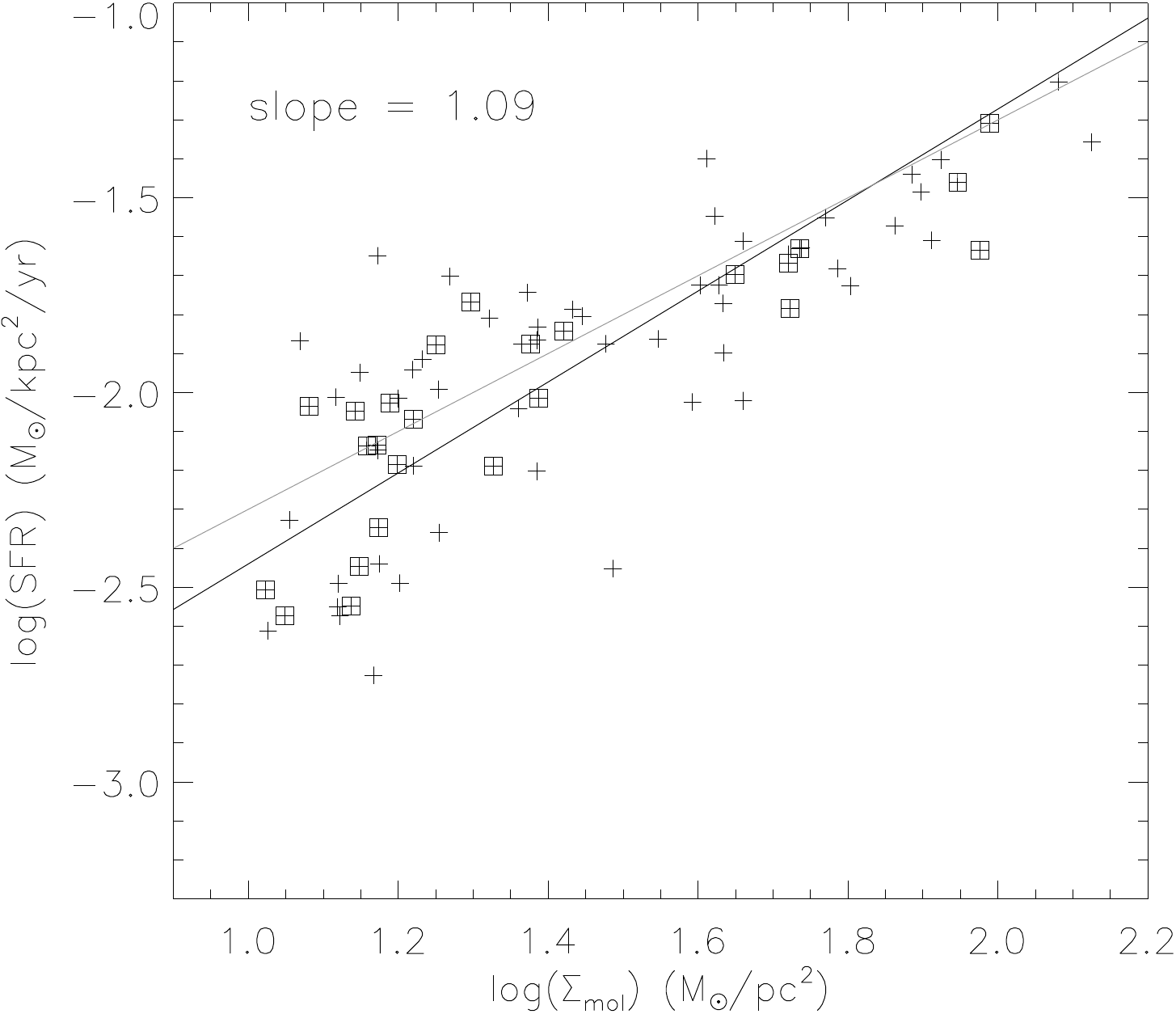}}
  \put(-80,35){\Large sim 20}
  \caption{Same as Fig.~\ref{fig:KSlaw} but for the models without adiabatic gas compression.
    A robust linear bisector fit is shown as a black solid line. 
  \label{fig:KSlaw1}}
\end{figure}

The relations between the model star formation rate and the molecular gas surface density of the three models including adiabatic gas compression
are presented in Fig.~\ref{fig:KSlaw}. Fig.~\ref{fig:KSlaw1} shows the star formation efficiency (SFE=SFR/M$_{\rm H_2}$) of
models~19 and 20 without adiabatic gas compression. The SFE is approximately constant and the gas located
in the bridge has only a marginally lower ($0.1$~dex) SFE than the disks.
In the models with adiabatic gas compression the bridge SFE is $\sim 3$ times lower whereas the disk SFE remains the same.
This is comparable to the observed decrease of the star formation efficiency in the Taffy bridge
region (Fig.~21 of Vollmer et al. 2012).


\begin{thebibliography}{}


\bibitem[Appleton et al.(2015)]{2015ApJ...812..118A} Appleton, P.~N., Lanz, L., Bitsakis, T., et al.\ 2015, ApJ, 812, 118

\bibitem[Appleton et al.(2017)]{2017ApJ...836...76A} Appleton, P.~N., Guillard, P., Togi, A., et al.\ 2017, ApJ, 836, 76

\bibitem[Braine et al.(2003)]{2003A&A...408L..13B} Braine, J., Davoust, E., Zhu, M., et al.\ 2003, A\&A, 408, L13 


\bibitem[Combes \& Gerin(1985)]{1985A&A...150..327C} Combes, F. \& Gerin, M.\ 1985, A\&A, 150, 327

\bibitem[Condon et al.(1993)]{1993AJ....105.1730C} Condon, J.~J., Helou, G., Sanders, D.~B., \& Soifer, B.~T.\ 1993, AJ, 105, 1730 


\bibitem[Di Matteo et al.(2007)]{2007A&A...468...61D} Di Matteo, P., Combes, F., Melchior, A.-L., et al.\ 2007, A\&A, 468, 61


\bibitem[Elmegreen \& Falgarone(1996)]{1996ApJ...471..816E} Elmegreen, B.~G., \& Falgarone, E.\ 1996, ApJ, 471, 816

\bibitem[Elmegreen et al.(2003)]{2003ApJ...590..271E} Elmegreen, B.~G., Elmegreen, D.~M., \& Leitner, S.~N.\ 2003, ApJ, 590, 271


\bibitem[Faucher-Gigu{\`e}re et al.(2013)]{2013MNRAS.433.1970F} Faucher-Gigu{\`e}re, C.-A., Quataert, E., \& Hopkins, P.~F.\ 2013, MNRAS, 433, 1970

\bibitem[Federrath \& Klessen(2012)]{2012ApJ...761..156F} Federrath, C. \& Klessen, R.~S.\ 2012, ApJ, 761, 156


\bibitem[Gao et al.(2003)]{2003AJ....126.2171G} Gao, Y., Zhu, M., \& Seaquist, E.~R.\ 2003, AJ, 126, 2171


\bibitem[Gilbert \& Graham(2007)]{2007ApJ...668..168G} Gilbert, A.~M., \& Graham, J.~R.\ 2007, ApJ, 668, 168


\bibitem[Harwit et al.(1987)]{1987ApJ...315...28H} Harwit, M., Houck, J.~R., Soifer, B.~T., et al.\ 1987, ApJ, 315, 28

\bibitem[Heyer et al.(2009)]{2009ApJ...699.1092H} Heyer, M., Krawczyk, C., Duval, J., et al.\ 2009, ApJ, 699, 1092

\bibitem[Johnson et al.(2015)]{2015ApJ...806...35J} Johnson, K.~E., Leroy, A.~K., Indebetouw, R., et al.\ 2015, ApJ, 806, 35

\bibitem[Joshi et al.(2019)]{2019ApJ...878..161J} Joshi, B.~A., Appleton, P.~N., Blanc, G.~A., et al.\ 2019, ApJ, 878, 161


\bibitem[Krumholz \& McKee(2005)]{2005ApJ...630..250K} Krumholz, M.~R. \& McKee, C.~F.\ 2005, ApJ, 630, 250

\bibitem[Krumholz et al.(2012)]{2012ApJ...745...69K} Krumholz, M.~R., Dekel, A., \& McKee, C.~F.\ 2012, ApJ, 745, 69

\bibitem[Krumholz et al.(2018)]{2018MNRAS.477.2716K} Krumholz, M.~R., Burkhart, B., Forbes, J.~C., et al.\ 2018, MNRAS, 477, 2716

\bibitem[Leitherer et al.(1999)]{1999ApJS..123....3L} Leitherer, C., Schaerer, D., Goldader, J.~D., et al.\ 1999, ApJS, 123, 3

\bibitem[Leroy et al.(2008)]{2008AJ....136.2782L} Leroy, A.~K., Walter, F., Brinks, E., et al.\ 2008, AJ, 136, 2782

\bibitem[Leroy et al.(2017)]{2017ApJ...846...71L} Leroy, A.~K., Schinnerer, E., Hughes, A., et al.\ 2017, ApJ, 846, 71

\bibitem[Mac Low \& Klessen(2004)]{2004RvMP...76..125M} Mac Low, M.-M., \& Klessen, R.~S.\ 2004, Reviews of Modern Physics, 76, 125

\bibitem[Mandal et al.(2020)]{2020MNRAS.493.3098M} Mandal, A., Federrath, C., \& K{\"o}rtgen, B.\ 2020, MNRAS, 493, 3098

\bibitem[Meidt et al.(2015)]{2015ApJ...806...72M} Meidt, S.~E., Hughes, A., Dobbs, C.~L., et al.\ 2015, ApJ, 806, 72.


\bibitem[Oka et al.(1998)]{1998ApJ...493..730O} Oka, T., Hasegawa, T., Hayashi, M., et al.\ 1998, ApJ, 493, 730

\bibitem[Oka et al.(2001)]{2001ApJ...562..348O} Oka, T., Hasegawa, T., Sato, F., et al.\ 2001, ApJ, 562, 348

\bibitem[O'Connell et al.(1994)]{1994ApJ...433...65O} O'Connell, R.~W., Gallagher, J.~S., \& Hunter, D.~A.\ 1994, ApJ, 433, 65

\bibitem[Ostriker \& Shetty(2011)]{2011ApJ...731...41O} Ostriker, E.~C. \& Shetty, R.\ 2011, ApJ, 731, 41

\bibitem[Peterson et al.(2012)]{2012arXiv1203.4203P} Peterson, B.~W., Appleton, P.~N., Helou, G., et al.\ 2012, ApJ, 751, 11

\bibitem[Padoan et al.(2012)]{2012ApJ...759L..27P} Padoan, P., Haugb{\o}lle, T., \& Nordlund, {\r{A}}.\ 2012, ApJL, 759, L27

\bibitem[Padoan et al.(2017)]{2017ApJ...840...48P} Padoan, P., Haugb{\o}lle, T., Nordlund, {\r{A}}., et al.\ 2017, ApJ, 840, 48

\bibitem[Peterson et al.(2018)]{2018ApJ...855..141P} Peterson, B.~W., Appleton, P.~N., Bitsakis, T., et al.\ 2018, ApJ, 855, 141


\bibitem[Renaud et al.(2015)]{2015MNRAS.446.2038R} Renaud, F., Bournaud, F., \& Duc, P.-A.\ 2015, MNRAS, 446, 2038

\bibitem[Robertson \& Goldreich(2012)]{2012ApJ...750L..31R} Robertson, B. \& Goldreich, P.\ 2012, ApJL, 750, L31

\bibitem[Rosolowsky, \& Leroy(2006)]{2006PASP..118..590R} Rosolowsky, E., \& Leroy, A.\ 2006, PASP, 118, 590


\bibitem[Springel et al.(2001)]{2001NewA....6...79S} Springel, V., Yoshida, N., \& White, S.~D.~M.\ 2001, NewA, 6, 79 


\bibitem[Vollmer \& Beckert(2003)]{2003A&A...404...21V} Vollmer, B., \& Beckert, T.\ 2003, A\&A, 404, 21

\bibitem[Vollmer et al.(2005)]{2005A&A...441..473V} Vollmer, B., Braine, J., Combes, F., \& Sofue, Y.\ 2005, A\&A, 441, 473

\bibitem[Vollmer et al.(2009)]{2009A&A...496..669V} Vollmer, B., Soida, M., Chung, A., et al.\ 2009, A\&A, 496, 669

\bibitem[Vollmer, \& Leroy(2011)]{2011AJ....141...24V} Vollmer, B., \& Leroy, A.~K.\ 2011, AJ, 141, 24

\bibitem[Vollmer et al.(2012)]{2012A&A...547A..39V} Vollmer, B., Braine, J., \& Soida, M.\ 2012, A\&A, 547, A39

\bibitem[Vollmer et al.(2012a)]{2012A&A...537A.143V} Vollmer, B., Soida, M., Braine, J., et al.\ 2012a, A\&A, 537, A143 

\bibitem[Vollmer et al.(2017)]{2017A&A...602A..51V} Vollmer, B., Gratier, P., Braine, J., et al.\ 2017, A\&A, 602, A51


\bibitem[Wiegel 1994]{q1} Wiegel W. 1994, Diploma Thesis, University of Heidelberg

\bibitem[Williams et al.(1994)]{1994ApJ...428..693W} Williams, J.~P., de Geus, E.~J., \& Blitz, L.\ 1994, ApJ, 428, 693

\bibitem[Wolfire et al.(2003)]{2003ApJ...587..278W} Wolfire, M.~G., McKee, C.~F., Hollenbach, D., et al.\ 2003, ApJ, 587, 278

\bibitem[Yeager, \& Struck(2019)]{2019MNRAS.486.2660Y} Yeager, T.~R., \& Struck, C.\ 2019, MNRAS, 486, 2660

\bibitem[Yeager \& Struck(2020)]{2020MNRAS.492.4892Y} Yeager, T.~R. \& Struck, C.\ 2020, MNRAS, 492, 4892

\bibitem[Zamora-Avil{\'e}s \& V{\'a}zquez-Semadeni(2014)]{2014ApJ...793...84Z} Zamora-Avil{\'e}s, M. \& V{\'a}zquez-Semadeni, E.\ 2014, ApJ, 793, 84

\bibitem[Zhu et al.(2007)]{2007AJ....134..118Z} Zhu, M., Gao, Y., Seaquist, E.~R., \& Dunne, L.\ 2007, AJ, 134, 118 

\bibitem[Zink et al.(2000)]{2000ApJS..131..413Z} Zink, E.~C., Lester, D.~F., Doppmann, G., \& Harvey, P.~M.\ 2000, ApJS, 131, 413 


\end{thebibliography}
\end{document}